\documentclass[floats,floatfix, amssymb, prd,nofootinbib,superscriptaddress]{revtex4-2}

\usepackage{amssymb,amsmath,verbatim,mathtools,needspace,enumitem,etoolbox,graphicx,physics,microtype,afterpage,xspace,tabularx,lmodern,multirow}
\usepackage{subcaption} 
\usepackage{gensymb}
\usepackage[dvipsnames, usenames]{xcolor}
\usepackage[unicode, colorlinks=true, linkcolor=linkcolor, citecolor=linkcolor, filecolor=linkcolor, urlcolor=linkcolor, linktocpage, breaklinks]{hyperref}
\usepackage[nolist,nohyperlinks]{acronym}
\usepackage{tikz}

\newcommand{\orcid}[1]{\href{https://orcid.org/#1}{\includegraphics[width=10pt]{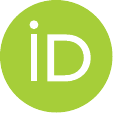}}}

\def\i{{\rm i}}

\definecolor{linkcolor}{rgb}{0.6, 0, 0.5}

\usetikzlibrary{decorations.markings}
\usetikzlibrary{decorations.pathmorphing}

\allowdisplaybreaks

\makeatletter 
\renewcommand\onecolumngrid{
\do@columngrid{one}{\@ne}
\def\set@footnotewidth{\onecolumngrid}
\def\footnoterule{\kern-6pt\hrule width 1.5in\kern6pt}
}
\makeatother

\begin{document}

\pagenumbering{arabic}

\title{Dynamical quasinormal mode excitation}

\author{Marina De Amicis \orcid{0000-0003-0808-3026}}
\email{mdeamicis@perimeterinstitute.ca}
\affiliation{Center of Gravity, Niels Bohr Institute, Blegdamsvej 17, 2100 Copenhagen, Denmark}
\author{Enrico Cannizzaro \orcid{0000-0002-9109-0675}}
\affiliation{CENTRA, Departamento de Fisica, Instituto Superior T\'ecnico – IST,
Universidade de Lisboa – UL, Avenida Rovisco Pais 1, 1049 Lisboa, Portugal}
\author{Gregorio Carullo \orcid{0000-0001-9090-1862}}
\email{g.carullo@bham.ac.uk}
\affiliation{Center of Gravity, Niels Bohr Institute, Blegdamsvej 17, 2100 Copenhagen, Denmark}
\affiliation{School of Physics and Astronomy and Institute for Gravitational Wave Astronomy, University of Birmingham, Edgbaston, Birmingham, B15 2TT, United Kingdom}
\author{Laura Sberna \orcid{0000-0002-8751-9889}}
\affiliation{School of Mathematical Sciences, University of Nottingham, University Park, Nottingham NG7 2RD, United Kingdom}

\begin{abstract}

We study the dynamical excitation of quasinormal modes (QNMs) through the plunge, merger and ringdown of an extreme-mass-ratio-inspiral into a Schwarzschild black hole, for generic orbital configurations.
We work out the QNM causality condition, crucial to eliminate amplitude divergences and to incorporate horizon redshift effects.
We then use it to derive a model of the time-dependent QNM excitation via a Green's function approach, driven by the point-particle source on a given trajectory.
Our model predicts that:
i) QNMs propagates along hyperboloidal slices in the minimal gauge;
ii) the signal is composed of an ``activation'' term, depending on the source past history,
and a local ``impulsive'' term;
iii) amplitudes grow in time in an ``activation function'' fashion, and the waveform displays a stationary ringdown regime at times $\sim 10-20M$ after its peak;
iv) at these late times, an infinite tower of non-oscillatory, exponentially-damped terms appear: the redshift terms.
The model is in good agreement with numerical solutions, capturing the main waveform features after the peak.
Additional components of the Green's function are required to complement the QNM description and reproduce the plunge-merger waveform.
We predict the late-time, stationary amplitude of the quadrupolar mode as a function of eccentricity, in agreement with accurate numerical solutions, marking the first time that QNM amplitudes are predicted for generic binary configurations.
Our work provides a first solid step towards analytically modeling the inspiral's imprint onto ringdown signals, generalizable to include higher orders in the mass ratio, black hole spin, non-vacuum configurations and corrections to the Einstein-Hilbert action.

\end{abstract}

\maketitle

\twocolumngrid
{\tableofcontents}
\onecolumngrid

\section{Introduction} %
%
\subsection{Background}   %
\label{subsec:background} %
%
%
Gravitational waves (GWs) emitted by merging compact objects are routinely observed through the LIGO-Virgo-Kagra interferometric network~\cite{LIGOScientific:2014pky, VIRGO:2014yos, KAGRA:2020tym}, with improved detectors in construction or planned~\cite{Saleem:2021iwi,Pandey:2024mlo, Colpi:2024xhw,Abac:2025saz}.
These signals can be leveraged to learn a great deal on binary properties~\cite{KAGRA:2021vkt}, on the formation, evolution and host environments of compact objects~\cite{KAGRA:2021duu, Cardoso:2021wlq}, and even on the expansion of the Universe~\cite{LIGOScientific:2021aug, Chen:2024gdn}.
Thanks to black hole (BH) uniqueness properties in general relativity (GR), binary BH mergers can additionally be used as fundamental physics probes, allowing to investigate the predictions of GR in the strong-field and dynamical regime~\cite{LIGOScientific:2021sio}, search for new fundamental fields~\cite{Berti:2015itd} or exotic states of matter in highly compact configurations~\cite{Cardoso:2019rvt}.

%
%
BH mergers proceed through long inspiralling orbits, gradually losing energy via GW radiation and terminating in a rapid plunge.
After common horizon formation, the resulting perturbed BH relaxes towards equilibrium during a ``ringdown'' phase.
Observations of these mergers critically rely on accurate models of the entire emitted waveform.
For mergers of comparable-mass BHs, complete waveforms can be predicted numerically solving Einstein's field equations by means of Numerical Relativity (NR)~\cite{10.1093/acprof:oso/9780199205677.001.0001}.
Machine-learning methods or insights from analytical expansions can then be used to construct rapid phenomenological models interpolating between numerical results, see~\cite{Chatziioannou:2024hju} for a review.
Despite their accuracy, such approaches often lack interpretability, providing only a limited understanding of the physical mechanisms generating the signal.
For example, it is common to investigate specific features of these signals to verify whether a given phenomenon (horizon absorption, spin-induced quadrupole moment, resonant spectra, etc.) can be inferred from these observations.
Phenomenological fits are sub-optimal for this purpose, due to the many overlapping effects affecting each phenomenological parameter~\cite{Baibhav:2023clw, Johnson-McDaniel:2021yge}.
Further, investigating BHs merging in non-vacuum environments or within modified theories of gravity through numerical approaches rapidly becomes computationally unfeasible due to the vast parameter space to be covered.

In principle, analytical closed-form predictions offer a solution in all these cases, faithfully representing the underlying physical generation mechanism, but are challenging to construct.
First-principles models of the inspiral signal have been extensively explored by means of perturbative expansions~\cite{Blanchet:2013haa, Bini:2025vuk, Damour:2025uka,Bjerrum-Bohr:2022blt, Pound:2021qin}, often augmented through resummation techniques within the Effective One Body (EOB) formalism~\cite{Buonanno:1998gg, Damour:2008gu}.
On the other hand, there is a surprising lack of first-principles predictions for the merger and post-merger phases describing the remnant BH formation and relaxation.
This is the focus of our work.\\

%
%
BH perturbation theory shows that the Green's function (GF) of a BH geometry, determining its response to perturbations, has infinite simple poles in the complex Fourier-domain: the quasinormal frequencies (QNFs)~\cite{Berti:2025hly}.
The GF depends only on the background and, as a consequence of the no-hair conjecture, the QNF spectrum depends solely on the BH mass $M$, spin $a$, electric and magnetic charges $Q, P$~\cite{Israel:1967wq,Robinson:1975bv,Carter:1971zc,Mazur:1982db}. 
However, the \textit{appearance} of these QNFs in the emitted time-domain waveform depends on the initial data~\cite{Chavda:2024awq}.
Gaussian-like perturbations give rise to a ``ringdown'' signal~\cite{Vishveshwara:1970zz}, composed of a superposition of damped sinusoids with complex frequencies coinciding with the QNFs, and whose amplitudes depend on the initial-data: the quasinormal modes (QNMs).
%

%
%
A ringdown signal is also excited by BH binary mergers~\cite{Anninos:1993zj, Pretorius:2005gq}.
%
%
After the remnant formation, a ``stationary'' relaxation regime ensues (typically $\sim 10-20M$ past the emission peak), with the corresponding waveform well-approximated by a superposition of constant amplitude QNMs~\cite{Chandrasekhar:1975zza,Andersson:1996cm,leaver1985analytic,Leaver:1986gd,Berti:2009kk, Berti:2025hly}, followed at later times by a power-law ``tail''~\cite{Price:1971fb, Zenginoglu:2012us}.
A posteriori fits of NR waveforms indicate that the stationary QNMs amplitudes carry an imprint of the preceding inspiral configuration~\cite{Kamaretsos:2011um,Kamaretsos:2012bs,London:2014cma,Baibhav:2017jhs,London:2018gaq,Borhanian:2019kxt,Cheung:2023vki,Zhu:2023fnf,MaganaZertuche:2024ajz,Carullo:2024smg,Pacilio:2024tdl,Nobili:2025ydt,Mitman:2025hgy}: amplitudes can be accurately parametrized in terms of progenitors' binary properties (combinations of mass ratio and spins), with close functional resemblance to the post-Newtonian inspiral multipoles~\cite{Kamaretsos:2011um,Kamaretsos:2012bs,London:2014cma,Borhanian:2019kxt}.
This suggests that inspiral (post-Newtonian) and late-time ringdown modes share a common ``source''.
Further, an accurate numerical mapping between the plunge geometry and Kerr QNM stationary amplitudes was constructed in~\cite{Hughes:2019zmt,Lim:2019xrb}, for extreme-mass-ratio inspirals with misaligned orbits.
At earlier times, during the near-merger regime, a ``transient'' dynamical phase takes place, with the waveform instantaneous frequency smoothly interpolating between the binary orbital frequency and the asymptotic QNF (e.g. Fig.~2 of~\cite{Damour:2025uka}).
Several studies have provided insights into this process, presenting phenomenological~\cite{Baker:2008mj,Damour:2014yha} or toy models~\cite{Damour:2007xr,Price:2015gia,Albanesi:2023bgi, Oshita:2024wgt}.
To describe the QNMs transient, Refs.~\cite{Baker:2008mj,Damour:2014yha} proposed an ``activation function'' ansatz accurately capturing the entire post-merger, variations of which are still in use today in both test-particle and comparable-mass waveform models~\cite{Nagar:2020pcj, Albanesi:2023bgi, Pompili:2023tna, Planas:2025feq}.
In Ref.~\cite{Damour:2007xr}, the QNM excitation was heuristically described as a resonant process driven by the inspiralling particle, evolving due to radiation-reaction modeled with the EOB formalism. 
Building on this idea, a ``driven harmonic oscillator'' toy model was recently explored in~\cite{Albanesi:2023bgi}, qualitatively capturing the excitation hierarchy between the two-families of co- and counter-rotating ringdown modes~\cite{Berti:2025hly} in the small mass-ratio limit, and providing another strong indication of the fundamental role played by the particle source term.
Because the source varies with time during the inspiral, QNMs naturally acquire a time-dependent excitation.
Although these studies provide useful insights, a first-principles understanding of the dynamical excitation of the QNMs is still missing.
An alternative formal framework is the complex angular momentum approach~\cite{Folacci:2018sef}, which expresses the GF as a sum over Regge poles in the complex-$\ell$ plane -- rather than QNMs.
Applied to Schwarzschild and plunging sources, Regge poles capture the qualitative features of the signal even at early times~\cite{Folacci:2018sef}. 
However, this approach is presently limited to simplified configurations (geodesic radial infall in Schwarzschild), and does not provide direct access to individual modes, making it less readily applicable in standard waveform modelling and data analysis.
Lack of explicit expressions for the QNMs dynamical excitation limits physical interpretation as well.

%
The difficulty in extending QNM amplitudes computations to the dynamical regime lies in the partial knowledge on how they propagate within the light-cone, namely the QNMs \textit{causality condition}.
Refs.~\cite{Leaver:1986gd,Sun:1988tz,Andersson:1996cm,Berti:2006wq} provided a causality prescription based on the scattering of the QNMs off the potential peak. Interestingly, this prescription gives rise to time-dependent QNM amplitudes, but is limited to compact sources localized far away from a Schwarzschild BH, becoming ill-defined for sources extending inside the light-ring (LR).
For a toy-model geometry and generic sources, an exact condition was recently given in~\cite{Chavda:2024awq}, and used to show that the amplitudes' time-dependence arises from the finite propagation time of the perturbations~\cite{Lagos:2022otp}. 
Another known obstacle to computing QNM amplitudes is the divergence of the QNM radial functions in Schwarzschild coordinates. These divergences are commonly cured by including suitable boundary terms to the integration over the source~\cite{Leaver:1986gd,Hadar:2009ip,Zhang:2013ksa,Kuchler:2025hwx}. However, the use of the hyperboloidal foliation~\cite{Zenginoglu:2011jz,Warnick:2013hba,Ansorg:2016ztf,Gajic:2024xrn,PanossoMacedo:2024nkw} and studies on other physical systems exhibiting QNMs, such as optical cavities~\cite{Colom_2018,Abdelrahman:18,Wu:2025sjt}, have shown that divergences appear only when one key ingredient is overlooked: causality.
Using the hyperboloidal foliation, Ref.~\cite{Besson:2024adi}, building on \cite{Ansorg:2016ztf}, recently proposed an asymptotic QNM expansion based on Keldysh's non-selfadjoint spectral theory. For Gaussian initial data on Schwarzschild, the expansion reproduces the late-time behaviour accurately, but still shows a loss of accuracy at early times, attributed to transient, non-modal growth.
Understanding QNMs excitations in the test particle limit is a promising avenue to improve modeling even for binaries with comparable masses -- a fact that has long been exploited in the construction of accurate EOB models~\cite{Nagar:2006xv,Damour:2007xr,Barausse:2011kb}.
Numerical studies (e.g. Fig.\,1 of~\cite{DeAmicis:2024eoy} for radial infalls, Fig.\,2 of~\cite{Nagar:2022icd} and Fig.\,12 of~\cite{Kuchler:2025hwx} for quasi-circular binaries) show that perturbative and non-linear waveforms display remarkably similar structures during the merger-ringdown regime.
The latter work was the first to analytically compute QNM stationary amplitudes following a quasi-circular inspiral, extending past results on radial infalls~\cite{Zhang:2013ksa} and on geodesic plunges from the innermost stable circular orbit~\cite{Hadar:2009ip}.
Their expansions showed how constant amplitude superpositions are well-convergent $\gtrsim 10M$ past the peak, confirming previous numerical studies~\cite{London:2014cma, Carullo:2018sfu, Baibhav:2023clw, Nee:2023osy, Zhu:2023mzv, Cheung:2023vki, Carullo:2024smg, Mitman:2025hgy}, while diverging at earlier times.
Extending QNMs expansions towards the merger regime requires incorporating their excitation induced by the inspiral motion, the problem on which we focus here.
%

\subsection{Summary}         %
\label{subsec:intro_summary} %
%
In this work, we construct a first-principles model for the dynamical excitation of QNMs during the inspiral-plunge of extreme mass-ratio inspirals on generic planar orbits around Schwarzschild.
Our set-up and conventions are described in Sec.~\ref{sec:conventions_methods}.
The resulting model is based on solving the sourced, linearized Einstein's equation with the GF method, focusing on the QNM contribution. 

From the QNM GF, in Sec.~\ref{subsec:QNMs_propagation_and_regularity} we derive the QNMs causality condition for generic sources.
We find that QNMs propagate along hyperboloidal slices in the minimal gauge~\cite{PanossoMacedo:2018hab,PanossoMacedo:2023qzp,PanossoMacedo:2024nkw}, smoothly interpolating between the standard scattering condition (far away from the BH) and light-cone propagation (close to the horizon).
We show that this condition automatically resolves the GF's divergence for a source extending towards the horizon, a common obstacle in the analytical calculation of QNMs amplitudes. 
The convolution of the QNM GF with the source is investigated in Sec.~\ref{subsec:QNM_signal}, where we find the signal to be divided into two components: the ``activation''
and ``impulsive'' contributions.
The impulsive contribution is ``local'': it only depends on the source configuration at the time it is emitted.
The activation contribution consists of a local component and a ``hereditary'' one, which depends on the past trajectory and at late-times yields the ``stationary'' (constant amplitude) response of the fundamental QNMs.
This dependence is akin to hereditary tail terms, which arise from the branch-cut propagator. However, the latter has a much slower decay than the QNM propagator, leading to the tails' much stronger dependence on past history, see~\cite{DeAmicis:2024not}, while the stationary ringdown signal is mainly influenced by the final stages of the plunge, in agreement with the toy-model analysis of Ref.~\cite{Price:2015gia}.

Since for external observers the particle never stops falling towards the horizon (see also~\cite{Zimmerman:2011dx,Kuchler:2025hwx}), the source influences the signal even at late times (i.e., after the peak of the waveform). 
To gain insight on this regime, in Sec.~\ref{subsec:NearH+_divergence} we apply a near-horizon expansion to our predictions.
After LR crossing, the source contribution is gradually quenched by the infinite horizon redshift, and it stops contributing to the QNM amplitudes, which saturate to constant values.
However, we find that in this limit the source also gives rise to an infinite tower of zero-frequency, exponentially damped terms, with decay rates given by multiples of the BH surface gravity $\kappa_{\mathcal{H}^+}$. We denote these as \textit{redshift terms}.  
These modes originate from both the impulsive contribution to the signal and the local activation component.
The existence of these redshifted contributions is in agreement with past literature~\cite{Price:1971fb,Dafermos:2005eh,Mino:2008at,Zimmerman:2011dx,Laeuger:2025zgb}, where they were named ``horizon modes"\footnote{Here, we choose to drop the ``horizon mode'' nomenclature to underline the fact that these terms arise only in the presence of a source or initial data which extend towards the horizon. The redshift terms, in fact, are not eigenfunctions of the Regge-Wheeler/Zerilli differential operators.}.
In Sec.~\ref{sec:time_dependent_early_ringdown} we construct the full QNM waveform, discussing in detail our predictions for a quasi-circular plunge and a radial infall. 
We first investigate the time dependence of the activation and impulsive coefficients (Secs.~\ref{subsec:exc_imp_coeffs} and \ref{subsec:exc_imp_contribs}).
We find that, after LR crossing, the overtone contributions follow a simple decreasing excitation hierarchy and are dominated at late times by redshift terms. We discuss redshift terms in more detail in Sec.~\ref{subsec:redshift_mode}.

In Sec.~\ref{subsec:full_waveform} we compare our first-principles QNM signal with the full waveform, obtained by numerically solving for the sourced linear perturbations around Schwarzschild. 
At early times, the instantaneous frequency of the predicted full QNM response is driven by the test-particle orbital frequency, encoded in the time-dependence of the modes amplitudes, behaving as activation functions.
This is because the QNM propagator is a superposition of terms oscillating at the QNFs: each of these frequencies is quasi-resonantly excited by the source motion during the inspiral-plunge, in agreement with the intuition developed in~\cite{Damour:2007xr,Albanesi:2023bgi}. Due to the short lifetime of higher overtones, this accumulation process is more effective for smaller overtone numbers. 
This should again be contrasted to hereditary tails, characterized by a zero-frequency propagator, onto which the effect of an oscillating source translates in destructive interference between different terms~\cite{DeAmicis:2024not}.
We find good agreement with the numerical results approximately $\sim 10 M$ after the apparent location of the source has crossed the LR. 
During the plunge-merger regime the QNM response qualitatively tracks the waveform morphology, but cannot fully describe the signal.
However -- somewhat surprisingly -- its magnitude is close to the full signal even $25 M$ before LR crossing.
We find that high overtones contribute the most to the signal during these last stages of the plunge, before the LR crossing. 

As a by-product of our analysis, in Sec.~\ref{sec:inspiral_imprints} we provide for the first time an analytical prediction of the amplitude and phase of the fundamental quadrupolar mode, sourced by particles in arbitrarily eccentric orbits. 
This prediction shows that eccentric corrections can increase the amplitude by up to 25\%, in agreement with numerical results, see also~\cite{Carullo:2024smg}.
Note that the algorithm discussed in Sec.~\ref{subsec:NearH+_divergence} can be applied to compute the late-times constant amplitudes of generic QNMs.
Our findings are discussed in Sec.~\ref{sec:discussion} and future directions exposed in Sec.~\ref{sec:future_directions}, including prospects to extend these results to finite mass-ratios, Kerr and beyond-vacuum/GR scenarios.
Explicit expressions of the source functions and the quasinormal (QN) eigenfunctions are provided in App.~\ref{app:source_functions} and~\ref{app:Chandra_transf}, while a more detailed derivation of the causality condition is in App.~\ref{app:QNMs_GF}.
The latter circumvents the need for standard QNM regularization procedures, which we review in App.~\ref{app:regularization}.
In App.~\ref{app:additional_orbits}, we present results for intermediate and highly eccentricity orbits, displaying the same qualitative features encountered for quasi-circular orbits.
Finally, in App.~\ref{app:redshift} we give a simple argument for the appearance of redshift terms, while in App.~\ref{app:Kerr_causality} we derive the causality condition for rotating BHs.

\section{Conventions and methods} %
\label{sec:conventions_methods}   %
%
We work in geometric units $c=G=1$ and we will rescale dimensional quantities with respect to $M$.
Our analysis focuses on non-spinning black holes binaries with small mass-ratios, thus we linearize Einstein's equations and discard higher-order corrections. 
The background metric is the Schwarzschild metric,
\begin{equation}
    ds^2=-A(r)dt^2+\frac{dr^2}{A(r)}+r^2d\Omega^2 \ , \  A(r)=1-\frac{r_h}{r} \, ,
    \label{eq:Schwarzschild_metric}
\end{equation}
where we have denoted $r_h=2$ the location of the event horizon.

Since at first perturbative order the mass and spin of the BHs are unchanged, the final BH will not be spinning. 
For this reason, we can expand the strain in spin-weighted spherical harmonics modes $_{-2}Y_{\ell m}(\theta,\varphi)$, retaining consistency with the space-time symmetries both in the inspiral and the ringdown phase (i.e. no mode-mixing~\cite{Berti:2014fga} arises):
\begin{equation}
    h(t,r,\theta,\varphi) =  \sum_{\ell \geq 2,|m|\leq \ell} h_{\ell m}(t,r) _{-2}Y_{\ell m}(\theta,\varphi)
\end{equation}
The even (odd) components of the strain are obtained by solving the Zerilli (Regge-Wheeler) equation 
\begin{equation}
        \left[\partial_t^2 -\partial_{r_*}^2+V^{e/o}_{\ell m}(r_*)\right]\Psi_{\ell m}(t,r_*) = \mathcal{S}_{\ell m}(t,r_*) \, ,
    \label{eq:RWZ_equation}
\end{equation}
where we have introduced the tortoise coordinate with the convention $r_*=r+r_h \, \log\left(r/r_h-1\right)$.
Our derivation will be insensitive to the even or odd parity nature of the multipoles, hence below we drop the subscript from $V^{e/o}$.
The strain multipoles $h_{\ell m}$ can be computed from the Regge-Wheeler/Zerilli eigenfunctions $\Psi_{\ell m}$ through
\begin{equation}
    \Psi_{\ell m}=\frac{h_{\ell m}}{\sqrt{\left(\ell+2\right)\left(\ell+2\right)\ell\left(\ell-1\right)}} \, .
\end{equation}
We divide the source function on the right-hand side of Eq.~\eqref{eq:RWZ_equation} into two different components
\begin{equation}
    \mathcal{S}_{\ell m}(t,r_*)=S_{\ell m}(t,r_*)+S^{\rm ID}_{\ell m}(t,r_*) \, .
    \label{eq:source_function}
\end{equation}
$S_{\ell m}(t,r_*)$
has a generic time dependence, and is related to the stress-energy tensor generating the perturbations~\cite{Nagar:2005ea,Nagar:2006xv}. $S^{\rm ID}_{\ell m}$ is the initial-data source, i.e.
\begin{equation}
    \begin{split}
        S^{\rm ID}_{\ell m}(t,r_*)=\Psi_{\ell m}(t_0,r_*)\partial_t\delta(t-t_0)+
        \partial_t\Psi_{\ell m}(t_0,r_*)\delta(t-t_0) \, ,
    \end{split}
    \label{eq:ID_source}
\end{equation}
This is another way to write the initial-data problem. Another possibility would be to consider only the source $S_{\ell m}$ in Eq.~\eqref{eq:RWZ_equation} and to impose that the RWZ solutions coincide with $\Psi(t_0,r_*),\,\partial_t\Psi(t_0,r_*)$ at $t_0$. As argued in Ref.~\cite{Leaver:1986gd} and references therein, imposing this constraint is equivalent to adding the source $S^{\rm ID}_{\ell m}(t,r_*)$ to Eq.~\eqref{eq:RWZ_equation}. 
The equivalence can be proven by comparing the convolution of the time-domain GF with the time-domain source in Eq.~\eqref{eq:ID_source}, to the anti-Fourier transform of the frequency-domain GF with the frequency-domain initial-data source $-i\omega\Psi(t_0,r_*)+\partial_t\Psi(t_0,r_*) $, which appears on the right-hand side of the frequency-domain RWZ equation, by taking into account the initial-data constraint while performing the Fourier transform.

In this work, we will study perturbations of a Schwarzschild BH driven by a test-particle infalling into it. Since we initialize the test-particle far away from the BH, we can assume null data of the perturbation field on the initial Cauchy hypersurface $t=t_0=0$, i.e. 
\begin{equation}
\Psi_{\ell m}(t=0,r_*)=\partial_t\Psi_{\ell m}(t=0,r_*)=0 \, .
\label{eq:null_initial_data}
\end{equation}
This condition implies $S^{\rm ID}_{\ell m}\equiv 0$, hence Eq.~\eqref{eq:source_function} depends only on the source function $S_{\ell m}$.
The latter is localized on the test-particle trajectory and can be decomposed in two pieces (as derived in Refs.~\cite{Nagar:2005ea} and~\cite{Nagar:2006xv})
\begin{equation}
\begin{split}
        S_{\ell m}(t,r_*)=f_{\ell m}\left(t,r_*\right)
    \delta\left(r_*-r_*\left(t\right)\right)+
    g_{\ell m}\left(t,r_*\right)\partial_{r_*}\delta\left(r_*-r_*\left(t\right)\right) \, .
\end{split}
\label{eq:source_generic_expression}
\end{equation}
Explicit expressions for the source functions $f_{\ell m} ,\, g_{\ell m}$ can be found in App.~\ref{app:source_functions}.
In the following derivation, we drop the subscript $(\ell m)$ identifying the multipole under study.
The trajectory of the test-particle $(r(t),\varphi(t),\theta(t)=\pi/2)$, with $\varphi(t)$ and $\theta(t)$ respectively azimuthal and polar angle, is computed solving the Hamiltonian equations of motion~\cite{Nagar:2006xv}
\begin{equation}
\begin{split}
    &\dot{r}=\frac{A}{\hat{H}}p_{r_*} \, ,\\
    &\dot{\varphi}=\frac{A}{r^2\hat{H}}p_{\varphi} ,\\
    &\dot{p}_{r_*}=A\hat{\mathcal{F}}_{r}-\frac{A}{r^2\hat{H}}\left(p^2_{\varphi}\frac{3-r}{r^2}+1 \right) \, ,\\
    &\dot{p}_{\varphi}=\hat{\mathcal{F}}_{\varphi} \, ,
\end{split}
    \label{eq:Hamiltonian_Eqs}
\end{equation}
with $(p_{r_*},p_{\varphi})$ as the $\mu$-rescaled momenta conjugate to $(r_*,\varphi)$,
and $\hat{H}$ as the $\mu$-rescaled Hamiltonian
\begin{equation}
    \hat{H}=\sqrt{A\left(1+\frac{p_{\varphi}^2}{r^2}\right)+p_{r_*}^2} \, .
    \label{eq:energy_unit_mu}
\end{equation}
The quantities $\hat{\mathcal{F}}_{r}$ and $\hat{\mathcal{F}}_{\varphi}$ are radiation-reaction dissipative forces that drive the dynamics.
The radiation-reaction forces are analytical functions of the trajectory, and were computed through a resummed PN analytical expansion for the fluxes of energy and angular momentum observed at infinity in~\cite{Chiaramello:2020ehz,Albanesi:2021rby}.
In Ref.~\cite{Albanesi:2023bgi}, it was shown that the analytical fluxes are consistent with perturbative numerical ones.
In the following, we will denote with $E_0$ the initial test-particle energy.
We solve Eq.~\eqref{eq:Hamiltonian_Eqs} using the time domain code \textsc{RWZHyp}~\cite{Bernuzzi:2010ty,Bernuzzi:2011aj}, for a set of initial data of the test-particle. Details on the trajectory of each simulation can be found in Table~\ref{tab:sims_ecc}.
The software \textsc{RWZHyp} uses an homogeneous grid in $r_*$ which for negative radii ends at a finite value, in order to keep the horizon outside of the computational domain.
At large distances, outside the region inside which the test-particle trajectory evolves, an hyperboloidal layer is attached to the computational domain.
The coordinates of the layer are the retarded time $\tau$ and the compactified spatial coordinate $\rho$. We refer to~\cite{Bernuzzi:2011aj} for an explicit expression of the coordinate $\rho$ in terms of $(t,r_*)$, while for $\tau$ it holds
\begin{equation}
    \tau-\rho=t-r_* \, .
    \label{eq:RWZcoordinates}
\end{equation}
The code uses double precision and the test particle is approximated by a narrow gaussian packet.
We use the same resolution as in Ref.~\cite{DeAmicis:2024not} (radial step $\Delta\rho=0.015$) -- we refer to this work for a detailed investigation of the code convergence and error budget quantification.
The code \textsc{RWZHyp} also solves for the gravitational perturbation $\Psi$ at future null infinity $\mathcal{I}^+$.
We will use the numerical waveform to test our analytical predictions.
It is important to stress that our analytical waveform prediction will be only informed by the numerical trajectory, which is independent from the numerical waveform.

\begin{table}[t]
\begin{tabular}{cccccccc}
\hline \hline
$e_0$ & $E_0$ & $p_{\varphi,0}$ &  $r_0$ & $b_{\rm LR}$ & $e_{\rm sep}$\\ 
\hline 
0.9 & 0.9890 & 3.9170 & 83.000 & 3.9315 & 0.869 \\ 
0.8 & 0.9791 & 3.8313 & 40.000 & 3.8881 & 0.778 \\ 
0.7 & 0.9713 & 3.7671 & 26.667 & 3.8429 & 0.670 \\ 
0.6 & 0.9649 & 3.7139 & 20.000 & 3.7998 & 0.563\\ 
0.5 & 0.9587 & 3.6502 & 15.400 & 3.7699 & 0.483\\ 
0.4 & 0.9538 & 3.6001 & 12.500 & 3.7400 & 0.393\\ 
0.3 & 0.9525 & 3.6103 & 11.429 & 3.7075 & 0.276\\ 
0.2 & 0.9484 & 3.5514 & 9.375 & 3.6909 & 0.201\\ 
0.1 & 0.9453 & 3.5044 & 7.778 & 3.6766 & 0.114\\ 
0.0 & 0.9449 & 3.5000 & 7.000 & 3.6693 & 0.000\\  
\hline \hline
\end{tabular}
\caption{Test particle trajectories.
From left to the right: initial eccentricity, initial energy and angular momentum, initial radius, impact parameter at the LR crossing and eccentricity at the separatrix crossing.
Note that the eccentricity decreases during the inspiral, but because of its definition in terms of radial turning points, it can increase close to the separatrix, as exemplified in Fig.~1 of~\cite{Albanesi:2023bgi}.
}
\label{tab:sims_ecc}
\end{table}

Due to the inclusion of radiation reaction, the trajectory incorporates corrections beyond linear order. 
Hence, in our treatment we are only including partial non-linear corrections, namely the ones we expect to be the leading contribution.
This approach was validated in Ref.~\cite{Nagar:2006xv}, where it was introduced, and later applied in a wide range of investigations~\cite{Nagar:2006xv,Damour:2007xr,Albanesi:2021rby,Albanesi:2023bgi}.
Such approach is close in spirit with the formal treatment of, e.g., Ref.~\cite{Hughes:2021exa}.
In Ref.~\cite{Nagar:2006xv}, the approach was shown to be accurate by finding good agreement between the numerical angular momentum flux at future null infinity and the analytical prediction of the binary’s angular momentum loss, computed through a resummed post-Newtonian expansion (see Fig. 4 of~\cite{Nagar:2006xv}).
%
%
A mismatch between the two quantities only appears in the final stages of the plunge~\cite{Nagar:2006xv}. However, in this limit, the motion becomes approximately geodesic and the numerical framework becomes consistent with first-order perturbation theory.
This result suggests that the error introduced by our approximation stays small throughout the orbital evolution.

Moreover, Ref.~\cite{Nagar:2022icd} directly investigated the test-particle numerical waveforms obtained using our framework with the full non-linear solution of Einstein’s equation, for different mass ratios in the intermediate regime, the largest being $q = 128$.
In particular, Ref.~\cite{Nagar:2022icd} compared the amplitudes at merger, where non-linear effects are expected to be most relevant. As shown in their Fig. 2, the test-particle and fully non-linear results are in excellent agreement for several multipoles (including the fundamental, the focus of this paper), suggesting that non-linearities in wave-generation are sub-dominant.

\section{Analytical predictions}   %
\label{sec:analytical_predictions} %

\subsection{QNM propagation and regularity}    %
\label{subsec:QNMs_propagation_and_regularity} %
%
The general solution of Eq.~\eqref{eq:RWZ_equation} can be computed by means of the GF method as
\begin{equation}
    \Psi(t,r_*)=\int_{-\infty}^{\infty}dt'\int_{-\infty}^{\infty}dr_*'G(t-t';r_*,r_*')\mathcal{S}(t',r_*') \, ,
    \label{eq:GreensFun_source_convolution}
\end{equation}
where the GF is a solution to the following impulsive problem, with zero boundary data
\begin{equation}
    \left[\partial_t^2-\partial_{r_*}^2+V(r)\right]G(t-t';r_*,r_*')=\delta(t-t')\delta(r_*-r_*') \, .
    \label{eq:GreensFun_definition}
\end{equation}
In Eq.~\eqref{eq:GreensFun_source_convolution} the integral does not have support on the entire $t'\in(-\infty,\infty)$ domain: the choice of GF will self-consistently impose a causality condition describing the signal propagation within the light-cone. 
To describe a physical signal we will use the retarded GF, enforcing the propagation inside/on the light-cone.

%
The solution of Eq.~\eqref{eq:GreensFun_definition} was investigated in detail by Leaver~\cite{Leaver:1986gd}, who found possible to separate the GF in three different components: an initial ``prompt'' response, a late-times ``tail'' and, at intermediate times, a ``ringing'' signal.
The latter is a superposition of exponentially decaying oscillatory modes, the QNMs.
Here, we focus on the portion of the signal \eqref{eq:GreensFun_source_convolution} propagated by the QNM GF. \\
%

\begin{figure}[t]
\centering
\includegraphics[width=0.49\textwidth]{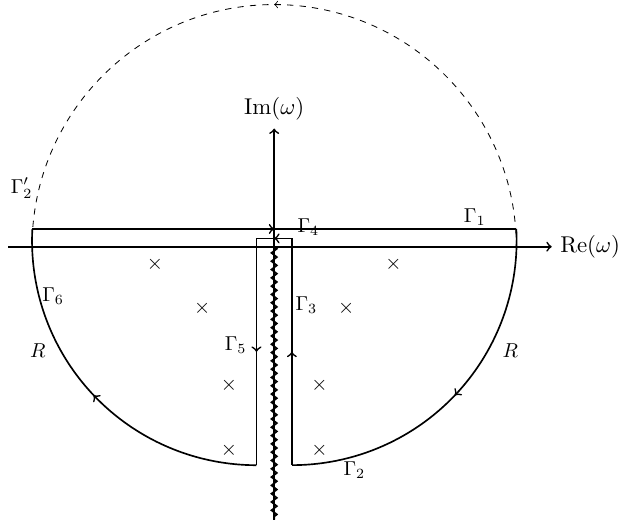}
\includegraphics[width=0.49\textwidth]{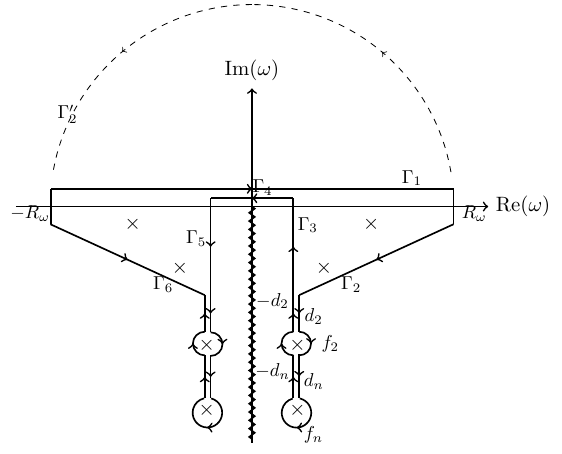}
\caption{
\textit{Left}: Standard contour used in the QNM literature~\cite{Leaver:1986gd,Andersson:1996cm}. \textit{Right}: Smooth deformation of the contour on the left (closed contour, thick line). As the deformation is performed without crossing any singularity, integration along the two closed contours yields the same result.
On the left, the radius of the two quarters of circumference is taken to $R\rightarrow\infty$ in the lower-half plane. On the right, we take $R_{\omega}\rightarrow\infty$ along the real line, and include an infinite number of overtones $n\rightarrow\infty$ along the imaginary axis.}
\label{fig:contour_plot2}
\end{figure}

%
%
First, we investigate over which portion of the light-cone the QNM GF propagates a perturbation.
Explicit computations can be found in App.~\ref{app:QNMs_GF}.
There, after a review of Leaver's computation in frequency domain, we move to the time domain.
Here, we only briefly summarize the derivation and focus on the key results.
Under the assumption of an observer placed far away from the GW source $r_*\gg r_*'$
and $\omega r_*\gg 1$, the retarded GF in Fourier domain can be written as follows
\begin{equation}
\tilde{G}(\omega;r_*,r_*')=\frac{i \, e^{i\omega r_*}}{2\omega A_{\rm in}(\omega)} u^{\rm in}(\omega,r_*') \, ,
\end{equation}
where $u^{\rm in}(\omega,r_*')$ is a solution of the homogeneous RWZ equation in the frequency domain, Eq.~\eqref{eq:homogeneous_RWZ_omega_Sch}, which reduces to a \textit{unitary ingoing} plane wave $e^{-i\omega r_*'}$ at the horizon, in the limit $r_*'\rightarrow-\infty$, as in Eq.~\eqref{eq:u_in_asympt_expr}.
The factor $e^{i\omega r_*}$ is another solution of this equation, independent from $u^{\rm in}$, and valid in the limit of an observer located at very large distances. The factor $(2\omega A_{\rm in}(\omega))^{-1}$ is the Wronskian of these two solutions.
The solution $u^{\rm in}$ can be written as \cite{Leaver:1986gd}
\begin{equation}\label{eq:u_in_two_terms}
u^{\rm in}(\omega,r_*)=A_{\rm in}(\omega)u^{\infty-}(\omega,r_*)+A_{\rm out}(\omega)u^{\rm out}(\omega,r_*) \, ,
\end{equation}
with $u^{\rm out}, u^{\infty-}$ respectively satisfying $u^{\rm out}\rightarrow e^{i\omega r_*}$ and $u^{\infty -}\rightarrow e^{-i\omega r_*}$ in the limit $r_*\rightarrow\infty$ (see Eqs~\eqref{eq:u_out_asympt_expr},\eqref{eq:u_inf-_asympt_expr}).
The QNFs are the values $\omega=\omega_n$ for which $A_{\rm in}(\omega_n)=0$. 
At these frequencies, $u^{\rm in}$ and $u^{\rm out}$ are linearly dependent and reduce to solutions $\propto u^h$ behaving as a purely ingoing (outgoing) plane wave at the horizon (infinity), i.e. $\propto e^{\pm i\omega r_*}$ for $r_*\rightarrow\pm \infty$. 
The explicit expression for the QN eigenfunctions $u^h$ (normalized as a unitary plane wave at the horizon) can be found in Eq.~\eqref{eq:u_in_near_QNFs}.
From Eq.~\eqref{eq:u_in_two_terms}, the GF in Fourier domain can be written as a sum of two contributions
\begin{equation}
\tilde{G}(\omega;r_*,r_*')=\tilde{G}^{(1)}(\omega;r_*,r_*')+\tilde{G}^{(2)}(\omega;r_*,r_*') \, , 
\label{eq:tilde_GF_separating_QNMs}
\end{equation}
with
\begin{align}
\tilde{G}^{(1)}(\omega;r_*,r_*')=& \frac{i}{2\omega} \, e^{i\omega r_*}u^{\infty -}(\omega,r_*')\, , \\
\tilde{G}^{(2)}(\omega;r_*,r_*')=&
\frac{i\, A_{\rm out}(\omega)}{2\omega A_{\rm in}(\omega)}e^{i\omega r_*}u^{\rm out}(\omega,r_*') \, .
\label{eq:tildeG2_def}
\end{align}
The term $\tilde{G}^{(2)}$ has isolated poles at the QNFs, yielding the time-domain QNM response, while $\tilde{G}^{(1)}$ is regular at the QNFs. 
As a consequence, only $\tilde{G}^{(2)}$ propagates the QNMs: in this work we will focus only on this component. 
To compute the time-domain retarded GF responsible for the QNMs propagation, we need to solve the integral
\begin{equation}
G^{(2)}(t-t';r_*,r_*')\equiv\int_{-\infty}^{\infty}d\omega \frac{e^{-i\omega(t-t')}}{2\pi}\, \tilde{G}^{(2)}(\omega;r_*,r_*')\, .
\label{eq:antitransf_tildeG2}
\end{equation}
The integrand is singular in $\omega=0$ since, for small $\omega$, 
$u^{\rm out}$ contains a piece $\propto \ln\omega$.
As a consequence, to compute the integral along the real $\omega$ axis, we analytically continue $\omega$ to the complex plane.
The general structure of the integrand in the complex plane is well known \cite{Leaver:1986gd} and shown on the left panel of Fig.~\ref{fig:contour_plot2}. 
In particular, there is a branch cut originating from $\omega=0$, which we fix on the negative imaginary axis in order to select the \textit{retarded} GF. 
The integration in Eq.~\eqref{eq:antitransf_tildeG2} is then performed on an axis parallel to the real line shifted at $\mathrm{Im}(\omega)=\epsilon>0$ with $\epsilon\ll1$.
In the lower half-plane $\mathrm{Im}(\omega)<0$, there is an infinite number of isolated simple poles (the QNFs) which in the limit $\mathrm{Im}(\omega)\to-\infty$ share the same real component $\mathrm{Re}(\omega_n)\ll1$~\cite{leaver1985analytic}.

The integral along the (shifted) real axis in Eq.~\eqref{eq:antitransf_tildeG2} can be computed by means of the residue theorem, once we choose a prescription to close the complex contour.
There are two options to close the contour. Both share the line parallel to the real axis and extend to infinity.
One contour is closed on the upper half-plane and the other on the lower half.
In the upper half-plane the integrand in Eq.~\eqref{eq:antitransf_tildeG2} has no poles nor branch cuts, while in the lower half-plane the contour is closed to avoid the branch cut and contains all the QNFs, as in the left panel of Fig.~\ref{fig:contour_plot2}.
We choose the contour closed on the upper (lower) half plane based on the regularity of the integrand in the limit $\mathrm{Im}(\omega)\rightarrow+ (-)\infty$.
As represented schematically in the right panel of Fig.~\ref{fig:contour_plot2}, it is possible to deform the contour closed on the lower-half plane shown in the left panel (the standard one used in the literature) to encircle each QNF in the limit $\mathrm{Im}(\omega)\rightarrow-\infty$. 
The deformation is smooth, i.e. performed without crossing any singularity, hence integration along the two closed contours in the lower-half plane of Fig.~\ref{fig:contour_plot2} (showed in thick lines), yields the same result.
Since the integrals along the lines connecting the QNFs ($d_i$ on the right panel of Fig.~\ref{fig:contour_plot2}) cancel out, for $\mathrm{Im}(\omega)\rightarrow-\infty$, inside the contour, we can approximate the homogeneous mode $u^{\rm out}$ with the Leaver's solutions at the QNFs \cite{Leaver:1985ax}
\begin{equation}
\begin{split}
u^{\rm out}(\omega,r_*')\simeq   e^{i\omega \left[r_*'-2r_h\log\left(\frac{r'-r_h}{r'}\right)\right]}\hat{a}(\omega,r')\, ,
\end{split}
\end{equation}
where we have defined
\begin{equation}
\hat{a}(\omega,r)\equiv \frac{ \sum_k a_k(\omega) A^k(r')}{\sum_ka_k(\omega)}\, .
\label{eq:hata_def}
\end{equation}
Substituting the expression above into Eq.~\eqref{eq:tildeG2_def}, we can rewrite the integrand in Eq.~\eqref{eq:antitransf_tildeG2} as
\begin{equation}
\left[\frac{i\, \hat{a}(\omega,r')\, A_{\rm out}(\omega)}{4\pi\omega A_{\rm in}(\omega)}\right]\cdot e^{-i\omega\left[t-r_*-\mathcal{C}(t',r_*')\right]}\, ,
\label{eq:antitransf_tildeG2_integrand_near_QNFs}
\end{equation}
with $\mathcal{C}(t',r'_*)$ defined as
\begin{equation}
\mathcal{C}(t',r_*')\equiv t'+r_*'-2r_h\log\left(\frac{r'-r_h}{r'}\right)\, .
\label{eq:QNMs_causality}
\end{equation}
In App.~\ref{app:QNMs_GF}, we argue that the absolute value of the square brackets term decays slower than an exponential in $\mathrm{Im}(\omega)$ and does not behave as $e^{|\mathrm{Im}(\omega)| a}$ for any real $a\neq 0$.
The behavior of the integrand in Eq.~\eqref{eq:antitransf_tildeG2} for $\mathrm{Im}(\omega)\rightarrow- \infty$ is thus dictated by the exponent in Eq.~\eqref{eq:antitransf_tildeG2_integrand_near_QNFs}.
For $t-r_*\geq \mathcal{C}(t',r_*')$, the integrand is exponentially suppressed as $\mathrm{Im}(\omega)\rightarrow- \infty$, while for $t-r_*< \mathcal{C}(t',r_*')$ it diverges exponentially in this limit.
We conclude that for $t-r_*\geq \mathcal{C}(t',r_*')$ the integral in Eq.~\eqref{eq:antitransf_tildeG2} should be computed closing the contour in the lower half complex $\omega$ plane, where it picks up the QNF contributions.
Instead, for $t-r_*< \mathcal{C}(t',r_*')$ the contour should be closed on the upper half plane, where $u^{\rm in,out}$ are analytical. 
Hence, the QNM response only comes from the lower-half plane integration, yielding 
\begin{equation}
\begin{split}
G^{\rm QNM}_{\ell m}(t-t';r_*,r_*')=\theta\left[t-r_*-\mathcal{C}(t',r_*')\right] \cdot \sum_{n=0}^{\infty}\sum_{p=\pm} B_{\ell mn p}\, e^{-i\omega_{\ell mnp}[t-r_*-\mathcal{C}(t',r_*')]}\hat{a}(\omega_{\ell mnp},r') \, .
\end{split} 
\label{eq:QNMs_Retarded_GreensFun_time}
\end{equation}
This is the fundamental result that we will use in the remainder of the work: it combines Leaver's prediction~\cite{Leaver:1986gd} with a new causality constrain prescribing how the QNMs travel in time domain, for generic $(t',r_*')$. 
To the best of our knowledge, this is the first time such condition is derived, with past literature focusing on the large $r_*'\gg 1 $ limit.
Details on the derivation, and an explicit definition of the \textit{geometrical excitation factors} $B_{\ell mnp}$, which depend only on the background geometry, are discussed in App.~\ref{app:QNMs_GF}.
Now that the QNFs appear explicitly, we have re-introduced $\ell m$ indices for clarity.
In Eq.~\eqref{eq:QNMs_Retarded_GreensFun_time}, we have introduced the notation $p=\pm$ to indicate the modes with $\mathrm{Re}(\omega_{\ell mn+})>0$ and the mirror modes $\mathrm{Re}(\omega_{\ell mn-})<0$;
as discussed in Ref.~\cite{leaver1985analytic}, for each QNF of a Schwarzschild BH there exist another QNF with the same imaginary but oppposite real component, denoted as \textit{mirror} mode.
In Eq.~\eqref{eq:QNMs_Retarded_GreensFun_time}, the Heaviside function determines how the QNMs response is propagated on the curved BH spacetime, i.e. the section of the light-cone interior $t-r_*\geq t'-r_*'$ on which the QNMs travel, 
\begin{equation}
t-r_*\geq\mathcal{C}(t',r_*') \, .
\label{eq:QNMs_propagation_lightcone_section}
\end{equation}
Given its role, we thus refer to $\mathcal{C}(t',r_*')$ as the QNM \textit{causality condition function}.
\begin{figure}[t]
\centering
\includegraphics[width=0.55\textwidth]{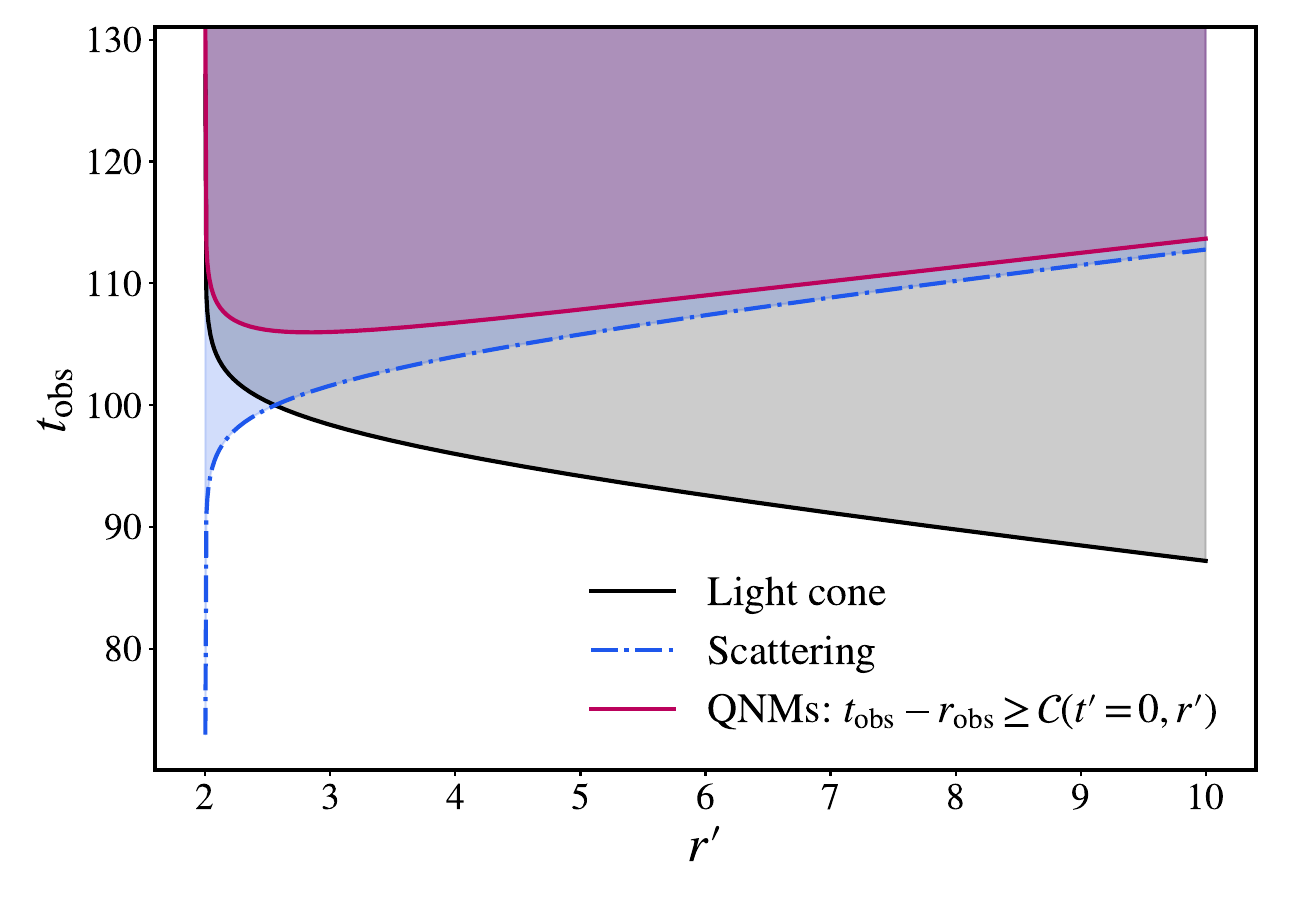}
\caption{
Time $t = t_{\rm obs}$ at which an observer located at $r_{*,\rm obs} = 100$ sees an impulsive signal emitted at $r_*'$ and $t'=0$. 
In black, a signal traveling on the light-cone $t_{\rm obs}-r_{*,\rm obs}=t'-r_*'$. 
In blue, a perturbation scattering at $r_*=0$, i.e. $t_{\rm obs}-r_{*,\rm obs}=t'+r_*'$. 
In purple, the QNMs propagation $t_{\rm obs}-r_{*,\rm obs}=\mathcal{C}(t',r_*')$, with $\mathcal{C}(t',r_*')$ defined in Eq.~\eqref{eq:QNMs_causality}.
\label{fig:light_cone_diagram}}
\end{figure}
In Fig.~\ref{fig:light_cone_diagram} we show the light-cone portion selected by the above condition, for different radii $r_*'$ of the initial pulse in Eq.~\eqref{eq:GreensFun_definition} on a $t'=0$ slice.
We compare the QNMs condition with the ``scattering" condition usually considered in the literature~\cite{Leaver:1986gd,Sun:1988tz,Andersson:1996cm,Berti:2006wq} when discussing QNMs propagation, i.e. $t-r_*>t'+r_*'$. 
Our condition~\eqref{eq:QNMs_causality} has the following asymptotic behavior
\begin{equation}
\begin{split}
&\mathcal{C}(t',r_*')\simeq t'-r_*' \ \, \ \ r'\rightarrow r_h\, ,\\
&\mathcal{C}(t',r_*')\simeq t'+r_*' \ \, \ \ r'\gg r_h\,  .
\end{split}
\label{eq:QNMs_causality_asymptotic_behaviors}
\end{equation}
Substituting into Eq.~\eqref{eq:QNMs_propagation_lightcone_section}, yields 
\begin{equation}
\begin{split}
&t-r_*\geq t'-r_*' \ \, \ \ r'\rightarrow r_h\, ,\\
&t-r_*\geq t'+r_*' \ \, \ \ r'\gg r_h\,  .
\end{split}
\end{equation}
Both limits are captured in Fig.~\ref{fig:light_cone_diagram}.
For initial-data far from the BH, the ``scattering" causality condition well approximates Eq.~\eqref{eq:QNMs_propagation_lightcone_section}. However, when approaching the LR, the ``scattering" prescription becomes ``ill-defined": for $r_*'<0$ it prescribes a signal traveling outside the light-cone (going below the black line in Fig.~\ref{fig:light_cone_diagram}).
For data well inside the LR, we find that QNMs propagate along the light-cone, and are thus affected by the horizon redshift, as we are going to explicitly show below.
This result is to be expected: all signals traveling marginally close to the horizon must approach the light-cone propagation due to causality. 
Interestingly, Eq.~\eqref{eq:QNMs_causality} corresponds to the hyperboloidal time coordinate in the minimal gauge~\cite{PanossoMacedo:2018hab,PanossoMacedo:2023qzp,PanossoMacedo:2024nkw}.
This is our first prediction: the QNM GF propagates perturbations along hyperboloidal slices in the minimal gauge.
The causality condition in Eq.~\eqref{eq:QNMs_Retarded_GreensFun_time} guarantees that the GF is regular for all $(t',r_*')$.
If we consider an initial pulse approaching the horizon, it holds $\mathcal{C}(t',r_*'\rightarrow-\infty)\rightarrow\infty$. 
As a consequence, the exponential piece in Eq.~\eqref{eq:QNMs_Retarded_GreensFun_time} diverges as $e^{|\omega_{\ell mn}^{\rm Im}|\mathcal{C}(t',r_*')}$.\footnote{The $p$ index is suppressed when indicating the imaginary part, as the latter is identical for $p =\pm$ modes.}
However, the amplitude observed at $\mathcal{I}^+$ stays finite: the causality condition in Eq.~\eqref{eq:QNMs_Retarded_GreensFun_time} shifts the time at which an observer at $\mathcal{I}^+$ can see the perturbations towards infinity: $t-r_*=\mathcal{C}(t',r_*'\rightarrow-\infty)\rightarrow\infty$, canceling the divergence.
The physical interpretation is that a signal emitted at the horizon undergoes an infinite redshift, reaching the observer only at infinitely late times. Hence, the divergent piece $e^{|\omega_{\ell mn}^{\rm Im}| \mathcal{C}(t',r_*')}$ in Eq.~\eqref{eq:QNMs_Retarded_GreensFun_time} is physical and does not require any regularization procedure: one should simply compute the full observable.
We will discuss this point in more detail in Sec.~\ref{subsec:NearH+_divergence}, comparing our results with regularization techniques proposed in the literature.
The QNM GF in Eq.~\eqref{eq:QNMs_Retarded_GreensFun_time} is only a portion of the full GF, given by the contribution of the simple poles (the QNFs) in $\tilde{G}^{(2)}$ to the Fourier transform in Eq.~\eqref{eq:antitransf_tildeG2}.
The full Fourier transform of $\tilde{G}^{(2)}$ receives contributions also from the branch cut and from the $\mathrm{Re}(\omega)\gg 1$ contour.
Since the branch cut is present only in the lower-half $\omega$-plane, and we close the contour in this portion of the plane for $t-r_*\geq\mathcal{C}(t',r_*')$, the branch cut contribution coming from $\tilde{G}^{(2)}$ will also carry a factor $\theta\left[t-r_*-\mathcal{C}(t',r_*')\right]$ in time domain.
The response propagated (in time domain) by the Fourier anti-transform of $\tilde{G}^{(1)}$ is, instead, unaffected by this causality condition, since $\tilde{G}^{(1)}$ is analytical at the QNFs. We argue that this piece originates the prompt response, propagating signals on the curved light-cone $t-r_*\geq t'-r_*'$ \cite{Andersson:1996cm}. 
In this work, we focus on the QNM component, and leave the treatment of the other two contributions to future studies.
\subsection{QNM signal}
\label{subsec:QNM_signal}
We now compute the QNM response to a test-particle falling into a Schwarzschild BH on a generic trajectory, by solving Eq.~\eqref{eq:RWZ_equation} with null initial data.
Substituting the source in Eq.~\eqref{eq:source_generic_expression} and the QNM GF, Eq.~\eqref{eq:QNMs_Retarded_GreensFun_time}, into the general solution Eq.~\eqref{eq:GreensFun_source_convolution}, yields 
\begin{equation}
\begin{split}
\Psi_{\ell m}(t,r_*)=\sum_{n,p} B_{\ell mnp} e^{-i\omega_{\ell mnp}(t-r_*)} \, \,  \left[c_{\ell m np}(t-r_*) + i_{\ell m np}(t-r_*) \right] \, ,
\end{split}
\label{eq:signal_exc_imp_coeffs}
\end{equation}
where we defined 
\begin{align}
c_{\ell mnp}(t-r_*)=&\int_{-\infty}^{\infty}dt'\int_{-\infty}^{\infty}dr_*'\delta(r_*'-r_*(t'))\theta\left[t-r_*-\mathcal{C}(t',r_*')\right]\,\left[u_{\ell m np}(t',r_*')f_{\ell m}(t',r_*')-\partial_{r_*'}\left(u_{\ell m np}(t',r_*')g_{\ell m}(t',r_*')\right)\right] \, , 
\label{eq:clmn_def}\\
i_{\ell mnp}(t-r_*)=&\int_{-\infty}^{\infty}dt'\int_{-\infty}^{\infty}dr_*'\delta(r_*'-r_*(t'))\delta\left[t-r_*-\mathcal{C}(t',r_*')\right] \frac{\partial \mathcal{C}(t',r_*')}{\partial r_*'} \, u_{\ell m np}(t',r_*')g_{\ell m}(t',r_*')  \, .
\label{eq:ilmn_def}
\end{align}
We compute Regge-Wheeler modes ($\ell+m$ odd) through a Mathematica notebook implementing Leaver's algorithm \cite{leaver1985analytic}.
The Zerilli modes ($\ell+m$ even) are computed from the Regge-Wheeler ones through the Chandrasekhar transformations (see Chapter 4 of Ref.~\cite{Chandrasekhar:1985kt}), reviewed in App.~\ref{app:Chandra_transf}.
In order to track the role of the causality condition, we write $u_{\ell m np}$ in Eqs.~\eqref{eq:clmn_def},~\eqref{eq:ilmn_def} as
\begin{equation}
u_{\ell m np}^{\rm odd}(t',r_*') = e^{i\omega_{\ell m np} \mathcal{C}(t',r_*')} \, \hat{a}(\omega_{\ell mnp},r') \, ,
\label{eq:rw_mode_general_expr}
\end{equation}
with $\hat{a}(\omega_{\ell mnp},r')$ as in Eq.~\eqref{eq:hata_def} for Regge-Wheeler modes and
\begin{equation}
u_{\ell m np}^{\rm even}(t',r_*') = e^{i\omega_{\ell m np} \mathcal{C}(t',r_*')} \, \hat{z}(\omega_{\ell mnp},r') \, ,
\label{eq:zerilli_mode_general_expr}
\end{equation}
for Zerilli modes. 
The expression for $\hat{z}(\omega,r)$ is lengthy, and shown in App.~\ref{app:Chandra_transf}.

The two terms in Eqs.~\eqref{eq:clmn_def},~\eqref{eq:ilmn_def} are labeled ``activation'' $c_{\ell m np}$ and ``impulsive'' $i_{\ell m np}$ coefficients, respectively.
The activation coefficients originate from the $\delta(r_*'-r_*(t'))$ piece in the source, and yield the constant QNMs amplitudes at late-times. 
These coefficients are an integral over the past history of the source, hence they accumulate in time more or less efficiently depending on the overtone number (i.e. on the decay rate of the eigenmode).
The coefficients $i_{\ell mns}$ come from the $\partial_{r_*}\delta(r_*'-r_*(t'))$ portion of the source in Eq.~\eqref{eq:source_generic_expression}, after integrating by parts in $dr_*'$.
This integration is justified since the function convoluted with the Dirac delta vanishes at the boundaries (as discussed in Sec.~\ref{subsec:QNMs_propagation_and_regularity}, the causality condition regularize the diverging QNMs eigenmodes at $\mathcal{H}^+$).
The $i_{\ell mnp}$ are local terms, hence do not depend on the past history of the source, contrary to the $c_{\ell mnp}$, and do not appear in the absence of a persistent source (e.g., in the response to initial data confined to one slice).
For this reason, we have denoted the $i_{\ell mnp}$ as impulsive coefficients.
We will discuss these coefficients in more detail in Sections~\ref{subsec:NearH+_divergence},~\ref{sec:time_dependent_early_ringdown}.
The double integral in the definition of $i_{\ell mnp}$ can be solved using the properties of the Dirac delta. We first use $\delta(r_*'-r_*(t'))$ to solve the integral in $r_*'$ and evaluate the integrand in $dt'$ on the trajectory $r_*(t')$.
Then, we use the following Dirac delta property to compute the integral on $t'$ 
\begin{equation}
\delta\left[t-r_*-\mathcal{C}(t',r_*(t'))\right]=\frac{r^2(\bar{t})\delta(t'-\bar{t})}{\dot{r}_*(\bar{t})\left[2r_h^2-r^2(\bar{t})\right]-r^2(\bar{t})} \, ,
\label{eq:delta_simplified}
\end{equation}
where we have introduced $\bar{t}=\bar{t}(t-r_*)$ solution of
\begin{equation}
t-r_*-\mathcal{C}(\bar{t},r_*(\bar{t}))=0 \, .
\end{equation}
Substituting Eq.~\eqref{eq:delta_simplified} into Eq.~\eqref{eq:ilmn_def}, and performing the integration in $t'$, we find the expression for the impulsive coefficients
\begin{equation}
i_{\ell mnp}(t-r_*)=\frac{\left[r^2(\bar{t})-8\right] u_{\ell mnp}(\bar{t},r_*(\bar{t})) g_{\ell m}(\bar{t},r_*(\bar{t}))}{\dot{r}_*(\bar{t})\left[2r_h^2-r^2(\bar{t})\right]-r^2(\bar{t})} \, .
\label{eq:ilmn_expression}
\end{equation}
To compute the activation coefficients $c_{\ell mnp}$, we use the Dirac delta to solve the integral in $t'$ through the property
\begin{equation}
    \delta(r_*'-r_*(t'))=-\frac{\delta(t'-t(r_*'))}{\dot{r}_*(t(r_*'))}
\end{equation}
We are then left with an integral in $r_*'$, with integrand computed on the trajectory $t(r_*')$
\begin{equation}
\begin{split}
c_{\ell mnp}(t-r_*)=-\int_{\bar{r}_*}^{\infty}dr'_* \frac{1}{\dot{r}_*(t(r'_*))}
\left[u_{\ell m np}f_{\ell m}-\partial_{r_*'}\left(u_{\ell m np}g_{\ell m}\right)\right]_{(t(r_*'),r_*')} \, ,
\end{split}
\label{eq:clmn_expression}
\end{equation}
where we have defined $\bar{r}_*=\bar{r}_*(t-r_*)$ solution of
\begin{equation}
t-r_*-\mathcal{C}(t(\bar{r}_*),\bar{r}_*)=0 \, .
\label{eq:causality_trajectory}
\end{equation}
We will solve the integral Eq.~\eqref{eq:clmn_expression} numerically in Section~\ref{sec:time_dependent_early_ringdown}.
It is useful to introduce two new functions $\psi_{\ell mnp},\,\zeta_{\ell mnp}$ denoted respectively as \textit{activation} and \textit{impulsive contributions} to the full QNMs signal in Eq.~\eqref{eq:signal_exc_imp_coeffs}, as 
\begin{align}
\psi_{\ell mnp}(t-r_*)\equiv& B_{\ell mnp} \, c_{\ell mnp}(t-r_*)\, e^{-i\omega_{\ell mnp}(t-r_*)} \, , \label{eq:activation_contribution}\\
\zeta_{\ell mnp}(t-r_*)\equiv& B_{\ell mnp}\,i_{\ell mnp}(t-r_*) \, e^{-i\omega_{\ell mnp}(t-r_*)} \, .
\label{eq:impulsive_contribution}
\end{align}
So that we can rewrite Eq.~\eqref{eq:signal_exc_imp_coeffs} as
\begin{equation}
\Psi_{\ell m}(t,r_*)=\sum_{n,p}\left[\psi_{\ell mnp}(t-r_*)+\zeta_{\ell mnp}(t-r_*)\right] \, .
\label{eq:signal_exc_imp_contribs}
\end{equation}
%

\subsection{QNM signal after light-ring crossing} %
\label{subsec:NearH+_divergence}                  %
%
To gain insight into the QNM excitation, we now analyze the behavior of the activation coefficients, Eq.~\eqref{eq:clmn_expression}, after the particle apparent location ($\bar{r}$ in Eq.~\eqref{eq:causality_trajectory}) has crossed the LR and is falling towards the horizon $\mathcal{H}^+$, i.e. for values of the integrand variable in the interval $r'\in(r_h,r_{\rm LR}]$, with $r_{\rm LR}=3$ in Schwarzschild.
Near $\mathcal{H}^+$, at leading order, the source functions and QNMs have the following behavior
\begin{align}
f_{\ell m}(t(r_*'),r_*'),g_{\ell m}(t(r_*'),r_*')\propto & \, (r'-r_h) +\mathcal{O}[(r'-r_h)^2] \,  ,
\label{eq:source_funs_NearH+}\\
u_{\ell m np}(t(r_*'),r_*')\propto   &\, e^{i\omega_{\ell m np}t(r')}(r'-r_h)^{-i\omega_{\ell mnp}r_h}\left[1+\mathcal{O}(r'-r_h)\right] \, . 
\label{eq:QNMs_mode_NearH+}
\end{align}
Inside the LR, the motion of a test-particle is quasi-geodesic~\cite{Buonanno:2000ef,Ori:2000zn} and it holds, in the limit $r\rightarrow r_h$
\begin{equation}
t(r')\propto-r_h\log(r'-r_h)+\mathcal{O}(r'-r_h) \, , 
\label{eq:traj_NearH+}
\end{equation} 
i.e.,
\begin{equation}
e^{i\omega_{\ell mn p}t(r')}\propto (r'-r_h)^{-i\omega_{\ell mnp}r_h}\left[1+\mathcal{O}(r'-r_h)\right] \, .
\label{eq:traj_NearH+_II}
\end{equation} 
This follows from an expansion around $r=r_h$ of the orbit, using e.g.~Eq.~(10.27) of~\cite{Ferrari:2020}.

Since we focus on retarded times $t-r_*$ such that $\bar{r}(t-r_*)< r_{\rm LR}$, we can split the integral in Eq.~\eqref{eq:clmn_expression}, in an integral over $r'\in[r_{\rm LR},\infty)$ and one over $r_*'\in[\bar{r},r_{\rm LR}]$
\begin{equation}
c_{\ell mnp}^{\bar{r}<r_{\rm LR}}=c_{\ell mnp}(\bar{r}=r_{\rm LR})-\int_{\bar{r}}^{r_{\rm LR}}dr' \frac{A^{-1}(r')}{\dot{r}_*(t(r'_*))}
\left[u_{\ell m np}f_{\ell m}-\partial_{r_*'}\left(u_{\ell m np}g_{\ell m}\right)\right]_{(t(r_*'),r_*')} \, .
\label{eq:clmn_expression_inside_LR}
\end{equation}
We consistently expand the source functions $f_{\ell m}(t(r_*'),r_*'),\, g_{\ell m}(t(r_*'),r_*')$ and the QN eigenmodes $u_{\ell mnp}(t(r_*'),r_*')$ near $\mathcal{H}^+$ at $r=r_h$, fixing the test-particle energy and angular momentum $\hat{H},\,p_{\varphi}$ to their values at the LR.
We then substitute these Taylor-expanded expressions in the second term on the right-hand side of Eq.~\eqref{eq:clmn_expression_inside_LR} and expand its integrand in the same limit.
Considering the leading order behavior in 
Eqs.~\eqref{eq:source_funs_NearH+},~\eqref{eq:QNMs_mode_NearH+} and~\eqref{eq:traj_NearH+_II}, the near-horizon contribution to $c_{\ell m np}$ can be approximated as
\begin{equation}
c_{\ell mnp}^{\bar{r}<r_{\rm LR}}(t-r_*)- c_{\ell mnp}(\bar{r}=r_{\rm LR})\simeq \sum_{j=0}^{\infty}\gamma_{j,\ell mnp}\left[1-\left(\bar{r}-r_h\right)^{j+1-2i\omega_{\ell mnp}r_h} \right]\, ,
\label{eq:clmn_NearH+}
\end{equation}
where we have defined
\begin{equation}
\gamma_{j,\ell mnp}\equiv\frac{\xi_{j,\ell mnp}}{j+1-2i\omega_{\ell mnp}r_h}\,,
\label{eq:gamma_klmn_def}
\end{equation}
with $\xi_{j,\ell mnp}$ constant coefficients which depend on the geodesic parameters (energy and angular momentum of the test-particle at the LR crossing).

In the above sum, terms for which $2r_h|\omega_{\ell mn}^{\rm Im}|-j>1$ diverge for $\bar{r}= r_h$.
For a Schwarzschild BH, $2r_h|\omega_{2 2n}^{\rm Im}|>1$ for $n>0$, hence at least one term in the $j$-sum is divergent for the overtones.
Only the activation coefficients of the fundamental mode and its mirror mode, $c_{220\pm}$, are regular as $\bar{r}\rightarrow r_h$.
However, this divergence does not appear in the observable waveform.
In fact, $\bar{r}$ is an \textit{apparent} trajectory, function of the observer retarded time $t-r_*$. 
It corresponds to the point of the test-particle trajectory at which the signal that reaches $\mathcal{I}^+$ at the retarded time $t-r_*$ is emitted, traveling on the light-cone portion selected by our QNM causality condition~\eqref{eq:QNMs_causality}.
An observer can never ``see" an object fall through the event horizon of a BH, since the signals emitted by the object are infinitely redshifted in this limit. 
This translates in $t-r_*\rightarrow\infty$ as $\bar{r}\rightarrow r_h$, which contributes a redshift factor that automatically regularizes the coefficient $c_{\ell mnp}$.
This can be seen explicitly as follows.
By definition of $\mathcal{C}(t,r)$ in Eq.~\eqref{eq:QNMs_causality}, along a geodesic trajectory inside the LR as in Eq.~\eqref{eq:traj_NearH+}, we can approximate
\begin{equation}
    \mathcal{C}(t(\bar{r}),\bar{r}_*)\simeq-2r_h\log\left(1-\frac{r_h}{\bar{r}}\right) \ \ , \ \ \bar{r}\rightarrow r_h \, .
    \label{eq:causality_NearH+}
\end{equation}
Substituting into Eq.~\eqref{eq:causality_trajectory} yields
\begin{equation}
\bar{r}-r_h=\bar{r} \cdot e^{-(t-r_*)/(2r_h)} \, .
\label{eq:apparent_trajectory_near_horizon}
\end{equation}
Note that we have recovered the well known redshift factor for a Schwarzschild black hole, given by its horizon surface gravity $\kappa_{h}$ (see e.g. Eq.\,(12.5.4) of~\cite{wald:1984})
\begin{equation}
\kappa_{h}=\frac{1}{2r_h}\, .
\end{equation}
When we insert the near-horizon apparent trajectory \eqref{eq:apparent_trajectory_near_horizon} into the expansion of the activation coefficient \eqref{eq:clmn_NearH+}, we find
\begin{equation}
\begin{split}
c^{\bar{r}<r_{\rm LR}}_{\ell mnp}(t-r_*)-c_{\ell mnp}(\bar{r}=r_{\rm LR})\simeq \sum_{j=0}^{\infty}\gamma_{j,\ell mnp}\, \left[1-\bar{r}^{j+1-2i\omega_{\ell mnp}r_h}e^{-\frac{j+1}{2r_h}(t-r_*)}e^{i\omega_{\ell mnp}(t-r_*)}\right] \, .
\end{split}
\label{eq:clmn_NearH+_fun_tau}
\end{equation}
The divergence is now isolated in the factor $e^{i\omega_{\ell mnp}(t-r_*)}$, for $t-r_*\rightarrow\infty$. 
Once the full signal is reconstructed, this divergent term cancels with the $e^{-i\omega_{\ell mnp}(t-r_*)}$ factor in Eq.~\eqref{eq:signal_exc_imp_coeffs}. 
In particular, 
the activation contribution of each QNM $\psi_{\ell mnp}$, defined in Eq.~\eqref{eq:activation_contribution},
is regular at all times and, for the apparent trajectory portion $\bar{r}<r_{\rm LR}$, it holds
\begin{equation}
\psi^{\bar{r}<r_{\rm LR}}_{\ell mnp}(t-r_*) \simeq  \, \chi_{\ell m np}e^{-i\omega_{\ell mnp}(t-r_*)}+  e^{-(t-r_*)\kappa_h}\sum_{j=0}^{\infty} \alpha_{j,\ell mnp}(t-r_*)e^{-j(t-r_*)\kappa_h}\, , 
\label{eq:psi_expression}
\end{equation}
where we have defined
\begin{equation}
\begin{split}
\chi_{\ell mnp} \equiv&  B_{\ell mnp}\,\left[c_{\ell mnp}(\bar{r}=r_{\rm LR})+\sum_{j=0}^{\infty}\gamma_{j,\ell m n p}\right]\, , \\
\alpha_{j,\ell mnp}(t-r_*)\equiv& -B_{\ell mnp} \, \gamma_{j,\ell mnp} \, \bar{r}^{j+1-2r_hi\omega_{\ell mnp}} \, .
\end{split}
\label{eq:QNMs_redshift_modes_amplitudes}
\end{equation}
In the limit $t-r_*\rightarrow\infty$, it holds that $\bar{r}\rightarrow r_h$, and we have $\alpha_{j,\ell mnp}\to-B_{\ell mnp}\, \gamma_{j,\ell mnp} \, r_h^{j+1-2r_hi\omega_{\ell mnp}}$.
The first piece of Eq.~\eqref{eq:psi_expression} is a complex exponential with frequency given by the QNFs $\omega_{\ell mnp}$ and constant amplitude $\chi_{\ell mnp}$, regular for all $n$.
The second line arises from the divergent component in the activation coefficient and is also regular for all $n$. 
This term is a superposition of an infinite number of exponentially decaying, non-oscillating terms that we denote as \textit{redshift terms}.
Each of these modes decays with a multiple of the horizon surface gravity $\kappa_{h}$, with the leading mode behaving as $e^{-\kappa_{h} (t-r_*)}$.
Each redshift term has a coefficient $\alpha_{j,\ell mns}$ that saturates to a constant at late times. 
For the quadrupole, it holds that $|\omega^{\rm Im}_{22 n>0}|>\kappa_h$. 
As a consequence, the redshift term will eventually dominate over all the overtones. 
The time of transition to a redshift-led decay depends on the relative amplitudes of the QNM and the redshift terms. 
We investigate this behavior for different orbital configurations in Sec.~\ref{sec:time_dependent_early_ringdown}.

In the mathematical relativity literature, it was already known that perturbations at the horizon measured by an observer at $\mathcal{I}^+$ must be exponentially redshifted, as discussed by Rodnianski and Dafermos~\cite{Dafermos:2005eh} and by Price \cite{Price:1971fb} in the context of a spherical collapse.
Refs.~\cite{Mino:2008at,Zimmerman:2011dx} found, respectively, the leading and sub-leading redshift terms for a test-particle plunging in a Kerr BH. In particular, Refs.~\cite{Mino:2008at,Zimmerman:2011dx} focused only on the near-horizon limit, computing the convolution of the source and the GF in frequency domain, later switching to time domain.
More recently, also Ref.~\cite{Laeuger:2025zgb} identified the dominant redshift term in the frequency domain by modeling a radial infall through the GF formalism, revealing it as a complex pole in the Fourier transform of the waveform.
In this work, we compute the convolution of source and the (full) QNM GF directly in time domain. The near-horizon expansion is a byproduct of our analysis and will be tested against the full result in Sec.~\ref{sec:time_dependent_early_ringdown}. As a result, we obtain an \textit{infinite tower} of faster-decaying redshift terms,  with rates given by multipoles of the horizon redshift, that was only suggested in Ref.~\cite{Mino:2008at}.
Further discussion on redshift terms, along with a simple argument clarifying their origin and features, is provided in App.~\ref{app:redshift}.
We repeat the same calculation for the impulsive coefficients $i_{\ell mnp}$ in Eq.~\eqref{eq:ilmn_expression} and their contribution to the QNM signal $\zeta_{\ell mnp}$, as defined in Eq.~\eqref{eq:impulsive_contribution}.
Expanding both source functions and the QN eigenmodes in the limit $\bar{r}\rightarrow r_h$, given the leading order behaviors in Eqs.~\eqref{eq:source_funs_NearH+},~\eqref{eq:QNMs_mode_NearH+} and~\eqref{eq:traj_NearH+}, we find
\begin{equation}
i_{\ell mnp}^{\bar{r}<r_{\rm LR}}(t-r_*)\simeq\left(\bar{r}-r_h\right)^{1-2r_hi\omega_{\ell mnp}}\sum_{j=0}^{\infty}\beta_{j,\ell mnp}\left(\bar{r}-r_h\right)^j \, ,
\label{eq:ilmn_NearH+}
\end{equation}
with $\beta_{j,\ell mnp}$ constant coefficients of the expansion, functions of the test-particle energy and angular momentum at the LR.
For the overtones $1-2r_h|\omega_{\ell mn>0}^{\mathrm{Im}}|<0$ and, as a result, there is at least one term in the $j$-sum above that is divergent at the horizon.
Exploiting the causality condition and Eq.~\eqref{eq:apparent_trajectory_near_horizon}, we can write the impulsive contribution $\zeta_{\ell mn}$ to the full waveform as
\begin{equation}
\zeta_{\ell mnp}^{\bar{r}<r_{\rm LR}}(t-r_*) \simeq e^{-(t-r_*)\kappa_h} \sum_{j=0}^{\infty}\delta_{j,\ell mnp}(t-r_*)\cdot e^{-j(t-r_*)\kappa_h} \, ,
\label{eq:zeta_NearH+}
\end{equation}
where we have introduced
\begin{equation}
\delta_{j,\ell mnp}\equiv B_{\ell mnp} \, \beta_{j,\ell mnp}\, \bar{r}^{j+1-2r_hi\omega_{\ell mnp}} \, ,
\label{eq:delta_klmns}
\end{equation}
so that at late times $t-r_*\rightarrow\infty$, we have $\delta_{j,\ell mnp}\to B_{\ell mnp} \, \beta_{j,\ell mnp}\, r_h^{j+1-2r_hi\omega_{\ell mnp}}$.
Again, the divergence is exactly canceled by the factor $e^{-i\omega(t-r_*)}$ in the full signal, so that all terms in Eq.~\eqref{eq:zeta_NearH+} are finite.

Past literature on QNMs excitation, to the best of our knowledge, did not account for causality; hence, the radial integral in Eq.~\eqref{eq:clmn_expression} is always extended to the horizon, $\bar{r}=r_h$, (see e.g.~\cite{Zhang:2013ksa,Kuchler:2025hwx} in the case of a test-particle source) resulting in a divergence when the source of the perturbation is not compact towards $\mathcal{H}^+$.
Refs.~\cite{Leaver:1986gd,Sun:1988tz} proposed 
an analytical continuation on a deformed contour in the complex $r$-plane to regularize the integral.
This is equivalent to adding a regularizing counter term~\cite{Sberna:2021eui, Cannizzaro:2023jle}, which removes the divergent piece in Eq.~\eqref{eq:clmn_NearH+}, as we show in App.~\ref{app:regularization}. 
We will refer to this approach as ``standard regularization".

In this work, we show that the activation coefficients are time-dependent due to causality: this dependence is encoded in the lower limit of the radial integral in Eq.~\eqref{eq:clmn_expression}, $\bar{r}$. 
This quantity, defined in Eq.~\eqref{eq:causality_trajectory}, is the apparent location of the test-particle for the observer at $\mathcal{I}^+$ and is a function of its retarded time $\bar{r}=\bar{r}(t-r_*)$.
Due to the horizon infinite redshift, the activation coefficients are evaluated at $\bar{r}\rightarrow r_h$ only in the limit $t-r_*\rightarrow\infty$.
By substituting this time-dependence into Eq.~\eqref{eq:clmn_NearH+}, it is possible to see that the divergent term, removed by the standard regularization, yields a regular \textit{observable} in the QNMs signal, decaying with the BH redshift. 
However, the two methods are not inconsistent with each other. The standard regularization technique is strictly valid only in the limit $t-r_*\rightarrow\infty$. At this time, all the information emitted by the source while falling towards the horizon has escaped to $\mathcal{I}^+$, and the redshift term, not captured by the standard regularization, has effectively vanished. 
%

\section{Time-dependent ringdown}         %
\label{sec:time_dependent_early_ringdown} %
%

In this section, we investigate the activation $c_{\ell mnp}$ and impulsive $i_{\ell mnp}$ coefficients and their contribution to the waveform multipole $\Psi_{\ell m}$, defined as  $\psi_{\ell mnp}$ in Eq.~\eqref{eq:activation_contribution} and $\zeta_{\ell mnp}$  in Eq.~\eqref{eq:impulsive_contribution} respectively. 
Then, we use these contributions to construct the predicted signal propagated by the QNM GF, Eq.~\eqref{eq:signal_exc_imp_contribs}, and compare this result against numerical perturbative simulations.
We focus on the quadrupole $(\ell m) = (22)$, and consider as trajectories the quasi-circular inspiral $e_0=0.0$ in Table~\ref{tab:sims_ecc} and a radial infall from $r_0=50$ with test-particle initial energy $E_0=1.00$. 
A brief description of the eccentric inspirals with $e_0=0.5,\,0.9$ in Table~\ref{tab:sims_ecc}, yielding qualitatively analogous results to the $e_0=0.0$ case, can be found in App.~\ref{app:additional_orbits}.
We solve numerically for the trajectory by means of the \textsc{RWZHyp} code, as detailed in Sec.~\ref{sec:conventions_methods}. This quantity is then fed into the integral in Eq.~\eqref{eq:clmn_expression} for $c_{\ell mnp}$ (performed after the change of variables $dr=A(r)dr_*'$) and into the expression of Eq.~\eqref{eq:ilmn_expression} for $i_{\ell mnp}$.
We use $\tau=t-r_*+\rho_+$, as in Eq.~\eqref{eq:RWZcoordinates}, to denote the observer's retarded time. 
We also translate all times by
\begin{equation}
    \tau_{\rm LR}\equiv \mathcal{C}(t(r=r_{\rm LR}),r=r_{\rm LR})\, .
    \label{eq:tau_LR}
\end{equation}
This is the retarded time at which the signal emitted when the test-particle crosses the LR, traveling on the QNM section of the light-cone~\eqref{eq:QNMs_propagation_lightcone_section}, reaches the observer at $\mathcal{I}^+$.

\subsection{Activation $c_{22n\pm}$ and impulsive $i_{22n\pm}$ coefficients}
\label{subsec:exc_imp_coeffs}
\begin{figure*}[t]
\captionsetup{justification=centering}
\caption*{$e_0=0.0$}
\captionsetup{justification=raggedright}
\includegraphics[width=1.0\textwidth]{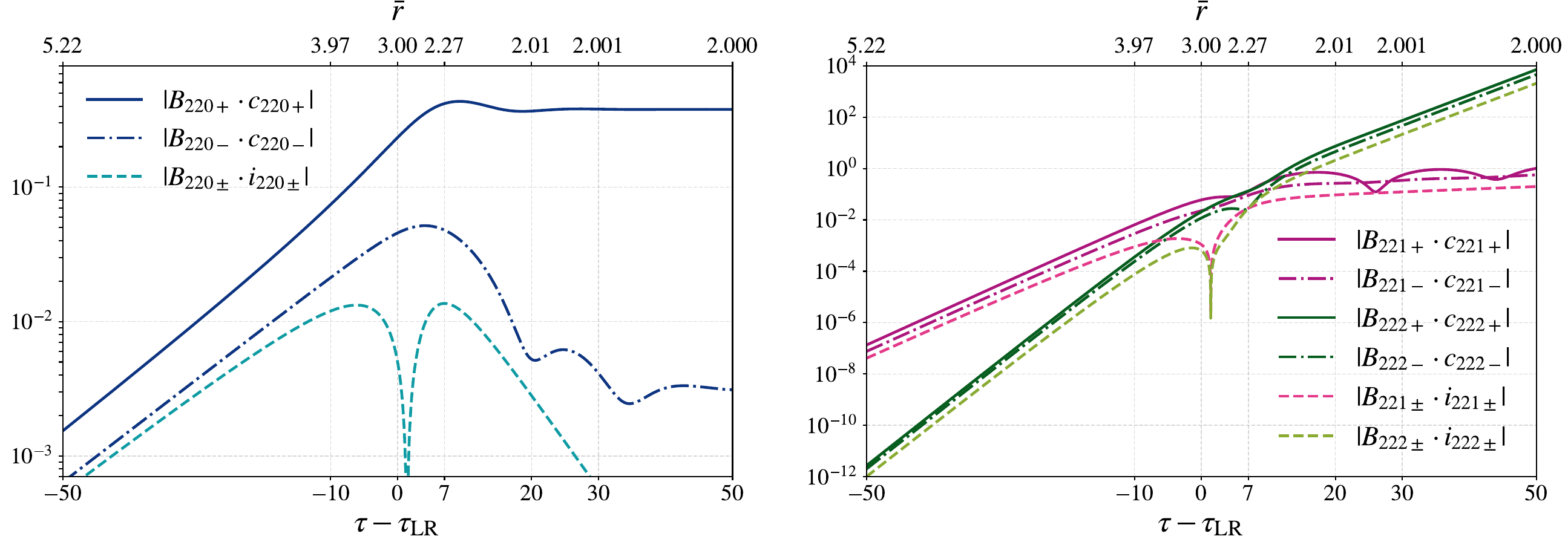}
\caption{Absolute value of the QNM activation $c_{22n\pm}$  
and impulsive $i_{22n\pm}$ coefficients
(weighted with the geometric excitation factors $B_{22n\pm}$) of the modes $(220\pm)$ (\textit{left}) and $(221\pm),\,(222\pm)$ (\textit{right}), vs the retarded time of the observer $\tau$ from the apparent LR crossing, $\tau_{\rm LR}$, Eq.~\eqref{eq:tau_LR}.
On the top horizontal axes we show the apparent location $\bar{r}$ of the test-particle, defined in Eq.~\eqref{eq:causality_trajectory}.
Results relative to a quasi-circular orbit, $e_0=0.0$ in Table~\ref{tab:sims_ecc}.
\label{fig:c22n_L22n_ID00mdea}}
\end{figure*}
\begin{figure*}[t]
\captionsetup{justification=centering}
\caption*{$r_0=50,\, E_0=1.00$}
\captionsetup{justification=raggedright}
\includegraphics[width=1.0\textwidth]{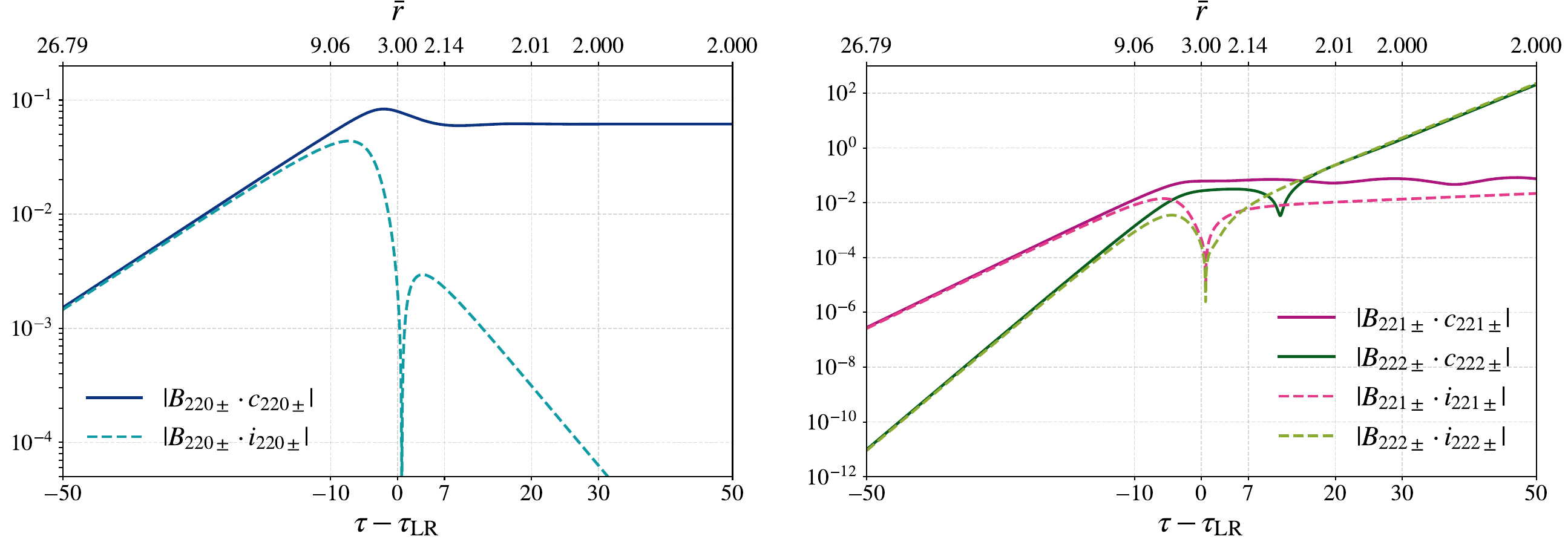}
\caption{Same as Fig.~\ref{fig:c22n_L22n_ID00mdea}, for a radial infall from $r_0=50$ with test-particle initial energy $E_0=1.00$.
Due to the symmetry of the problem, results for regular and mirror modes coincide in magnitude.
\label{fig:c22n_L22n_ID10mdea}}
\end{figure*}
In Fig.~\ref{fig:c22n_L22n_ID00mdea} we show the activation and impulsive coefficient of the fundamental mode $(220\pm)$ and the first two overtones $(221\pm),\,(222\pm)$ for the quasi-circular inspiral $e_0=0.0$ in Tab.~\ref{tab:sims_ecc}.
The $c_{220+}$ coefficient behaves approximately as an activation function: it grows until it reaches a maximum value around $\tau\approx \tau_{\rm LR}+7$,
and it saturates to a constant at late times. Interestingly, $c_{220+}$ grows even once the test-particle has crossed the LR. In particular, its maximum is reached when the test-particle is at $\bar{r}\approx 2.27$.
The mirror mode $c_{220-}$ appears to be more sensitive to the near-horizon motion: this quantity peaks around the same $\tau$ as $c_{220+}$, but saturates to a constant value at much later times, for $\tau\approx \tau_{\rm LR}+40 , \, \bar{r}\rightarrow r_h$.
The $i_{220\pm}$ contribute only close to the LR crossing and vanish at both early and late times, contrary to $c_{220\pm}$.
At late-times these results are in agreement with the expansions in Eqs.~\eqref{eq:clmn_NearH+},~\eqref{eq:ilmn_NearH+}.
Equation~\eqref{eq:clmn_NearH+} is dominated by a constant in the limit $\bar{r}\rightarrow r_h$, with all the other terms vanishing, yielding the constant amplitude of the $(220\pm)$. 
The expansion for the impulsive coefficients in Eq.~\eqref{eq:ilmn_NearH+}, instead, vanishes for $\bar{r}\rightarrow r_h$ as $(\bar{r}-r_h)^{1-2r_h|\omega^{\rm Im}_{220}|}$.
On the other hand, for the first overtone, both $c_{221\pm}$ and $i_{221\pm}$ grow (albeit slowly) in $\tau$ and do not saturate to a constant value nor vanish in the limit $\tau\gg\tau_{\rm LR},\,\bar{r}\rightarrow r_h$. 
This can be explained through the near-horizon expansions in Eqs.~\eqref{eq:clmn_NearH+},~\eqref{eq:ilmn_NearH+}, considering that $2r_h\omega_{221}^{\rm Im}\simeq 1.096\gtrsim 1 $: the leading order behavior in these expressions is given by $(\bar{r}-r_h)^{1-2r_h|\omega_{221}^{\rm Im}|}$, slowly diverging. 
Even if this divergent contribution has a smaller amplitude than the constant $\mathcal{O}[(\bar{r}-r_h)^0]$ term present in $c_{221\pm}$, it will eventually dominate the activation coefficient.
For the $(222\pm)$ modes, $2r_h|\omega^{\rm Im}_{222}|\simeq 1.91$, hence the coefficients display a faster growth.
In $c_{222\pm}$, the divergent behavior $(\bar{r}-r_h)^{1-2r_h|\omega_{222}^{\rm Im}|}$, completely swamps the constant contribution in Eq.~\eqref{eq:clmn_NearH+}. 
The retrograde mode $c_{222-}$ shows a very short time interval $\approx 5M$ in which it is approximately constant.
For higher overtones, our results are similar to what is already shown in Fig.~\ref{fig:c22n_L22n_ID00mdea} for $n>0$: $c_{22n>0\pm},\, i_{22n>0\pm}$ diverge at late times, with higher overtones displaying a faster growth.
From Eq.~\eqref{eq:clmn_expression} we see that the activation coefficients are integrals over the past history of the source.
During the plunge this dependence translates in a progressive accumulation, as shown in Fig.~\ref{fig:c22n_L22n_ID00mdea}. 
In particular, before the LR crossing the activation coefficients exhibit an exponential growth which is faster the higher is the overtone number.
From the near-horizon results of Sec.~\ref{subsec:NearH+_divergence}, in particular Eqs.~\eqref{eq:psi_expression},~\eqref{eq:QNMs_redshift_modes_amplitudes}, it then follows that the late-time constant amplitudes depend chiefly on the late stages of the plunge, near the LR crossing, in agreement with the discussion of Ref.~\cite{Price:2015gia}. The lower is the overtone number, the stronger is the dependence on the plunge at earlier times. 
In Fig.~\ref{fig:c22n_L22n_ID10mdea}, we repeat the same analysis for a radial infall from $r_0=50$ with initial energy $E_0=1.00$.
Note that the $(+),\,(-)$ modes are degenerate, as expected from the perturbation symmetry.
The $c_{220\pm}$ behavior is qualitatively similar to the quasi-circular plunge.
However, now $c_{220\pm}$ reach their maximum at $\tau\lesssim\tau_{\rm LR},$ (equivalently $\bar{r}\lesssim r_{\rm LR}$), earlier than in the quasi-circular case.

The overtone behavior is more interesting.
The $c_{221\pm}$ are increasing functions at early times but saturate to an approximate constant value just before the particle reaches the LR. 
The activation coefficient of the $n=2$ overtone and its mirror mode exhibit a similar behavior: $c_{222\pm}$ grow until a time $\tau\lesssim\tau_{\rm LR}$, then there is a transient during which $c_{222\pm}$ is approximately constant, lasting until $\tau\approx 10 +\tau_{\rm LR}$. 
Only after this time $c_{222\pm}$ start growing.
Overall, for radial infalls the amplitude of the term giving rise to the late time divergence in the overtones, $\mathcal{O}[(\bar{r}-r_h)^{1-2r_h|\omega_{22n}^{\rm Im}|}]$ in Eq.~\eqref{eq:clmn_NearH+}, is suppressed with respect to the constant term $\mathcal{O}[(\bar{r}-r_h)^0]$, giving rise to sizable intervals of approximately constant coefficients, unlike the quasi-circular case.
Finally, the impulsive coefficients $i_{22n\pm}$ are smaller than in the quasi-circular case for $\tau\gtrsim\tau_{\rm LR}$.
Overall, in the full signal propagated by the QNM GF, Eq.~\eqref{eq:signal_exc_imp_coeffs}, all terms behaving as $\propto \mathcal{O}[(\bar{r}-r_h)^{1-2r_h|\omega_{22n}^{\rm Im}|}]$ are suppressed for radial infalls. 
\begin{figure*}[t]
\captionsetup{justification=centering}
\caption*{$e_0=0.0$}
\captionsetup{justification=raggedright}
\includegraphics[width=1.0\textwidth]{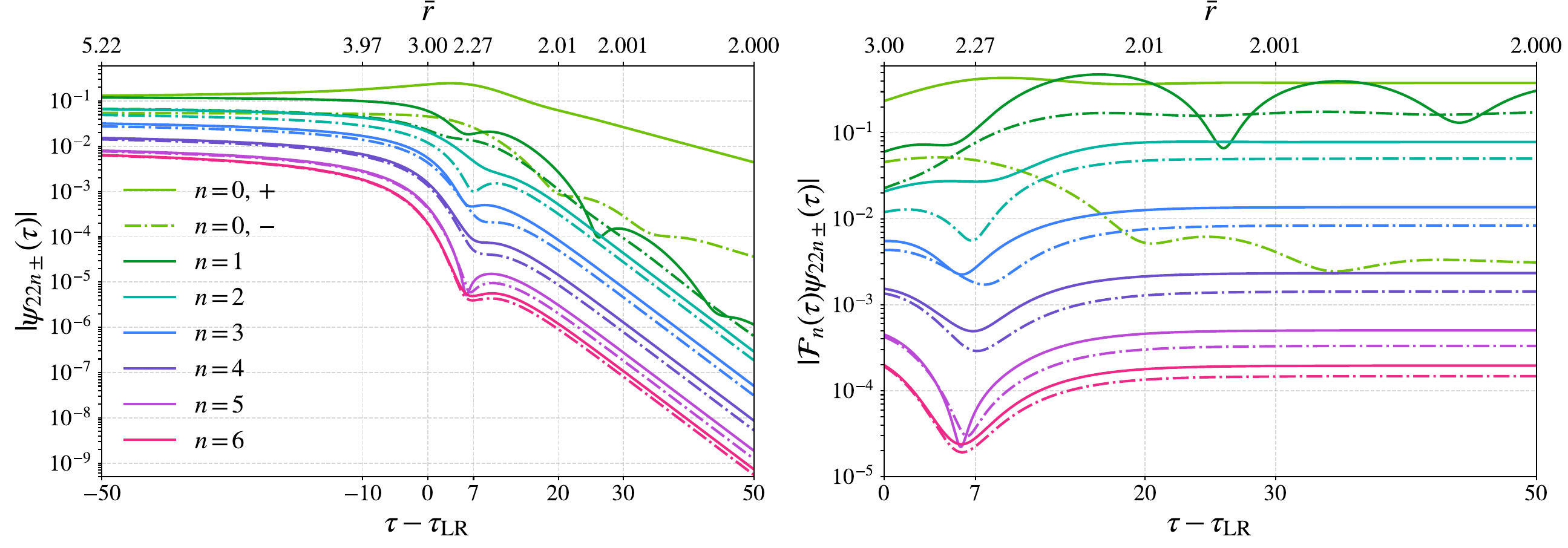}
\includegraphics[width=1.0\textwidth]{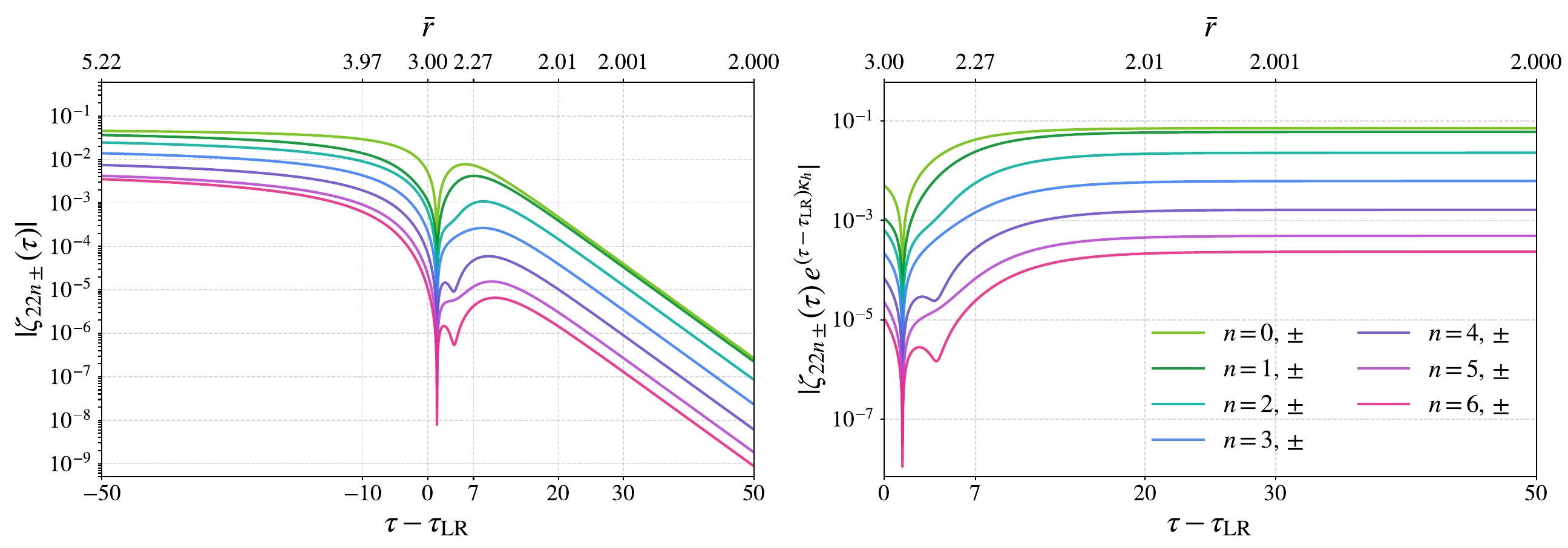}
\caption{\textit{Left}: Activation $\psi_{22n\pm}$ (top row, Eq.~\eqref{eq:activation_contribution}) and impulsive $\zeta_{22n\pm}$ (bottom row, Eq.~\eqref{eq:impulsive_contribution})  contribution to the full signal, vs.~the retarded time of the observer at $\mathcal{I}^+$, translated with the particle's apparent LR crossing $\tau_{LR}$, Eq.~\eqref{eq:tau_LR}. 
\textit{Right}: Activation and impulsive contributions rescaled by their late-time asymptotics, with $\mathcal{F}_n(\tau)$ defined in Eq.~\eqref{eq:rescaling_functions}.
We show several $(22n+)$ modes (thick lines) and mirror modes $(22n-)$ (dot-dashed), with overtones marked in different colors.
For the impulsive contribution $|\zeta_{22n+}|=|\zeta_{22n-}|$.
On the top horizontal axes the apparent location $\bar{r}$ of the test-particle emitting the signal observed at $\tau$, as defined in Eq.~\eqref{eq:causality_trajectory}.
Results relative to a quasi-circular inspiral-plunge, $e_0=0.0$ in Table~\ref{tab:sims_ecc}.
\label{fig:EV_clmn_Llmn_ID00mdea}
}
\end{figure*}
\subsection{Activation $|\psi_{22n\pm}|$ and impulsive $|\zeta_{22n\pm}|$ contributions to the waveform}
\label{subsec:exc_imp_contribs}
\begin{figure*}[t]
\captionsetup{justification=centering}
\caption*{$r_0=50,\, E_0=1.00$}
\captionsetup{justification=raggedright}
\includegraphics[width=1.0\textwidth]{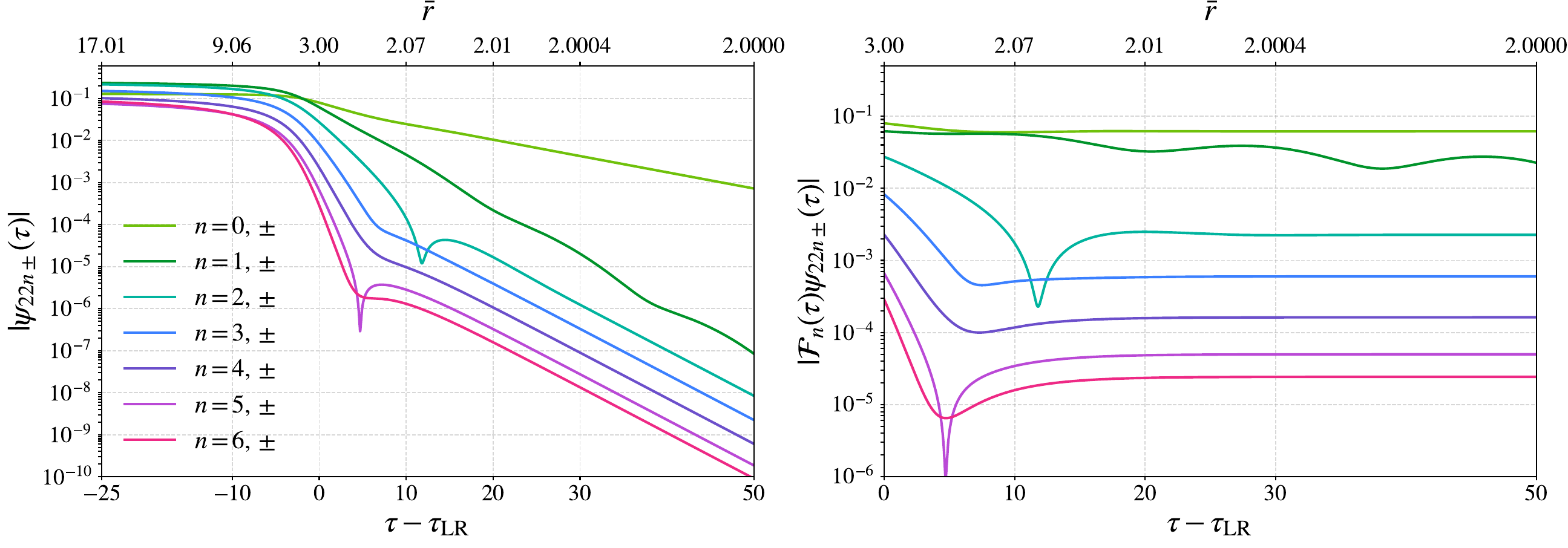}
\includegraphics[width=1.0\textwidth]{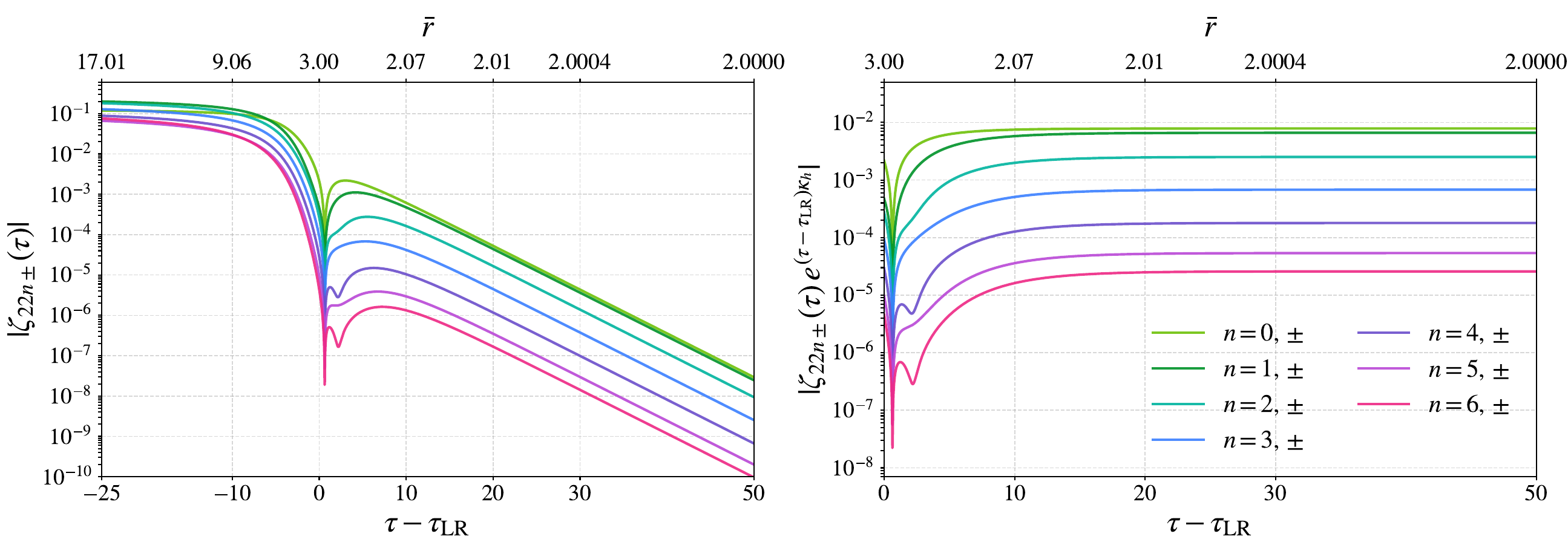}
\caption{Same as Fig.~\ref{fig:EV_clmn_Llmn_ID00mdea}, but for a radial infall from initial separation $r_0=50$ with initial energy $E_0=1.00$. 
In this case the results for regular and mirror modes are the same in absolute value.
\label{fig:EV_clmn_Llmn_ID10mdea}}
\end{figure*}
\begin{figure*}[t]
\includegraphics[width=0.49\textwidth]{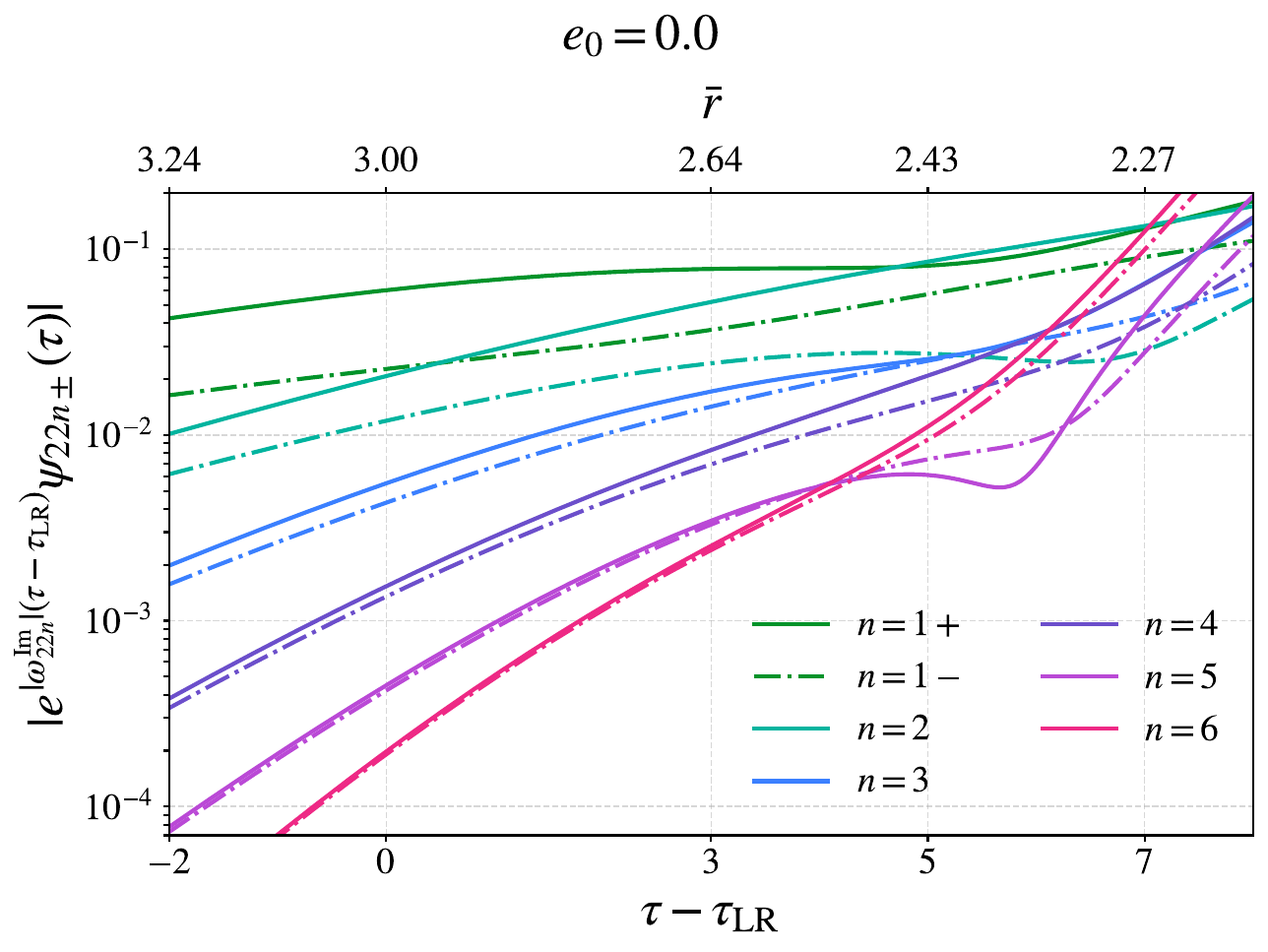}
\includegraphics[width=0.49\textwidth]{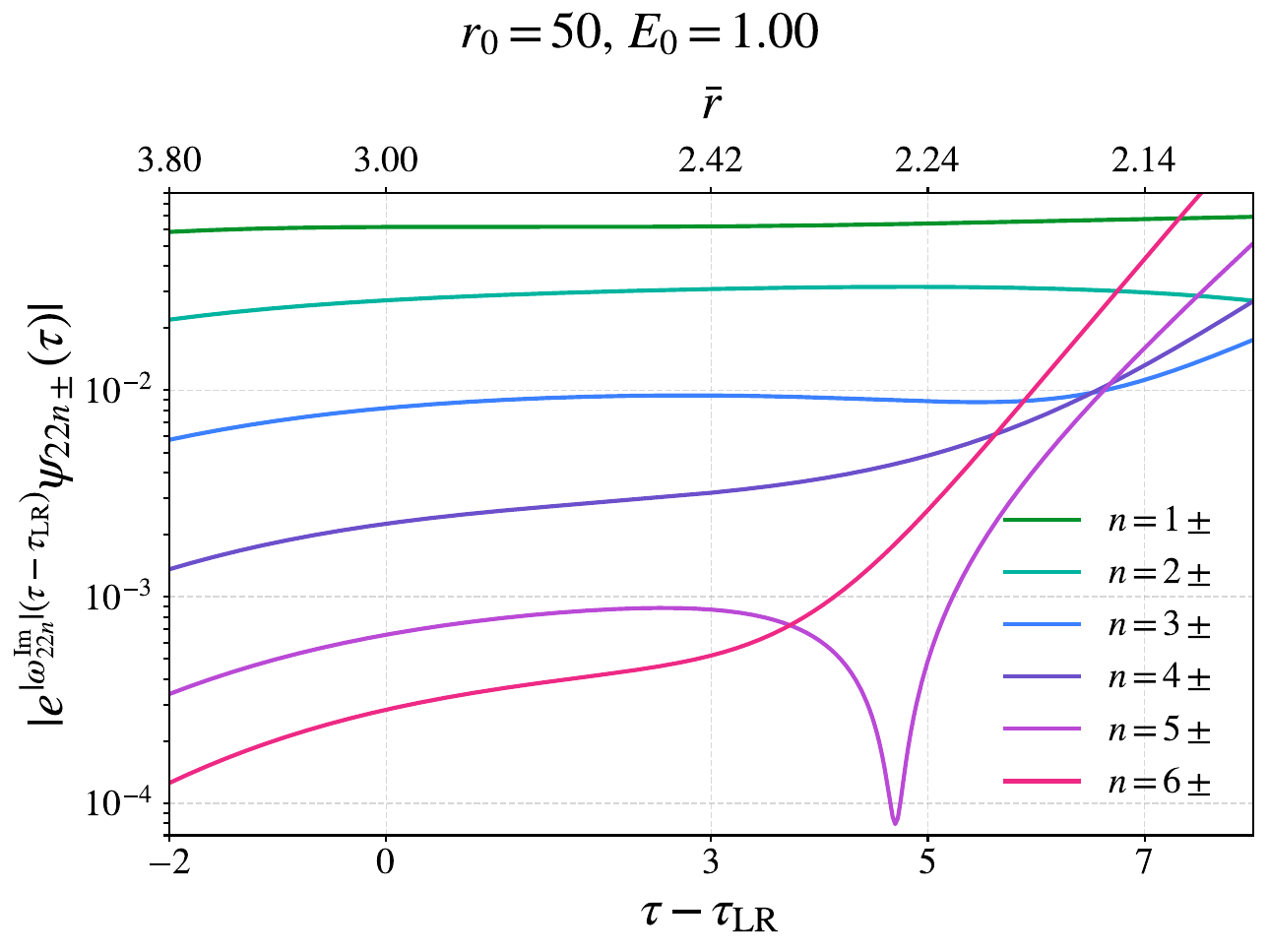}
\caption{
Activation $\psi_{22n\pm}$ contribution rescaled with the leading QN decay $e^{|\omega^{\rm Im}_{22n}|(\tau-\tau_{\rm LR})}$, vs.~the retarded time of the observer at $\mathcal{I}^+$, relative to the particle's apparent LR crossing $\tau_{LR}$, Eq.~\eqref{eq:tau_LR}. 
We show several $(22n+)$ modes (thick lines) and mirror modes $(22n-)$ (dot-dashed), with overtones marked in different colors.
On the top horizontal axes the apparent location $\bar{r}$ of the test-particle emitting the signal observed at $\tau$, as defined in Eq.~\eqref{eq:causality_trajectory}.
Results relative to a quasi-circular inspiral-plunge (\textit{left}), $e_0=0.0$ in Table~\ref{tab:sims_ecc}, and a radial infall (\textit{right}) from $r_0=50$ with initial energy $E_0=1.00$. 
Regions in which the curves are constant indicate the time interval for which the QN constant amplitude behavior is dominant in the activation contribution, before being swamped by the redshift terms.
\label{fig:OVresc_clmn_ID00-01mdea}
}
\end{figure*}
In the previous section, we have shown that for the overtones $n>0$ the coefficients diverge as $\tau\gg \tau_{\rm LR},\,\bar{r}\rightarrow r_h$, motivating this behavior with the near-horizon expansions in Eq.~\eqref{eq:clmn_NearH+}, \eqref{eq:ilmn_NearH+}.
As argued in Sec.~\ref{subsec:NearH+_divergence}, this divergence contributes to an observable and should not be regularized: even though the QN eigenfunctions are not regular at the horizon, a signal emitted at this location reaches $\mathcal{I}^+$ in an infinite amount of time, exactly canceling the divergence. 
Now we analyze the full activation $\psi_{\ell mn\pm}$ and impulsive $\zeta_{\ell mn\pm}$ contributions to the waveform for several QNMs.
We find that the pure \textit{redshift term} components coexist with the standard constant amplitude QNMs in $\psi_{\ell mn\pm}$, and are the only late-times contributions to $\zeta_{\ell mn\pm}$.
In Fig.~\ref{fig:EV_clmn_Llmn_ID00mdea}, we show the absolute value of $|\psi_{22 n\pm}|$, $|\zeta_{22 n\pm}|$ for $n\leq 6$ in a quasi-circular orbit.
The fundamental mode activation contribution (top left) grows until $ \tau\approx 7+\tau_{\rm LR} \, \,(\bar{r}\approx 2.27)$, then displays an exponential decay.
The behavior of $|\psi_{22n>1\pm}|, \, |\psi_{220-}|$ is different: these quantities do not have a peak near the LR, but are decreasing functions of $\tau$ starting from early times $\lesssim\tau_{\rm LR}-50$.
Interestingly, for the modes $(220-),\,(221)$, the activation contribution is influenced by the source even for $\bar{r}-r_h\ll1$, carrying information of the near-horizon region at late times, in agreement with the behavior of their activation coefficients shown in Fig.~\ref{fig:c22n_L22n_ID00mdea}.
For $n>0$ there is a transition near $ \tau\approx 7+\tau_{\rm LR} , \,\bar{r}\approx 2.27$: before, each $|\psi_{22 n>0\pm}|$ decays in $\tau-\tau_{\rm LR}$ at a different rate, while for $\tau\gtrsim7+\tau_{\rm LR}$ all $|\psi_{22 n>0 \pm}|$ decay with the same rate.
These results are consistent with those of Sec.~\ref{subsec:exc_imp_coeffs} and  the near-horizon expansion of Eq.~\eqref{eq:clmn_NearH+}, giving rise to the redshift term $\propto e^{-\tau \,\kappa_h}$ in Eq.~\eqref{eq:psi_expression}.
To better interpret this result, in the top right of Fig.~\ref{fig:EV_clmn_Llmn_ID00mdea} we show the behavior of the rescaled quantity $
|\mathcal{F}_n(\tau) \cdot \psi_{22n\pm}(\tau)|$, where
\begin{equation}
\begin{split}
&\mathcal{F}_{n=0}(\tau)\equiv  e^{|\omega_{220}^{\rm Im}|(\tau-\tau_{\rm LR})} \, , \\
&\mathcal{F}_{n>0}(\tau)\equiv e^{(\tau-\tau_{\rm LR})\, \kappa_h} \, .
\end{split}
\label{eq:rescaling_functions}
\end{equation}
For $\tau\gtrsim 7+\tau_{\rm LR}$, $|\mathcal{F}_n(\tau)\psi_{22n\pm}(\tau)|$ saturate to a constant value for all $n$: Eq.~\eqref{eq:rescaling_functions} correctly identifies the asymptotic behavior of the $|\psi_{22n\pm}|$ inverse at late times.
This result confirms the picture depicted by Eq.~\eqref{eq:psi_expression}: at late times, in each $\psi_{22n>0\pm}$, the overtone's characteristic rapid decay is \textit{swamped} by the leading redshift term.
In particular, in a quasi-circular plunge, the ratios between each QNM amplitude, $\chi_{\ell mn\pm}$, and each QNM contribution to the leading redshift term, $\alpha_{0,\ell mn \pm}$ in Eq.~\eqref{eq:QNMs_redshift_modes_amplitudes}, are such that the transition takes place close to the LR crossing, as shown in Fig.~\ref{fig:EV_clmn_Llmn_ID00mdea}. 
Note that the first overtone $(221\pm)$ has imaginary frequency $\omega_{221}^{\rm Im}\sim 0.274$ close to the redshift decay factor $\kappa_{h}= 0.25$. This explains the oscillations around a constant value for $|\mathcal{F}_n\psi_{221+}|$ as interference between the leading redshift term in $\psi_{221+}$ and its QNM decay.
These oscillations are not present for the $\psi_{221-}$ term: the QNM mirror mode $(221-)$ is less excited with respect to the $(221+)$ QNM and the leading redshift term in $n=1$.
In App.~\ref{app:additional_orbits} we show results for two eccentric configurations, $e_0=0.5,\,0.9$ in Table~\ref{tab:sims_ecc}, finding the same overall picture. 
In Fig.~\ref{fig:EV_clmn_Llmn_ID10mdea}, we repeat the analysis for the radial infall. 
The rescaled contributions $|\mathcal{F}_n \cdot \psi_{22n\pm}|$ saturate at late times towards a constant value for $n\geq2$, while $|\mathcal{F}_1 \cdot \psi_{2 21\pm}|$ is a decaying function even at late times $\tau\sim50+\tau_{\rm LR}$. 
This implies that $\psi_{221\pm}$ decays with its QNF behavior $\sim e^{-|\omega_{221}^{\rm Im}|\tau}$ even for $\tau\sim50+\tau_{\rm LR}$.
Moreover, the transition from a QN to a redshift term decay for the $n=2,3,4$ happens at later times compared to the quasi-circular and eccentric cases, that are shown in Figs.~\ref{fig:EV_clmn_Llmn_ID00mdea},~\ref{fig:EV_clmn_Llmn_ID05mdea} and~\ref{fig:EV_clmn_Llmn_ID09mdea}.
Consistently with results in Sec.~\ref{subsec:exc_imp_coeffs}, the redshift term (originating from the divergent piece in $c_{22n\pm}$)
is less excited in a radial infall compared to quasi-circular or eccentric plunges (even for large eccentricities).
We argue that this is due to the faster timescale of the radial infall, which allows less information emitted near the horizon to escape to infinity.
In Fig.~\ref{fig:OVresc_clmn_ID00-01mdea}, we show the overtones' activation contributions, this time rescaled by their QN decay $e^{|\omega^{\rm Im}_{22n}|(\tau-\tau_{\rm LR})}$. 
In the quasi-circular case, the rescaled contribution of the $(221+)$ mode is approximately constant for $2\lesssim \tau-\tau_{\rm LR}\lesssim 5$ signaling that 
this mode has already stabilized at this time, before being swamped by the redshift.
However, higher overtones and mirror modes do not exhibit a clear stationary phase. As a result, the QNM and redshift terms cannot be easily disentangled before the redshift contribution takes over, preventing us from identifying the onset of a constant-amplitude QNM phase for these modes.
In contrast, for a radial infall, the redshift is suppressed and QNMs appear with constant amplitudes
earlier in the activation contributions. For instance, for the $(221\pm)$ modes, $|e^{|\omega_{221}^{\rm Im}|(\tau - \tau_{\rm LR})} \psi_{221\pm}|$ remains approximately constant starting from $\tau - \tau_{\rm LR} \approx -2$ and until $\tau - \tau_{\rm LR} \approx 7$.
As the overtone number increases, the stationarity region narrows, and for the $n = 6,\, \pm$ modes, we cannot clearly distinguish a constant QNM amplitude before the redshift dominates.
On the bottom row of Figs.~\ref{fig:EV_clmn_Llmn_ID00mdea},\,\ref{fig:EV_clmn_Llmn_ID10mdea}, we show the contribution of the impulsive coefficients $|\zeta_{22n\pm}|$ for a quasi-circular plunge and a radial infall, see Figs.~\ref{fig:EV_clmn_Llmn_ID05mdea},\,\ref{fig:EV_clmn_Llmn_ID09mdea} of App.~\ref{app:additional_orbits} for $e_0=0.5,0.9$, qualitatively similar to $e_0=0.0$.
The $|\zeta_{22n\pm}|$ are slowly decreasing functions of $\tau, \,  \, \bar{r}$, displaying  different trends up until $\tau\sim \tau_{\rm LR} \, , \, \bar{r}\sim r_{\rm LR}$. 
From this time/apparent source location 
their behavior changes and  the $|\zeta_{22n\pm}|$ all follow the same exponential 
decay, regardless of $n$. 
To investigate this behavior, we again study
\begin{equation}
|\zeta_{22n\pm}\, e^{(\tau-\tau_{\rm LR})\,\kappa_h}| \, ,
\label{eq:rescaled_zeta}
\end{equation}
shown on the right bottom panels of Figs.~\ref{fig:EV_clmn_Llmn_ID00mdea}, \ref{fig:EV_clmn_Llmn_ID10mdea} (see also Figs.~\ref{fig:EV_clmn_Llmn_ID05mdea},~\ref{fig:EV_clmn_Llmn_ID09mdea} in App.~\ref{app:additional_orbits}).
At $\tau\gtrsim\tau_{\rm LR}$ or $\bar{r}\gtrsim 3$, this rescaled function saturates towards a constant, from which we infer the leading order coefficient of the expansion in Eq.~\eqref{eq:zeta_NearH+}.
\begin{figure*}[t]
\captionsetup{justification=centering}
\caption*{$e_0=0.0$}
\captionsetup{justification=raggedright}
\includegraphics[width=1.0\textwidth]{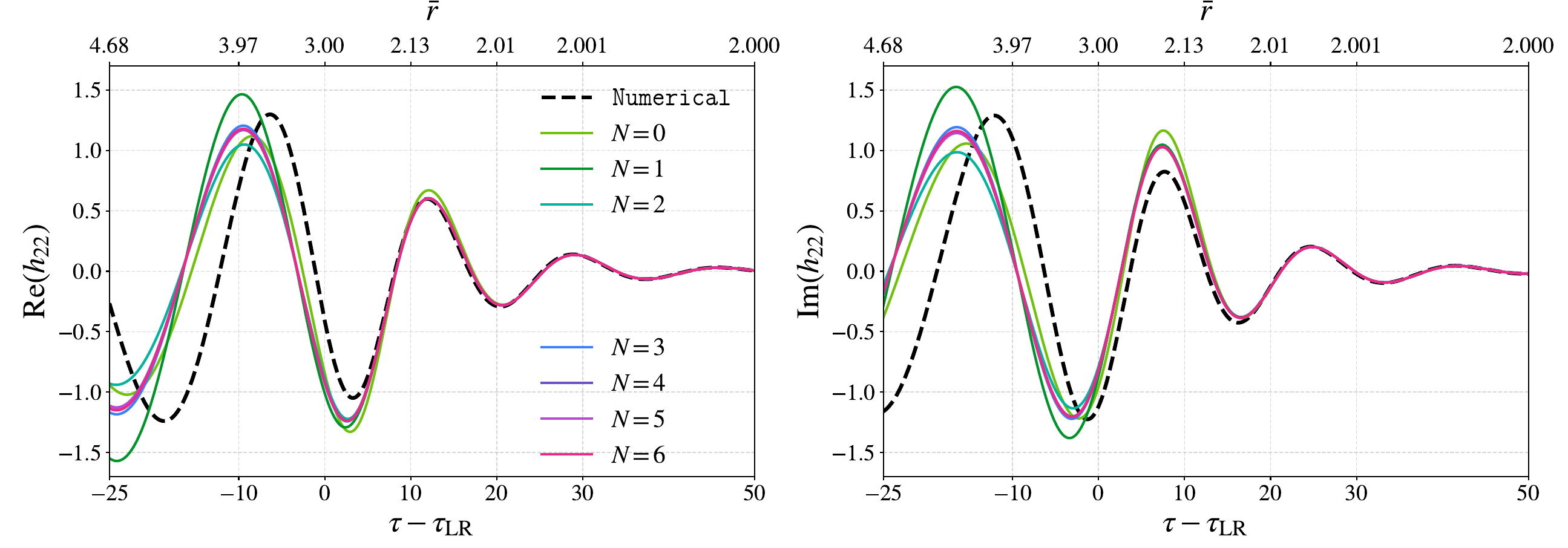}
\includegraphics[width=1.0\textwidth]{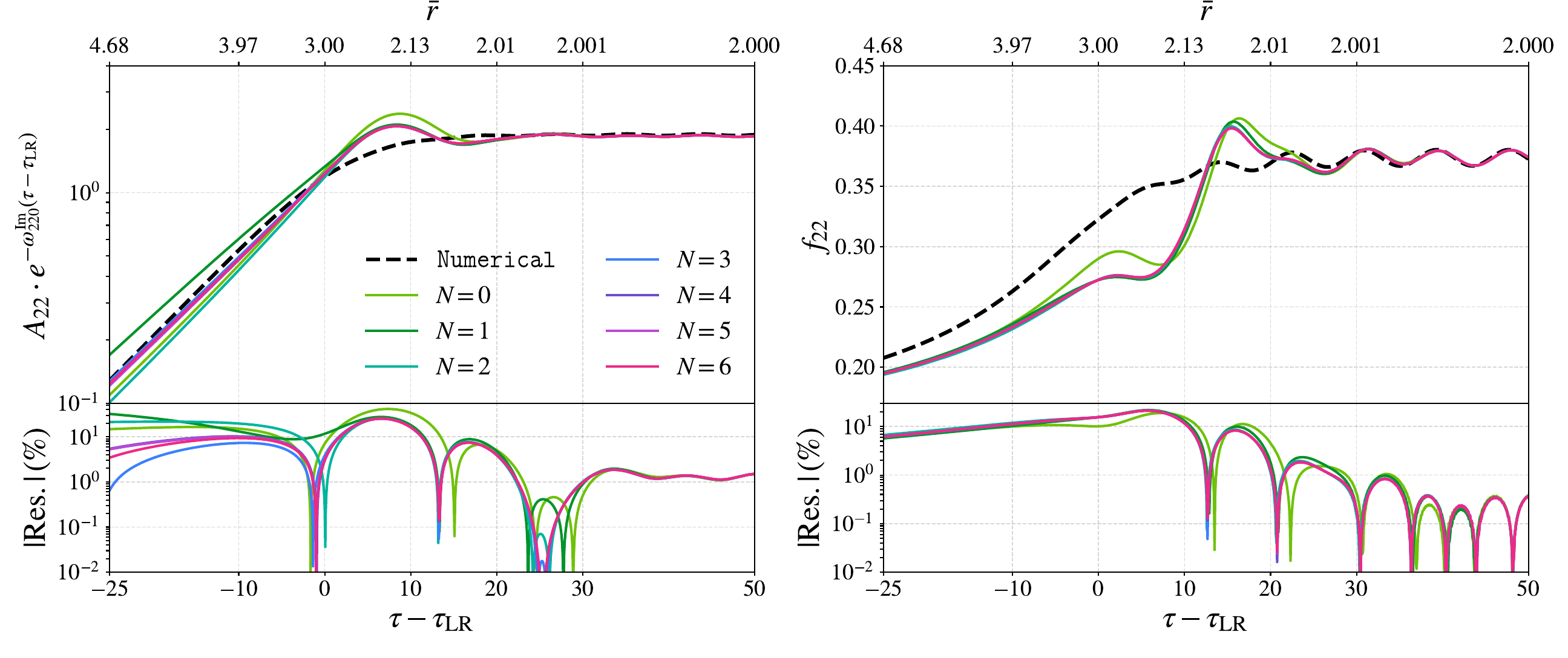}
\caption{\textit{Top}: Strain quadrupole vs the retarded time $\tau$ of the observer at $\mathcal{I}^+$.
\textit{Bottom}: Quadrupole amplitude $A_{22}$ rescaled with the factor $e^{-\omega_{220}^{\rm Im}(\tau-\tau_{LR})}$ (\textit{left}) and instantaneous frequency (\textit{right}) vs $\tau$.
Retarded times have been translated with $\tau_{\rm LR}$ as defined in Eq.~\eqref{eq:tau_LR}.
On the top horizontal axes the apparent location $\bar{r}$ of the test-particle emitting the signal observed at $\tau$, as in Eq.~\eqref{eq:causality_trajectory}.
In dot-dashed black, the numerical results obtained with the \textsc{RWZHyp} code. Thick coloured lines represent the analytical prediction Eq.~\eqref{eq:signal_exc_imp_coeffs} obtained summing over a number $N$ of overtones (colors) and their respective counter-rotating modes.
Results relative to a quasi-circular inspiral-plunge, $e_0=0.0$ in Table~\ref{tab:sims_ecc}.
In the bottom left panel, we also show the residuals between numerical and analytical predictions for different values of $N$, as defined in Eq.~\eqref{eq:residuals_definition_sourced_QNMs}.
\label{fig:Psi22_ID00mdea}}
\end{figure*}
Comparing the behavior of $|\zeta_{22n\pm}|$ in a quasi-circular or eccentric plunge, Figs.~\ref{fig:EV_clmn_Llmn_ID00mdea}, \ref{fig:EV_clmn_Llmn_ID05mdea} and \ref{fig:EV_clmn_Llmn_ID09mdea}, we see that the impulsive contributions are more excited for $\tau\lesssim\tau_{\rm LR}$ in the radial infall, Fig.~\ref{fig:EV_clmn_Llmn_ID10mdea}. 
The redshift decay post-LR crossing is instead suppressed for radial infalls also in the impulsive contributions $|\zeta_{22n\pm}|$. This is consistent with the results discussed in the previous section, where we observed a suppression of the leading divergent term in the near-horizon expansion for $i_{\ell mn\pm}$, Eq.~\eqref{eq:ilmn_NearH+}, originating the redshift contribution in the $\bar{r}\rightarrow r_h$ expansion of $\zeta_{\ell mn\pm}$, Eq.~\eqref{eq:zeta_NearH+}.
\subsection{Total QNM waveform and comparison with numerical solutions}
\label{subsec:full_waveform}

\begin{figure*}[h!]
\captionsetup{justification=centering}
\caption*{$r_0=50,\, E_0=1.00$}
\captionsetup{justification=raggedright}
\centering
\includegraphics[width=0.49\textwidth]{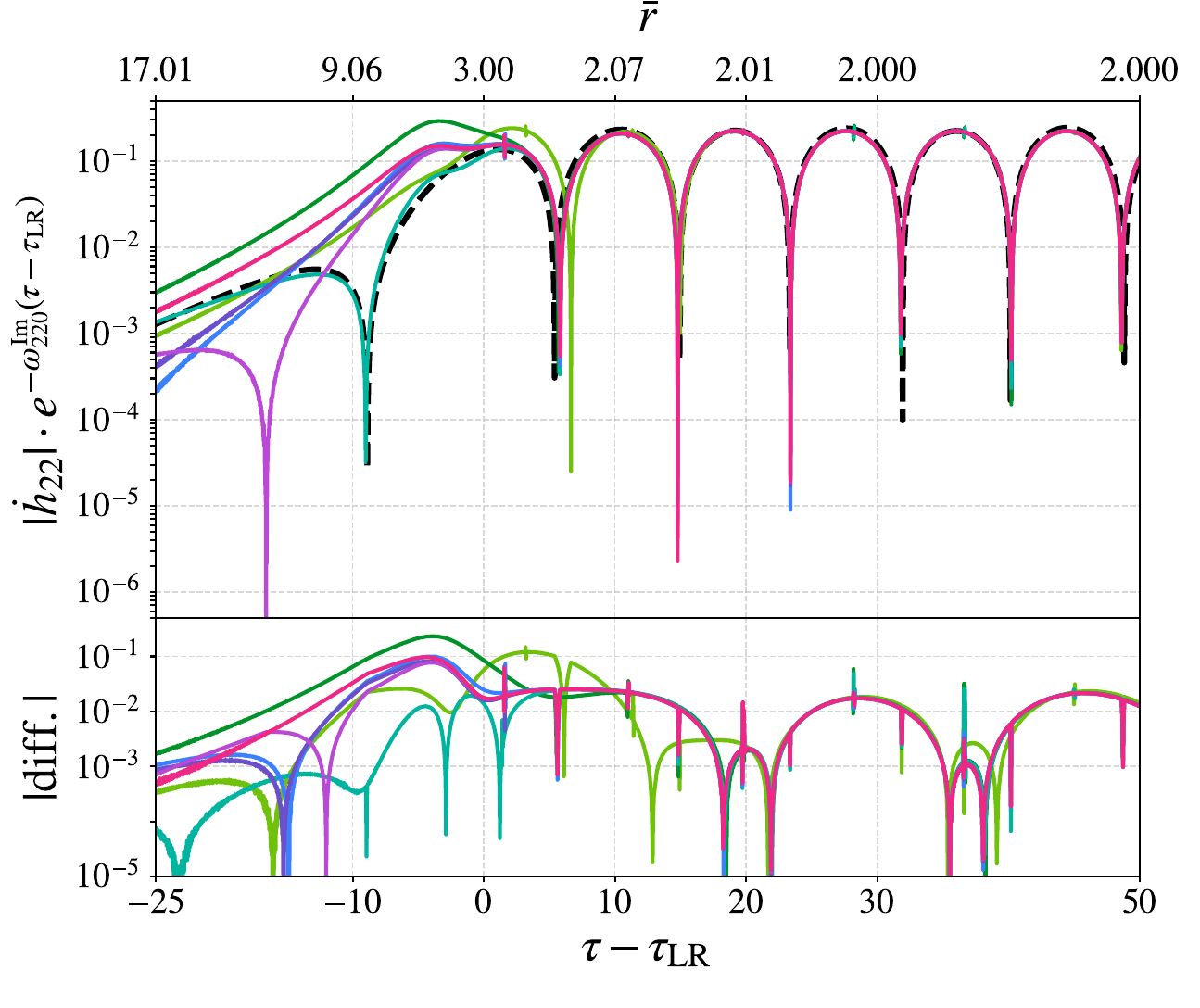}
\includegraphics[width=0.49\textwidth]{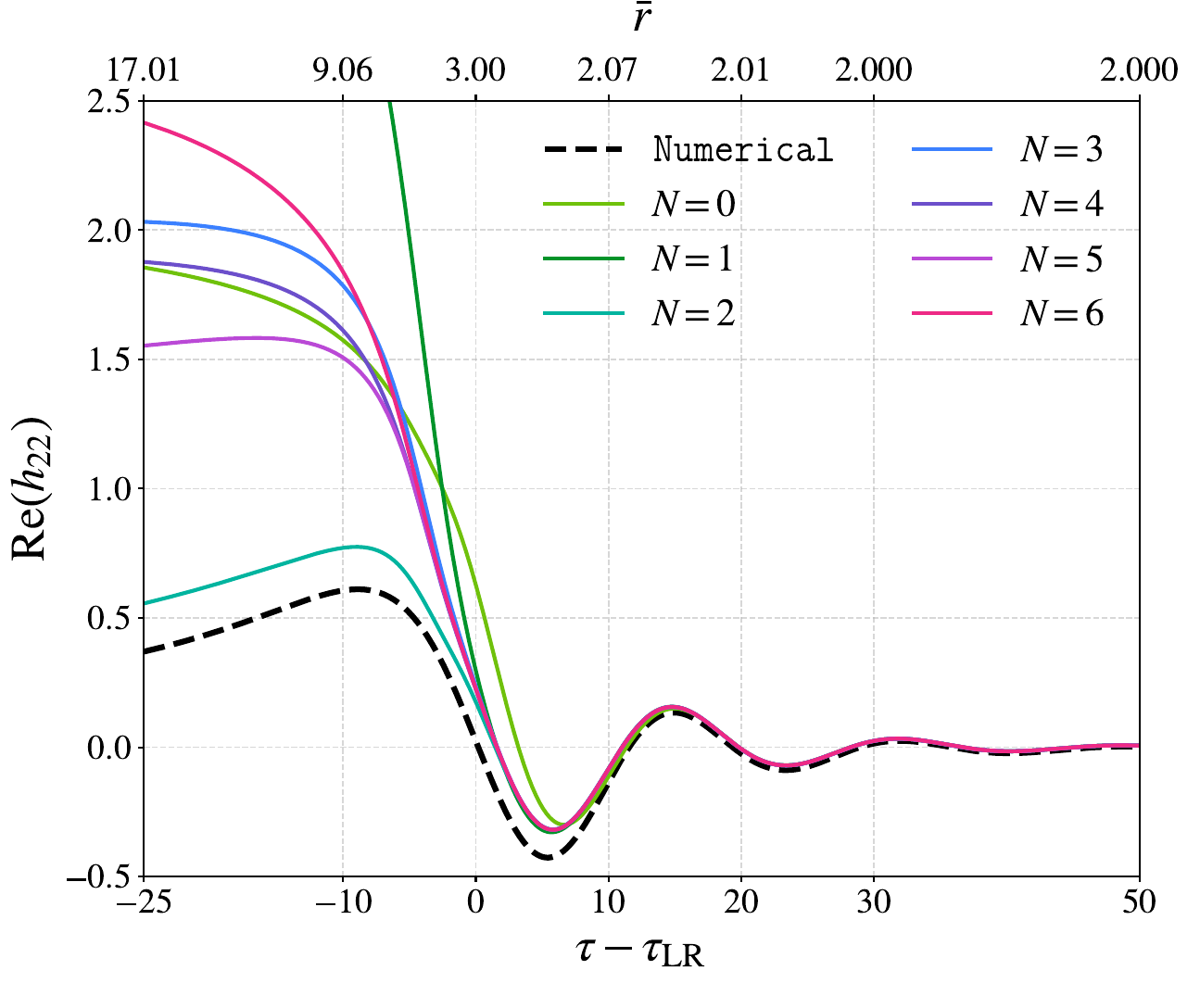}
\caption{
\textit{Left}: Quadrupolar news amplitude rescaled with the factor $e^{-\omega_{220}^{\rm Im}(\tau-\tau_{LR})}$ (top) vs the retarded time $\tau$ of the observer at $\mathcal{I}^+$  and absolute value of the difference between numerical and analytical results (bottom). 
\textit{Right}: real $(22)$ mode strain component vs $\tau$. 
The retarded time $\tau$ is translated with respect to $\tau_{LR}$, defined in Eq.~\eqref{eq:tau_LR}.
On the top horizontal axis the position of the source (test-particle) when emitting the signal later observed at $\tau$, defined in Eq.~\eqref{eq:causality_trajectory}. 
In dot-dashed grey, the numerical results obtained through the \textsc{RWZHyp} code. Thick coloured lines represent the analytical prediction~\eqref{eq:signal_exc_imp_coeffs} obtained summing over a different number $N$ of overtones (different colors) and their respective counter-rotating modes.
Results relative to a radial infall from $r_0=50$ with initial energy $E_0=1.0$.
Numerical and analytical waveforms have been aligned in phase and time.
Spikes are due to finite-resolution numerical noise amplified during the alignment procedure.
\label{fig:Psi22_ID10mdea}}
\end{figure*}
We now use the activation and impulsive contributions to construct the entire QNMs portion of the signal, Eq.~\eqref{eq:signal_exc_imp_contribs}, comparing it with the full numerical waveform obtained through the \textsc{RWZHyp} code.
We quantify the agreement between the analytical and the numerical waveforms through the percentage residuals
\begin{equation}
\mathrm{Res}.[\%]=100\cdot\frac{X_{\rm numerical}-X_{\rm analytical}}{X_{\rm numerical}} \, \, .
\label{eq:residuals_definition_sourced_QNMs}
\end{equation}
In Fig.~\ref{fig:Psi22_ID00mdea}, we show the analytical results obtained adding up to $N$ overtones and their mirror modes, contrasting them with the full numerical signal for the quasi-circular case. 
We show the strain polarizations, amplitude and instantaneous frequency observed at $\mathcal{I}^+$ as a function of the retarded time $\tau$. 
The amplitude is rescaled as $A_{22}e^{-\omega_{220}^{\rm Im}(\tau-\tau_{\rm LR})}$: since the fundamental mode dominates at late times, our calculations predict this rescaled variable should saturate to a constant.
The late-time signal at $\tau>\tau_{\rm LR}+20$ is indeed dominated by the fundamental mode, and in good agreement with our prediction, with residuals of $\sim 2\%$ in the amplitudes and $\sim 0.4\%$ in the frequencies.
At earlier times, overtones become progressively more relevant.
For $0\lesssim\tau-\tau_{\rm LR}\lesssim20$ the first two overtones must be included in the sum of Eq.~\eqref{eq:signal_exc_imp_contribs} to reach convergence in the analytical prediction.
For times $\tau\lesssim \tau_{\rm LR}$, instead, $n\approx 5$ overtones are necessary.
In this early region, residuals are much larger, with Fig.~\ref{fig:Psi22_ID00mdea} showing an amplitude overestimation in proximity of the LR crossing $\tau\lesssim 10+\tau_{\rm LR}, \, \bar{r}\lesssim 2.13$.
This result suggests that the signal propagated through the QNM GF is not sufficient to reproduce the full signal around LR crossing --- other contributions must be taken into account, coming from the prompt response and the branch-cut portion of the GF.
The instantaneous frequency of the predicted signal grows in time, saturating to the fundamental mode frequency, with beatings due to its mirror mode.
For times $\tau\lesssim\tau_{\rm LR}$, the predicted growth in frequency is slow and similar, albeit shifted, to the numerical waveform. 
However, while the latter smoothly connects to the fundamental mode real frequency, our prediction displays a local peak at the LR crossing, and is later rapidly saturating to the fundamental QNF. 
This result can be understood from the source behavior.
Before the LR crossing the RWZ source oscillates with $m\dot{\varphi}$, and $\dot{\varphi}$ grows in time until it reaches a maximum exactly at the LR crossing. Afterwards, the orbital frequency rapidly falls off and the source becomes non-oscillating. 
Hence, before the light-ring crossing, the QNMs are quasi-resonantly excited, while afterwards, the QNMs excitation behaves as a free oscillator. After the light-ring crossing, the particular solution associated to the non-oscillating source is given by the redshift terms.
These results formalize and extend the quasi-resonant picture of Refs.~\cite{Damour:2007xr,Albanesi:2023bgi} for the QNMs excitation during the plunge.
In Fig.~\ref{fig:Psi22_ID10mdea}, we repeat the analysis for the radial infall. We show the time derivative of the only non vanishing polarization $(+)$, rescaled by its late-time asymptotics, $|\dot{h}_{22}\, e^{-\omega_{220}^{\rm Im}(\tau-\tau_{\rm LR})}|$.
The time derivative serves to obtain a cleaner comparison by suppressing the tail part of the signal, prominent in radial infalls.
The qualitative picture is unchanged with respect to the quasi-circular case.
For this binary configuration, more overtones are needed at early times $\tau\lesssim\tau_{\rm LR}$  to achieve convergence in the QNM signal. 
Even though we are not able to reproduce the initial transient leading to the ringdown, our analytical prediction can be smoothly extended to earlier times yielding an important piece in the analytical description of the plunge signal. 
We predict that at $\tau\approx\tau_{\rm LR}-25$ the QNMs portion of the signal is comparable in magnitude with the perturbative numerical waveform, and cannot be neglected.
In contrast, the standard constant amplitude ringdown picture~\cite{Baibhav:2023clw} cannot be extended to the plunge, since it diverges for times earlier than $\tau\approx\tau_{\rm LR}+10$ (often crudely referred to as \textit{ringdown starting time}).
This is in agreement with past results~\cite{Zhang:2013ksa, Kuchler:2025hwx}.

\section{Stationary ringdown}

Here, we focus on the late-time portion of the signal, where a ``stationary'' ringdown description, namely a superposition of modes with constant amplitudes, is accurate.
We discuss the behavior of the leading redshift component, and the dependence of QNM amplitudes on the inspiral configuration.

\subsection{Leading redshift term}
\label{subsec:redshift_mode}

We now investigate the contribution of the redshift terms past the LR-crossing. 
As shown in Eq.~\eqref{eq:psi_expression}, for $\bar{r}\ll r_{\rm LR}$ the activation contribution $\psi_{\ell mn\pm}$ consists of two terms: a constant amplitude QNM and an infinite tower of redshift terms, the leading one behaving as $\sim e^{-\tau \, \kappa_h}$. 
The impulsive term $\zeta_{\ell mn\pm}$ only contributes to the redshift terms, Eq.~\eqref{eq:zeta_NearH+}.
Substituting Eqs.~\eqref{eq:psi_expression} and \eqref{eq:zeta_NearH+} into Eq.~\eqref{eq:signal_exc_imp_contribs}, we can write the full QNMs signal after the LR crossing as
\begin{equation}
\begin{split}
\Psi^{\bar{r}\ll r_{\rm LR} }_{\ell m}(t-r_*)=\sum_{n=0}^{\infty}\sum_{p=\pm}\chi_{\ell mnp}e^{-i\omega_{\ell mnp}(t-r_*)}+ \sum_{j=0}^{\infty}\kappa_{j,\ell m}e^{-(j+1)(t-r_*)\kappa_h} \, ,
\end{split}
\label{eq:post_merger_signal}
\end{equation}
where we have introduced the \textit{redshift amplitudes} $\kappa_{j,\ell m}$ 
\begin{equation}
\begin{split}
\kappa_{j,\ell m}\equiv\sum_{n=0}^{\infty}\sum_{p=\pm}(\alpha_{j,\ell mnp}+\delta_{j,\ell mnp}) \, ,
\end{split}
\label{eq:redshift_amplitude}
\end{equation}
with $\alpha_{j,\ell mn\pm},\,\delta_{j,\ell mn\pm}$ defined in Eqs.~\eqref{eq:QNMs_redshift_modes_amplitudes} and~\eqref{eq:delta_klmns}, in the limit $\bar{r}\to r_h$.
\begin{figure*}[t]
\includegraphics[width=0.49\textwidth]{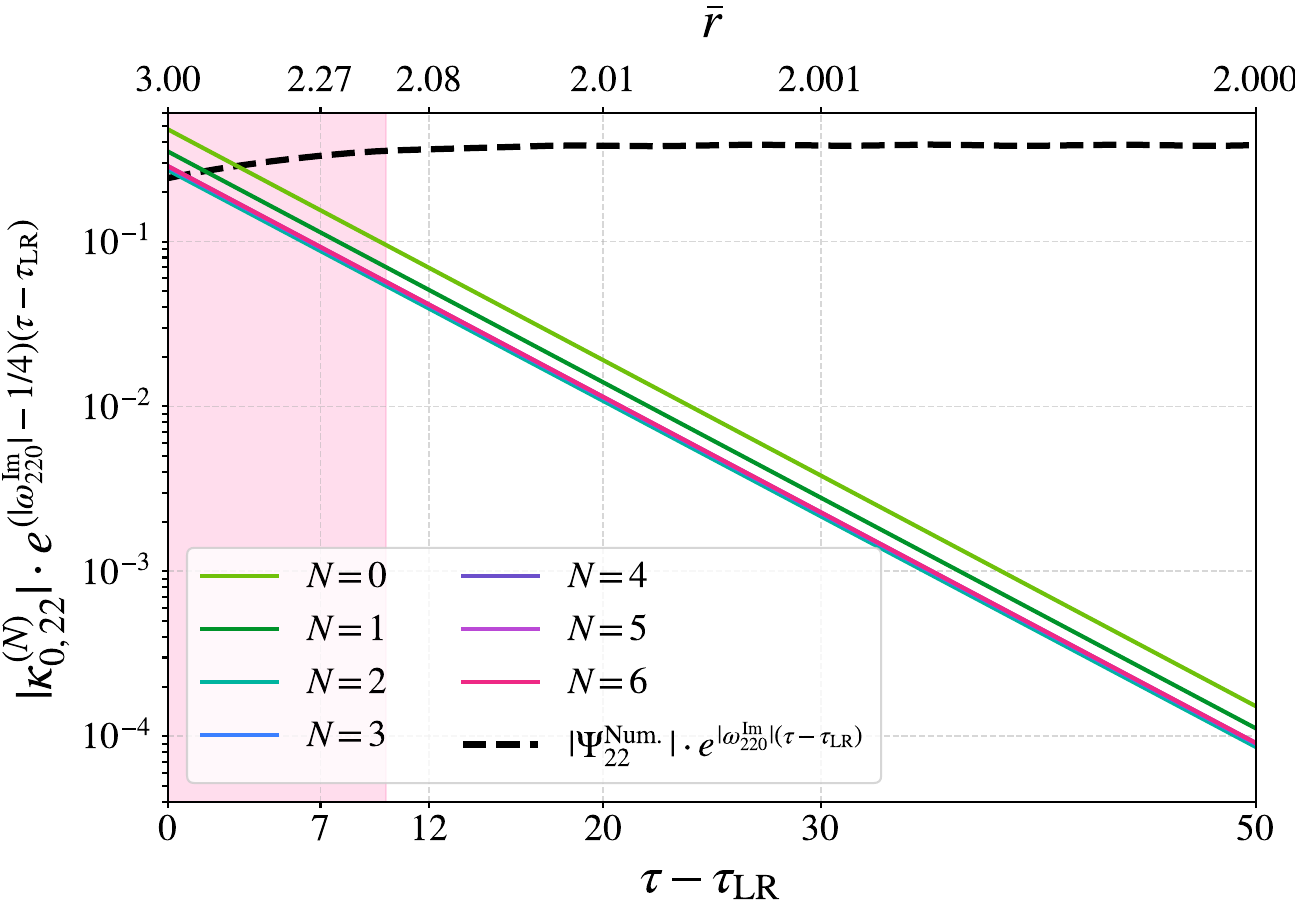}
\includegraphics[width=0.49\textwidth]{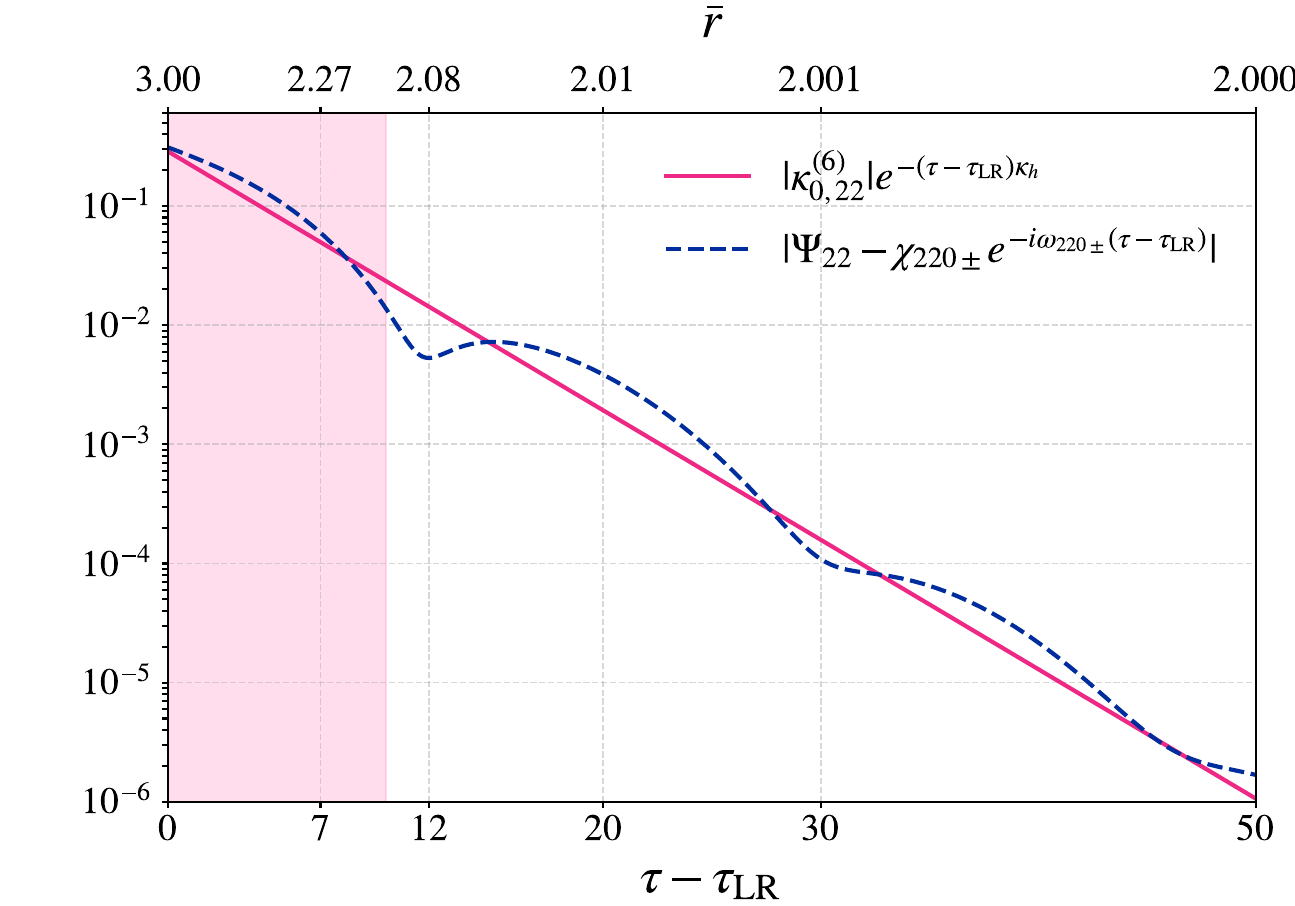}
\caption{
\textit{Left}: Leading redshift term, computed by summing the activation and impulsive contributions of both regular and mirror modes up to $N$ overtones, see Eq.~\eqref{eq:convergence_redshift_sum}, vs.~the retarded time $\tau$ of the observer at $\mathcal{I}^+$.
On the top axis, the apparent location of the test-particle, Eq.~\eqref{eq:causality_trajectory}.
In dot-dashed black, the quadrupole $\Psi^{\rm Num}_{22}$, computed numerically with the \textsc{RWZHyp} code. 
Quantities are rescaled by the fundamental mode decay.
\textit{Right}: Total leading redshift term, summing contributions up to $N=6$ in Eq.~\eqref{eq:convergence_redshift_sum} (pink, solid), and analytical prediction for the quadrupole strain $\Psi_{22}$, subtracting the $n=0$ activation contributions (violet, dashed).
Results relative to the quasi-circular inspiral $e_0=0.0$ in Table.~\ref{tab:sims_ecc}.
The redshift predictions are not valid in the shaded region.
\label{fig:redshift_plots}}
\end{figure*}
For each overtone $n$, the activation and impulsive contributions to the leading redshift term have already been analyzed in Figs.~\ref{fig:EV_clmn_Llmn_ID00mdea},\ref{fig:EV_clmn_Llmn_ID10mdea}. 
Here, we combine these contributions and study the dominant redshift amplitude $\kappa_{0,\ell m}$ and its impact on the total waveform $\Psi_{\ell m}$.
To gauge how many overtones significantly contribute to the total redshift amplitude, we introduce
\begin{equation}
\kappa^{(N)}_{j=0,\ell m}\equiv\sum_{n=0}^{N}\sum_{p=\pm}(\alpha_{0,\ell mnp}+\delta_{0,\ell mnp})\, ,
\label{eq:convergence_redshift_sum}
\end{equation}
and study its convergence when varying $N$.
We focus on the quasi-circular trajectory and
we compute $\alpha_{0,22n>0\pm}$ and $\delta_{0,22n\pm}$ by averaging the rescaled quantities shown in the right panels Fig.~\ref{fig:EV_clmn_Llmn_ID00mdea} in the range $\tau-\tau_{\rm LR}\in[40,70]$.
For the impulsive terms and the activation overtones, the redshift terms dominate at late times and this yields the desired result.
For the $n=0, \,p=\pm$ activation terms, the QNM decay dominates at late times, so we first average to compute the QNM amplitudes $\chi_{220\pm}$, and then extract $\alpha_{0,220\pm}$ by averaging
\begin{equation}
e^{(\tau-\tau_{\rm LR})\, \kappa_h}\left[\psi_{220 \pm}(\tau)-\chi_{220\pm}e^{-i\omega_{220\pm}(\tau-\tau_{\rm LR})}\right] \, .
\end{equation}
We show our results on the  redshift amplitude convergence in the left panel of Fig.~\ref{fig:redshift_plots}.
We plot the redshift term rescaled by the late-time decay of the total quadrupole, 
$|\kappa_{0,22}^{(N)}| \cdot e^{(|\omega_{220}^{\rm Im}|-\kappa_h)\cdot(\tau-\tau_{\rm LR})}$,
comparing it to the total amplitude of the quadrupole $\Psi_{22}$ obtained numerically, also rescaled.
Only the overtones $n<3$ contribute appreciably to the leading redshift term.
For $10\lesssim\tau-\tau_{\rm LR}\lesssim 25$ the redshift contribution is at most two order of magnitude smaller than the leading $220+$ mode (earlier times, where the near-horizon expansion is not valid, are shaded).
In Fig.~\ref{fig:EV_clmn_Llmn_ID00mdea}, the redshift component in the activation contribution $\psi_{22n>0\pm}$ was shown to swamp the respective overtone decay for $\tau\gtrsim 10+\tau_{\rm LR}$. 
This holds true in the total QNM signal as well, when summing over overtones $n$, mirror modes $p$, and activation plus impulsive coefficients.
In fact, the redshift amplitude only decreases by a factor of $\sim 2$ when adding together all the contributions, maintaining its dominance over overtones terms.
To visualize this result, in the right panel of Fig.~\ref{fig:redshift_plots}
we show
\begin{equation}
\bigg| \,\Psi_{22}(\tau)-\sum_{s=\pm}\chi_{220\pm}e^{-i\omega_{220}(\tau-\tau_{\rm LR})} \, \bigg| \, ,
\label{eq:quantity_invest_overt_swamp}
\end{equation}
removing the longest-lived mode contribution, where $\Psi_{22}$ is computed using Eq.~\eqref{eq:signal_exc_imp_contribs} up to $n=6$.
For $\tau\gtrsim \tau_{\rm LR}+ 15$, the decay of Eq.~\eqref{eq:quantity_invest_overt_swamp} is consistent with the leading redshift term $|\kappa_{0,22}|e^{-(\tau-\tau_{\rm LR})\, \kappa_h}$.

Analogous results are valid in the eccentric case.
For radial infalls, single redshift contributions are suppressed, hence their dominance on overtone contributions starts at later times.

%
\subsection{Dependence on the inspiral configuration} %
\label{sec:inspiral_imprints}                         %
%

%
\begin{figure*}[t!]
\includegraphics[width=1\textwidth]{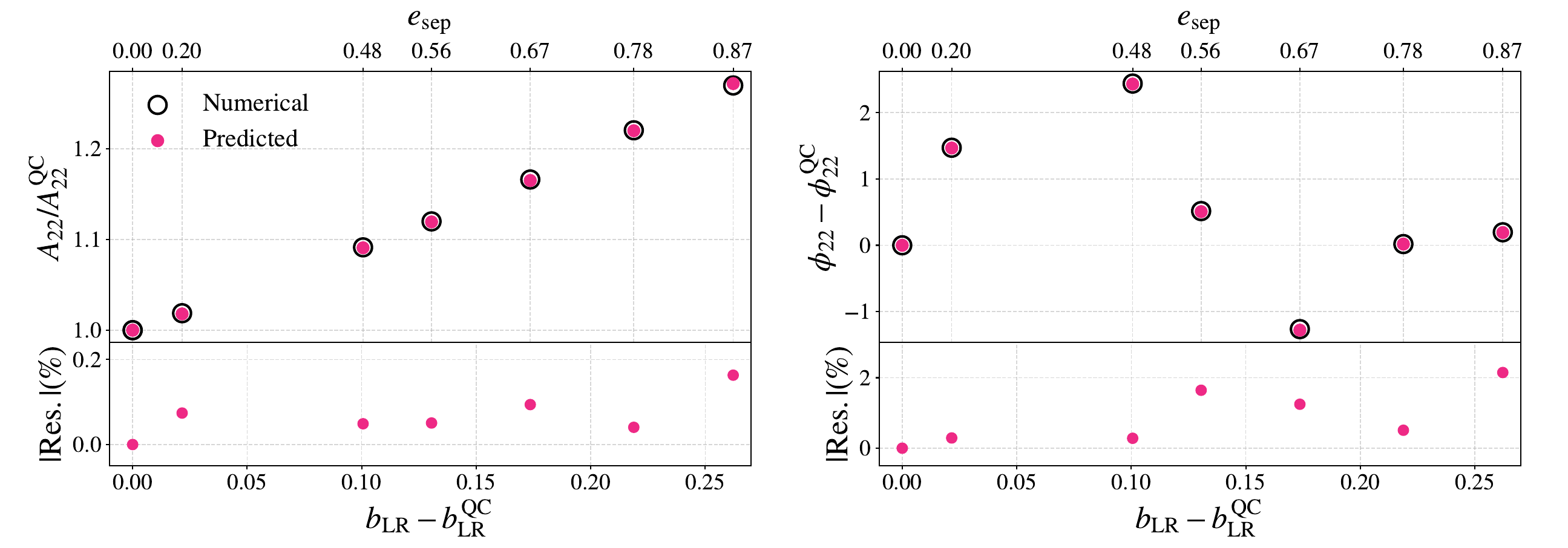}
\caption{Amplitude (\textit{left}) and phase (\textit{right}) of the quadrupole at $\tau=\tau_{\rm LR}+40$ for the eccentric orbits in Table~\ref{tab:sims_ecc}, as a function of the orbit's impact parameter at LR crossing (bottom $x$-axis) and the eccentricity at the separatrix (top $x$-axis). 
We show the numerical results obtained with the \textsc{RWZHyp} code (black empty circles) and the analytical prediction computed as in Eq.~\eqref{eq:signal_exc_imp_contribs} (filled dots).
Results are rescaled with respect to the quasi-circular (labeled as QC) case.
\label{fig:A22_phi22_vs_bLR_esep}}
\end{figure*}
In this section, we investigate the dependence of the late-time behavior of the leading $(\ell m)=(22)$ mode on the eccentricity of the test-particle orbit.
Specifically, we fix the observer at $\tau=40+\tau_{\rm LR}$, when the transient has long decayed out of the strain and the signal is well described by the near-horizon expansion Eq.~\eqref{eq:post_merger_signal}.
We define two parameters to assess the eccentricity of the inspiral: the eccentricity at the separatrix $e_{\rm sep}$ (see~\cite{DeAmicis:2024not} for more details) and the impact parameter evaluated at the LR crossing $b_{\rm LR}=E^{\rm LR}/p^{\rm LR}_{\varphi}$, a gauge-invariant quantity~\cite{Carullo:2023kvj}.
We have denoted as $E^{\rm LR},\,p_{\varphi}^{\rm LR}$ the test-particle energy and angular momentum at the LR crossing.

\begin{figure}[]
\includegraphics[width=0.5\textwidth]{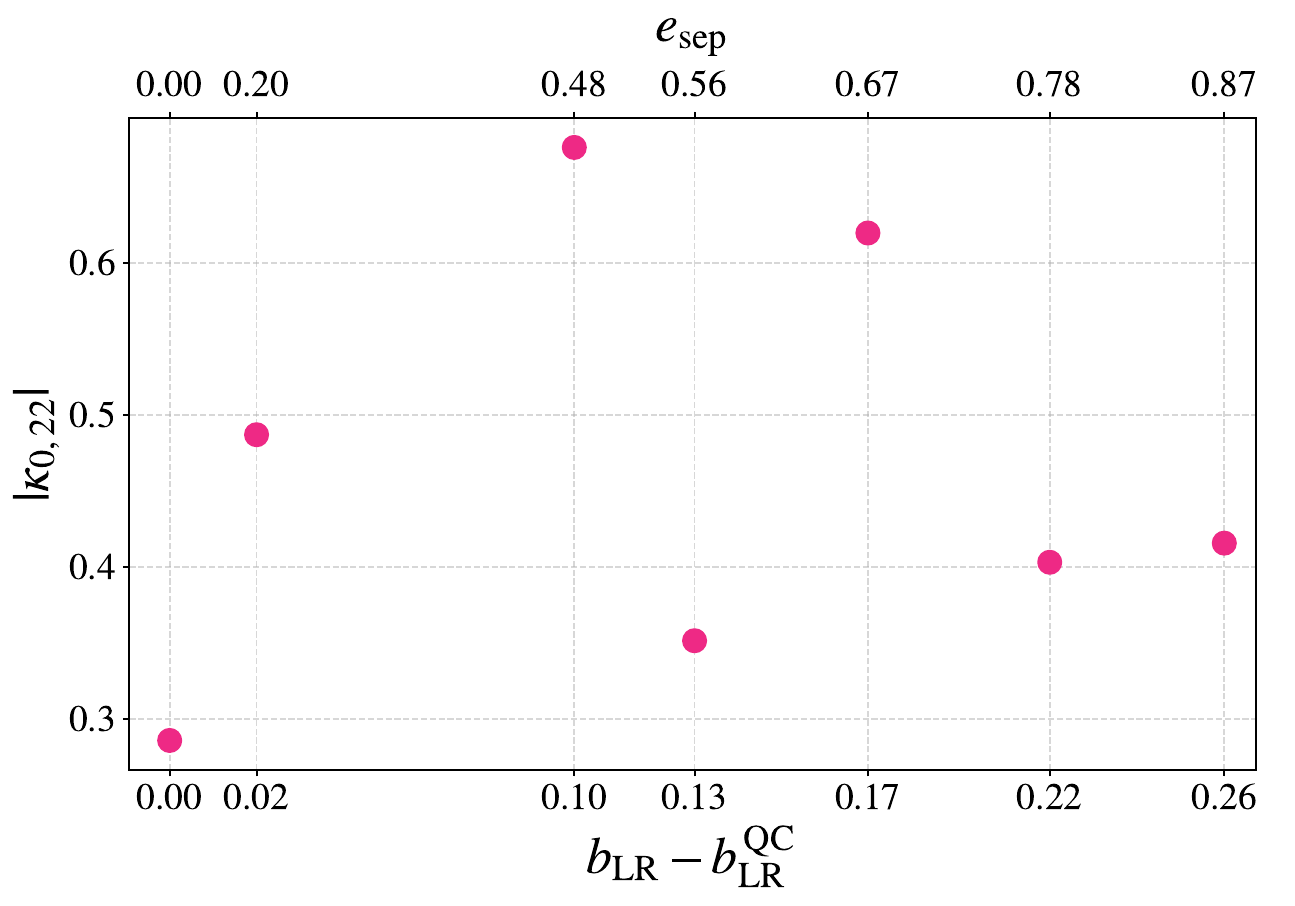}
\caption{Amplitude of the leading redshift term in the quadrupole, $\kappa_{0,22}$ defined in Eq.~\eqref{eq:redshift_amplitude} at $\tau=40+\tau_{\rm LR}$, for the simulations of Table~\ref{tab:sims_ecc}. The bottom $x$-axis shows the impact parameter at LR crossing; the top axis the eccentricity at separatrix crossing.
\label{fig:kappa022_vs_bLR_esep}}
\end{figure}

In Fig.~\ref{fig:A22_phi22_vs_bLR_esep} we show the amplitude $A_{22}$ and phase $\phi_{22}$ of the quadrupole at $\tau=40+\tau_{\rm LR}$ vs $b_{\rm LR}$ and $e_{\rm sep}$, rescaled and translated, respectively, with respect to their quasi-circular values.
We compare the results directly read from the numerical simulations with the analytical values computed through Eq.~\eqref{eq:signal_exc_imp_contribs}.
Our prediction is in agreement with the perturbative numerical waveform and shows a simple dependence of the post-merger quadrupole amplitude on the inspiral's eccentricity. 
In particular, the eccentricity can increase $A_{22}$ by more than $\sim 25\%$, in agreement with what found fitting numerical simulations in the comparable-mass case~\cite{Carullo:2024smg}.
Residuals are $< 0.2\%$ in the rescaled amplitude and $\lesssim 2\%$ in relative phase.
In the right panel of Fig.~\ref{fig:kappa022_vs_bLR_esep} we show the dependence of the leading redshift term amplitude in the quadrupole, $|\kappa_{0,22}|$, on the quantities $b_{\rm LR},\,e_{\rm sep}$. Similarly to $\phi_{22}$, it displays an oscillatory dependence as a function of the eccentricity.
%

\section{Discussion: The QNM signal}   %
\label{sec:discussion}                 %

%
When driven by a test-particle falling into the BH, the QNM contribution to the signal has a complex phenomenology.
It can be written in terms of retarded-time-dependent coefficients (the activation $c_{\ell mn\pm}$ and impulsive $i_{\ell mn\pm}$ coefficients), the standard geometrical excitation factors $B_{\ell mn\pm}$ and the complex exponential $e^{-i\omega_{\ell mn\pm}(t-r_*)}$.
The $c_{\ell mn\pm}$ accumulate in time, i.e.~they are integrals over the past history of the source, for a test-particle localized outside the LR, Eq.~\eqref{eq:clmn_def}. As the source approaches the horizon the $c_{\ell mn\pm}$ become purely local terms. 
The $i_{\ell mn\pm}$ are instead local for generic locations of the test-particle, Eq.~\eqref{eq:ilmn_def}.
For the longest-lived modes $(220\pm)$, the activation coefficients behave similarly to activation functions: they grow during the inspiral, have a maximum near the LR crossing and eventually saturate to a constant amplitude at late times, where the (stationary) ringdown dominates the signal. 
The impulsive coefficients of the $(220\pm)$ modes, instead, grow in amplitude until the LR crossing and then decay.
Instead, the activation and impulsive coefficients of overtones steadily grow at late-retarded times, for generic orbital configurations.
This growth can be explained through a near-horizon expansion in the apparent location of the test-particle $\bar{r}$ after LR crossing, as a divergence for $\bar{r}\rightarrow 2$.
In this limit, we found that when the divergence in the QN eigenfunction (computed at the source) is combined with the exponential decay of the mode at the observer's location along the QNM causal path, it automatically removes divergences, yielding a finite overtone contribution.
The same process gives rise to an infinite tower of \textit{redshift terms} in the observable signal.
These modes are non-oscillatory terms exponentially decaying in retarded time, with decay rates given by multiples of the BH surface gravity $\kappa_{\mathcal{H}^+}$. 
The full post-LR-crossing signal can thus be written as
\begin{equation}
\begin{split}
\Psi^{\bar{r}\ll r_{\rm LR}}_{\ell m}(\tau)=\sum_{n,p}\chi_{\ell mnp}e^{-i\omega_{\ell mnp}\tau}+\sum_{j=0}^{\infty}\kappa_{j,\ell m}e^{-(j+1)\tau\kappa_h} \, .
\end{split}
\label{eq:qnms_plus_redshift}
\end{equation}
The $\chi_{\ell mn\pm}$ are the asymptotic (constant) amplitudes of each QNM, receiving contributions only from the activation coefficients $c_{\ell mn\pm}$. The redshift terms, instead, receive contributions from both $c_{\ell mn\pm}$ and $i_{\ell mn\pm}$.
The leading redshift term decays slower than all overtones.
Hence, there always exist a retarded time $\tilde{\tau}$ after which the overtones' contribution to the signal is smaller than the redshift one, depending on the amplitude $\kappa_{0,\ell m}$.
In particular, the QNM GF contribution predicts that in quasi-circular and eccentric inspiral-plunges the overtones are swamped by the redshift at $\tilde{\tau}\approx\tau_{\rm LR}+15$.
Radial infalls represent a special limit, in which the redshift excitation is highly suppressed. As a consequence, the first overtone decay dominates over the leading redshift up until $\approx \tau_{\rm LR}+50$.
High overtones contribute the most to the signal during the last stages of the plunge, before the LR crossing. 
For circular and eccentric inspirals, overtones up to $n=5$ yield a non-negligible contribution to the QNM waveform at $\tau \lesssim\tau_{\rm LR}$, while for radial infalls, overtones beyond $n=6$ are necessary to achieve convergence in the QNM signal. 
However, at such early times the QNM contribution does not reproduce the full waveform, and other contributions should be taken into account.
As the LR is crossed, the overtones are less excited and, due to their short life-time, quickly decay out of the strain.
At these later times, convergence is typically achieved already when including up to $n=2$.
When comparing the complete QNM analytical prediction to the full numerical linear solution, the QNM response alone cannot completely describe the initial transient dominating the plunge-merger stage, signaling that prompt and tail contributions must be taken into account.
However, it constitutes a fundamental piece: the QNM response is comparable to the numerical waveform even at times $\tau\approx \tau_{\rm LR}-25$.
%

\section{Applications and future directions} %
\label{sec:future_directions}                %

Our results have several applications in GW waveform modelling, in the interpretation of numerical simulations, and searches of new physics using the ringdown signal, opening up several lines of future investigation.

\subsection{Stationary approximation and ``ringdown start time''} %
%
Standard templates for the stationary ringdown are constructed through a superposition of constant amplitudes QNMs, without information on how the modes are activated.
As a consequence, it is necessary to introduce the ad-hoc parameter $\tau_{\rm start}$, denoted as the start time of the ringdown
\begin{equation}
h_{\ell m}^{\rm ring}(\tau)=\theta(\tau-\tau_{\rm start})\sum_{n,p} A_{\ell mn p}\,  e^{-i\omega_{\ell mnp}(\tau-\tau_{\rm start})} \, .
\label{eq:ringdown_template}
\end{equation}
%
%
The $\theta$-function enforces regularity of the template, which would exponentially diverge for $\tau<\tau_{\rm start}$.
In this work, we provide a smooth analytical prediction for how the QNM response is activated in time, removing altogether the need to introduce $\tau_{\rm start}$.
Our results show that in the presence of a persistent source, this model accurately describes the waveform modal content only at late times, while before the LR the ringdown amplitude grows as an activation function.
Even at late times, we find that Eq.~\eqref{eq:ringdown_template} provides only a partial solution, as it doesn't account for redshift terms, see Eq.~\eqref{eq:qnms_plus_redshift}.

For applications where it is convenient to focus on a stationary ringdown description (e.g., searches of new physics, see~\cite{Carullo:2025oms}), the validity of this approximation for each QNM can be read off our results in Figs.~\ref{fig:EV_clmn_Llmn_ID00mdea}, \ref{fig:EV_clmn_Llmn_ID10mdea} for the fundamental and redshift terms, and Fig.~\ref{fig:OVresc_clmn_ID00-01mdea} for the overtones.
In the quasi-circular case, the one of greatest observational interest, a constant amplitude approximation for the fundamental mode is valid starting at $\Delta t \equiv \tau - \tau_{\rm LR} \approx 15 \, M$, with $ \tau_{\rm LR}$ marking LR crossing; retrograde modes saturate significantly later ($\Delta t \approx 35M$).
In this section, we have re-introduced factors of $M$ for clarity.
The first overtone $n=1$ saturation to a constant can be briefly observed within $\Delta t \approx [2, 5] M$, before being swamped by the redshift contribution.
Higher overtones $n >1$ amplitudes cannot be read directly from the activation contributions, as redshift terms become dominant before the overtone amplitude settles to a constant.
We leave an extensive tabulation of QNM and redshift stationary amplitudes to future work, but we note that Eqs.~\eqref{eq:QNMs_redshift_modes_amplitudes},\eqref{eq:delta_klmns} already provide an algorithm to compute all these quantities.
These results mark the first time an analytical prediction for the start of the stationary QNM behavior is achieved.

%
Further investigations are underway to compute and characterize, from first principles, how sensitive each mode is to the different phases of the plunge trajectory, akin to the results of~\cite{DeAmicis:2024not} on the branch-cut tails and~\cite{Price:2015gia} for a QNM toy model.
Given the rapid decay of higher overtones, we expect their excitation to accumulate predominantly near the LR crossing. 
Consequently, as $n \gg 1$, the corresponding amplitudes should become increasingly universal and determined primarily by geometric excitation coefficients, with little imprint from the earlier inspiral.
We leave to future work a detailed investigation of the signal convergence with overtone number~\cite{Oshita:2024wgt}.

\subsection{Redshift terms}

%
The study of near-horizon source contributions to the ringdown was initiated in Ref.~\cite{Mino:2008at}, which computed the leading redshift term (there termed “horizon mode”) for a test particle plunging into a Kerr black hole, and suggested the existence of an infinite hierarchy of sub-leading, increasingly redshifted contributions.
Subsequently, Ref.~\cite{Zimmerman:2011dx} argued that the leading-order term vanishes due to a cancellation, leaving the sub-leading mode as the dominant contribution.
In both these works, the convolution of the QNM GF with the test-particle source was computed in a near-horizon expansion.
More recently, however, work by some of the same authors identified a non-vanishing leading-order redshift term, finding no evidence of the earlier cancellation~\cite{Laeuger:2025zgb}.
All these results were obtained in the frequency domain, and then transformed to the time domain, without incorporating the causality condition.
In this work, we predict the existence of an infinite tower of redshift terms through a more generic framework: we first compute the signal propagated by the QNM GF in time domain and only afterwards perform a near-horizon expansion. 
Our results indicate the presence of the leading redshift term in the signal, in disagreement with~\cite{Zimmerman:2011dx}. 
We leave to future work the task of understanding the origin of this discrepancy and the impact of other components of the GF on the redshift terms.
%

%
Due to their physical origin, connected to the very nature of BHs (the horizon), we expect redshift terms to be present also in comparable-mass mergers.
However, more work is needed to assess their relevance in comparable-mass waveforms, for example via fitting techniques on NR simulations, and to identify the regime in which these modes are most relevant.
However, we expect fitting approaches to be subtle, given the purely-damped nature of this term, the fact that its decay rate is close to the first overtone's, and the need to simultaneously consider a large number of components to obtain an unbiased extraction. 

%
Redshift terms could provide unique information to test for the presence of an event horizon.
Standard horizon tests rely on ``echoes'' of the post-peak signal, typically appearing at much later times with suppressed amplitudes~\cite{Cardoso:2017cqb}.
Further work will be required to investigate the extent to which redshift terms will be sensitive to the nature of the remnant object, i.e., in the presence of a partially-absorbing membrane different from an horizon.

\subsection{Overtones extraction}

%
A significant amount of work has been dedicated to understand the overtone content of post-merger signals, particularly in the comparable-mass limit using NR simulations. 
Robustly identifying more than one overtone in the waveform is a non-trivial task that has sparked a wide debate in recent literature, see~\cite{Berti:2025hly} for a review.
Agnostic investigations show the presence of low-frequency features in the non-linear signal, hindering the identification of constant-amplitude overtones close to the peak~\cite{Baibhav:2023clw, May:2024rrg}.
More recent studies~\cite{Cheung:2023vki,Mitman:2025hgy} could robustly extract $n\leq3$ constant-amplitude overtones at later times when requiring stability over a wide parameter space, while Ref.~\cite{Ma:2024hzq} argued that late-time tails could act as a noise source for the extraction of overtones at intermediate times, but this has not been investigated in detail yet.

%
Our model shows that in the quasi-circular case, overtones are dominant over redshift terms only close to the peak, $\Delta t \lesssim 5 M$.
However, at these early times only $n=1$ displays an approximately constant amplitude, while QNMs are still undergoing dynamical growth, and a standard stationary approximation is invalid.
Further complicating their extraction, close to LR crossing the QNM signal component does not provide an accurate description of the waveform: the prompt and branch-cut components need to be included to avoid overfitting non-modal contributions.
Finally, redshift terms swamp the overtones after this intermediate $\Delta t \gtrsim 5 M$ time-window: if ignored when fitting, redshift terms could induce a spurious variation when extracting stationary overtones.
More work is necessary to assess the relevance of these features in the full, non-linear response of comparable-mass mergers.
All these effects pose a challenge for early-time spectroscopy involving stationary overtones, even in our simplified perturbative setting: they can be more confidently measured in a phase when other effects are prominent.
We hope our results will inspire the construction of more accurate time-dependent overtone models, together with informative priors for data-analysis applications.
%

%
In recent years, powerful analytical methods have been introduced to investigate BH perturbations. Refs.~\cite{Green:2022htq,Cannizzaro:2023jle} introduced a product under which QNMs are orthogonal, and which can be used to compute the QNM excitation from initial data. This product, however, is time-independent and cannot be readily used to investigate the transient signal due to the source. Further investigation is needed to incorporate causality and the source in the calculation of activation coefficients with QNM products.
Once time-dependence will be incorporated, scalar products could prove a useful tool to extract highly damped modes during the near-LR, highly dynamical regime.

\subsection{Implications for Kerr and comparable-mass binaries}
%
%
Numerical evidence indicates that the QNM excitation pattern uncovered in our work generalises to comparable-mass binaries.
In head-on BH collisions
-- starting at rest from infinity -- the vast majority of the non-linear merger-ringdown waveform is described by the source-driven perturbative waveforms considered here~\cite{DeAmicis:2024eoy}.
Quantitative agreement between the perturbative numerical waveforms we use here and NR was found also for quasi-circular orbits in~\cite{Nagar:2022icd}.
A comparison between the small-mass ratio expansion (in the self-force formalism) and quasi-circular comparable-mass simulations was also recently performed in Ref.~\cite{Kuchler:2025hwx}, finding good agreement and pointing to the need of including the remnant BH spin to further improve the overlap.
In future work, we plan to extend our findings to spinning BHs as well. 
As a first step, in App.~\ref{app:Kerr_causality} we provide the causality condition for Kerr QNMs, and inspect its near-horizon behavior. We expect the decay rate of Kerr redshift terms to be set by the Kerr horizon redshift, and to oscillate at the horizon frequency, as derived in \cite{Mino:2008at,Zimmerman:2011dx}.
Finally, merger-ringdown models in the EOB framework have long relied on phenomenological ansatzes with time-dependent QNM activation functions, representing the entire post-merger regime~\cite{Damour:2014yha}.
This same functional form is remarkably accurate in describing both comparable-mass and perturbative waveform (e.g.~\cite{Albanesi:2024fts}), with only a few parameters incorporating finite mass-ratio corrections required to bridge the two.
These are all strong indications that, while inclusion of appropriate mass-ratio corrections will be necessary to capture early-time non-linearities, we do not expect the QNM dynamical excitation structure to be altered when going from a perturbative setting to near equal-mass systems.
In the future, we expect our results to inform the construction of more accurate closed-formed models beyond phenomenological ansatzes.

%
Second-order perturbations and other nonlinear features in the ringdown will also be affected by the source and QNM causality condition. 
This could lead, for instance, to \emph{nonlinear redshift terms}, and to time-dependent amplitudes to sub-leading quadratic modes, as a result of the time-dependent source (see also \cite{Chavda:2024awq}).
Other relevant nonlinear effects include the absorption-driven evolution of the BH mass and spin~\cite{Sberna:2021eui, Redondo-Yuste:2023ipg,May:2024rrg, Capuano:2024qhv, Zhu:2024rej}, which can also induce time-dependence in the early ringdown.
%

%
As numerical simulations of higher-curvature gravity theories display a similar waveform morphology to GR, e.g.~\cite{Cayuso:2023aht, Corman:2022xqg,Corman:2024cdr}, we expect our dynamical picture to generalise beyond Einstein's gravity.
Construction of accurate perturbative templates beyond-GR would alleviate the curse of dimensionality when constructing merger-ringdown waveforms for different classes of alternative theories, with fewer simulations needed to inform a small number of parameters incorporating beyond-GR corrections.

%
Despite current limitations, our results (see also~\cite{Kuchler:2025hwx}) are the initial building-block of a systematic framework to achieve the long-sought computation of analytical merger-ringdown waveforms.
Next steps will focus on the inclusion of the BH spin, prompt response and branch-cut contributions, and will be reported in future work.

\acknowledgments
We are indebted to Alessandro Nagar and Sebastiano Bernuzzi for pointing us towards source-induced excitation.
M.DeA. is grateful to Thibault Damour, Maarten van der Meent, Rodrigo Panosso Macedo, Marica Minucci and David Pereniguez for careful feedback on some of the computations of this work.
We thank Vitor Cardoso, Sam Dolan, Jacopo Lestingi, Jaime Redondo-Yuste and Anil Zenginoglu for instructive discussions.
M.DeA. and G.C.~thank the Institut des Hautes Études Scientifiques for its warm hospitality, where part of this work was developed.
The present research was also partly supported by the ``\textit{2021 Balzan Prize for Gravitation: Physical and Astrophysical Aspects}'', awarded to Thibault Damour.
G.C.~acknowledges funding from the European Union’s Horizon 2020 research and innovation program under the Marie Sklodowska-Curie grant agreement No. 847523 ‘INTERACTIONS’.
L.S.~acknowledges support from the UKRI Horizon guarantee funding (project no. EP/Y023706/1). L.S.~is also supported by a University of Nottingham Anne McLaren Fellowship.
We acknowledge support from the Villum Investigator program by the VILLUM Foundation (grant no. VIL37766) and the DNRF Chair program (grant no. DNRF162) by the Danish National Research Foundation.
This project has received funding from the European Union's Horizon 2020 research and innovation programme under the Marie Sklodowska-Curie grant agreement No 101131233.\\

\clearpage
\appendix

\section{Point-particle source} %
\label{app:source_functions}    %

Following Ref.~\cite{Nagar:2006xv}, the source functions $f(t,r_*),\,g(t,r_*)$ defined in Eq.~\eqref{eq:source_function} are
\begin{equation}
\begin{split}
f(t,r_*)\equiv & -\frac{16\pi\mu A(r) Y_{\ell m}^*}{r\hat{H}\lambda \left[r(\lambda-2)+6\right]}\left\lbrace -2i m \,p_{\varphi}p_{r_*} + 5+\frac{12\hat{H}^2 r}{r(\lambda-2)+6}-\frac{r\lambda}{2}+\frac{2p^2_{\varphi}}{r^2}\right. \\
&\left. +\frac{p_{\varphi}^2}{r^2(\lambda-2)}\left[r(\lambda-2)(m^2-\lambda-1)+2(3m^2-\lambda-5)\right] \right\rbrace \, ,
\end{split}
\end{equation}
\begin{equation}
g(t,r_*)\equiv -\frac{16\pi\mu A(r) Y_{\ell m}^*}{r\hat{H}\lambda \left[r(\lambda-2)+6\right]} (p^2_{\varphi} + r^2) \, ,
\end{equation}
where we have defined $\lambda\equiv\ell(\ell+1)$.

\section{Chandrasekhar transformations between Regge-Wheeler and Zerilli modes} %
\label{app:Chandra_transf}                                                      %
%
Consider a Regge-Wheeler mode $u^{\rm RW}(\omega,r_*)$ and a Zerilli one $u^{\rm Z}(\omega,r_*)$. Following Chandrasekhar (Chapter 4 Ref.~\cite{Chandrasekhar:1985kt}), these modes are related through
\begin{equation}
u^{\rm Z}(\omega,r_*) = \frac{1}{\kappa_{\ell}+12i\omega }\left[ \left(\kappa_{\ell}+\frac{72}{F_{\ell}(r)}\right)u^{\rm RW}(\omega,r_*)+12A(r)\partial_r u^{\rm RW}(\omega,r_*)\right]\, ,
\label{eq:zerilli_from_rw_chandra}
\end{equation}
where we have introduced the following quantities
\begin{equation}
\kappa_{\ell}\equiv\ell^4+2\ell^3-\ell^2-2\ell \ \ , \ \ \ F_{\ell}(r)\equiv\frac{r^2\left[r\ell (\ell+1)-2r+6\right]}{r-2} \, .
\end{equation}
Substituting Eq.~\eqref{eq:rw_mode_general_expr} in the transformation Eq.~\eqref{eq:zerilli_from_rw_chandra}, we find $\hat{z}(\omega,r_*)$ defined in Eq.~\eqref{eq:zerilli_mode_general_expr},
\begin{equation}
\hat{z}(\omega,r_*)\equiv \frac{1}{\kappa_{\ell}+12i\omega}\left[\left(\kappa_{\ell}+\frac{72}{F(r)}+12i\omega\frac{r^2-8}{r^2}\right)\hat{a}(\omega,r_*)+12A(r)\partial_r\hat{a}(\omega,r_*)\right] \, .
\end{equation}
%

\section{Time-domain QNM Green's function} %
\label{app:QNMs_GF}                        %
%
The GF $G_{\ell m}(t-t';r_*,r_*')$ is a solution of the impulsive problem
\begin{equation}
    \left[\partial_t^2-\partial^2_{r_*}+V(r_*)\right]G_{\ell m}(t-t';r_*,r_*')=\delta(t-t')\delta(r_* - r_*')
    \label{eq:GF_def_t-domain}
\end{equation}
To the best of our knowledge, a closed-form expression of $G_{\ell m}(t-t';r_*,r_*')$ is not known.
Hence, to solve the above function, we move to (from) frequency domain through the (anti-)Fourier transform
\begin{equation}
\begin{split}
       & \tilde{G}_{\ell m}(\omega;r_*,r_*') = \int_{t'}^{\infty}dt \  e^{i\omega(t-t')}G_{\ell m}(t-t';r_*,r_*')\, , \\
       &G_{\ell m}(t-t';r_*,r_*')  = \frac{1}{2\pi}\int_{-\infty}^{\infty}d\omega \, e^{-i\omega(t-t')}\tilde{G}_{\ell m}(\omega;r_*,r_*')\, ,
\end{split}
\label{eq:transform_t}
\end{equation}
where the lower limit in the first integral is set by the property of the retarded GF: for $t<t'$, $G_{\ell m}(t-t';r_*,r_*')=0$.

In the $\omega$-domain, Eq.~\eqref{eq:GF_def_t-domain} becomes (assuming zero boundary conditions for the GF)
\begin{equation}
        \left[-\frac{d^2}{d{r^2_*}}-\omega^2+V(r_*)\right]\tilde{G}_{\ell m}(\omega;r_*,r_*')=\delta(r_* - r_*') \, .
    \label{eq:GF_def_om-domain}
\end{equation}
The general solution is~\cite{Maggiore:2018sht}
\begin{equation}
    \tilde{G}_{\ell m}(\omega;r_*,r_*') = \frac{1}{W(\omega)} \cdot \left[ \theta(r_*-r_*') \, u_{\ell m}^{\rm in}(\omega,r_*') \, u_{\ell m}^{\rm out}(\omega,r_*) + \theta(r_*'-r_*) \, u_{\ell m}^{\rm in}(\omega,r_*) \, u_{\ell m}^{\rm out}(\omega,r_*') \right] \, ,
    \label{eq:GF_definition_omega_domain_Sch}
\end{equation}
where the functions $u_{\ell m}^{\rm in}, \ u_{\ell m}^{\rm out}$ are independent solutions of the homogeneous equation
\begin{equation}
     \left[\frac{d^2}{d{r^2_*}}+\omega^2-V(r_*)\right]u_{\ell m}(\omega,r_*)=0 \, ,
     \label{eq:homogeneous_RWZ_omega_Sch}
\end{equation}
and $W(\omega)\equiv u^{\rm out} \partial_{r_*}u^{\rm in}-u^{\rm in} \partial_{r_*}u^{\rm out} $.
This function, known as Wronskian, does not depend on $r$ (see Eq.~(12.186) of~\cite{Maggiore:2018sht}).
In the limits $r_*\rightarrow\pm\infty$, the potential $V(r_*)\rightarrow 0$ and the two solutions $u^{\rm in,out}_{\ell m}$ reduce to plane waves.
We impose the following boundary conditions on the solutions $u^{\rm in,out}_{\ell m}$
\begin{equation}
u_{\ell m}^{\rm in}(\omega,r_*) =
\begin{cases}
& e^{-i\omega r_*} \ \ , \ \ r_*\rightarrow-\infty\\
& A_{\rm in}(\omega) e^{-i\omega r_*} + A_{\rm out}(\omega) e^{i\omega r_*} \ \ , \ \ r_*\rightarrow\infty
\end{cases} \, , 
\label{eq:u_in_asympt_expr}
\end{equation}
\begin{equation}
u_{\ell m}^{\rm out}(\omega,r_*)  =
\begin{cases}
& B_{\rm in}(\omega) e^{-i\omega r_*} + B_{\rm out}(\omega) e^{i\omega r_*} \ \ , \ \ r_*\rightarrow-\infty\\
& e^{i\omega r_*} \ \ , \ \ r_*\rightarrow\infty \, .
\end{cases}
\label{eq:u_out_asympt_expr}
\end{equation}
Since $W(\omega)$ is independent of $r$, we choose to compute it for $r_*\rightarrow\infty$.
Using the above expressions, we find
\begin{equation}
W(\omega) = -2i\omega A_{\rm in}(\omega) \, .
\label{eq:wronskian_expression}
\end{equation}
Following Leaver~\cite{Leaver:1986gd}, it is useful to introduce a third homogeneous solution defined as
\begin{equation}
u_{\ell m}^{\infty-}(\omega,r_*)  =
\begin{cases}
& C_{\rm in}(\omega) e^{-i\omega r_*} + C_{\rm out}(\omega) e^{i\omega r_*} \ \ , \ \ r_*\rightarrow-\infty\\
& e^{-i\omega r_*} \ \ , \ \ r_*\rightarrow\infty \, .
\end{cases}
\label{eq:u_inf-_asympt_expr}
\end{equation}
The solutions $u^{\rm in},\,u^{\rm out}$ and $u^{\infty-}$ are not all independent. In particular, given their behavior in the limit $r_*\rightarrow\infty$, it is possible to write
\begin{equation}
u^{\rm in}_{\ell m}(\omega,r_*)=A_{\rm in}(\omega)u^{\infty -}_{\ell m}(\omega,r_*)+A_{\rm out}(\omega)u^{\rm out}_{\ell m}(\omega,r_*) \, .
\label{eq:u_in_vs_u_out_u_inft-}
\end{equation}

Leaver~\cite{Leaver:1986gd} proposed the following explicit expressions for the solutions $u^{\rm out,\infty-}_{\ell m}$
\begin{equation}
u_{\ell m}^{\rm out,\infty-}(\omega,r_*)=\left(4\omega\right)^{\mp2i\omega}e^{\pm i\phi_{\pm}}\left(\frac{r-2}{r}\right)^{-2i\omega}\sum_{L=-\infty}^{\infty}b_L\left[G_{L+\nu}(-2\omega,\omega r)\pm iF_{L+\nu}(-2\omega,\omega r) \right] \, .
\label{eq:u_out_full}
\end{equation}
where $G$ and $F$ are Coulomb wave function, 
where $r = r(r_*)$, and where the upper (lower) sign relates to the solution $u^{\rm out}$ ($u^{\infty -}$) and the phases $\phi_{\rm \pm}(\omega)$ are functions of $\omega$ such that $u_{\ell m}^{\rm out,\infty-}$ are unitary plane waves in the limit $r_*\rightarrow\infty$, as in Eqs.~\eqref{eq:u_out_asympt_expr},~\eqref{eq:u_inf-_asympt_expr}.
Note that throughout this section we use the notation $r_h\rightarrow2$.
The coefficients $b_L$ in Eq.~\eqref{eq:u_out_full} are functions of $\omega$ that satisfy a three-term recurrence equation, see Refs.~\cite{leaver1985analytic,Leaver:1986gd} for more details.

We compute the time-domain GF by computing the anti-Fourier transform 
\begin{equation}
G_{\ell m}(t-t';r_*,r_*')=\frac{1}{2\pi} \int_{-\infty}^{\infty}d\omega\frac{i\, e^{-i\omega(t-t')}}{2\omega A_{\rm in}(\omega)}
\left[\theta(r_*-r_*')
u^{\rm in}(\omega,r_*')u^{\rm out} \, (\omega,r_*)+\theta(r_*'-r_*)u^{\rm in}(\omega,r_*)u^{\rm out}(\omega,r_*')\right]\, .
\label{eq:GF_time_anti_transform}
\end{equation}
We focus on systems in which the observer is far away from the source and the BH, $r_*\gg r_*'$, $ r_*\gg 1$. In particular, we consider the limit in which the observer is located at $\mathcal{I}^+$, hence
we can also assume $\omega r_*\gg1$ for all $\omega$.
Under this assumption, we focus on the first integral in Eq.~\eqref{eq:GF_time_anti_transform}. In the following, to simplify our notation, we drop the $\theta(r_*-r_*')$.
Then, we can approximate $u^{\rm out}$ with its asymptotic expression in the limit $\omega r_*\gg 1$, i.e., $u^{\rm out}(\omega,r_*)=e^{i\omega r_*}$.
Substituting Eq.~\eqref{eq:u_in_vs_u_out_u_inft-} into Eq.~\eqref{eq:GF_time_anti_transform}, we can rewrite the time domain GF as 
\begin{equation}
G_{\ell m}(t-t';r_*,r_*')=\frac{i}{2\pi}\left[\int_{-\infty}^{\infty}d\omega \, \frac{e^{-i\omega(t-r_*-t')}}{2\omega}u^{\infty-}(\omega,r_*')+\int_{-\infty}^{\infty}d\omega \, \frac{e^{-i\omega(t-r_*-t')}}{2\omega}\frac{A_{\rm out}(\omega)}{A_{\rm in}(\omega)}u^{\rm out}(\omega,r_*')\right]
\label{eq:GF_time_anti_transform_II}
\end{equation}
Both integrands are non-analytic at $\omega=0$ since the solutions $u^{\infty-,\rm out}$ are superpositions of the Coulomb wave functions, see Eq.~\eqref{eq:u_out_full}, and $G_{L+\nu}\propto \ln\omega$ for small $\omega$~\cite{abramowitz1968handbook}. Following Ref.~\cite{Leaver:1986gd}, to compute the integrals we perform an analytic continuation to complex $\omega$.
The complex logarithm is a multi-valued function, giving rise to a branch cut in the complex plane originating from the branch point $\omega=0$. To select the retarded GF, we fix this branch cut in the lower half plane, on the negative imaginary frequencies axis, $\mathrm{Re}(\omega)=0,\, \mathrm{Im}(\omega)<0$~\cite{Leaver:1986gd}.
In the lower-half plane, there is an infinite number of isolated points $\omega_n$ such that $A_{\rm in}(\omega_n)=0$ and the Wronskian Eq.~\eqref{eq:wronskian_expression} vanishes, yielding an infinite number of isolated simple poles in the second integral in Eq.~\eqref{eq:GF_time_anti_transform_II}.
We call these values as the \textit{quasinormal frequencies} (QNFs).
Near $\omega\simeq\omega_n$, we Taylor expand $A_{\rm in}(\omega\approx\omega_n)$ as
\begin{equation}
A_{\rm in}(\omega)\simeq(\omega-\omega_n)\frac{dA_{\rm in}(\omega)}{d\omega}\bigg|_{\omega=\omega_n} \equiv(\omega-\omega_n)\alpha_n \, .
\label{eq:wronskian_near_QNFs}
\end{equation}
In the present work, we use the $\alpha_n$ computed by Leaver~\cite{Leaver:1986gd}.
Note that Leaver~\cite{Leaver:1986gd} uses a variable $s$ which is connected to the frequency $\omega$ used in the present paper as $s=-2i\omega$. As a consequence, our definition of $\alpha_n$ differs with respect to Leaver's one, $\alpha_n^{\rm Leaver}$, by the factor $\alpha_n=-2i\alpha_n^{\rm Leaver}$.
At the QNFs, Eq.~\eqref{eq:u_in_vs_u_out_u_inft-} reduces to 
\begin{equation}
u^{\rm in}_{\ell m}(\omega_n,r_*)=A_{\rm out}(\omega_n)u^{\rm out}_{\ell m}(\omega_n,r_*)\equiv u^{h}_{\ell m}(\omega_n,r_*)\, .
\label{eq:u_in_vs_out_inf-_at_QNFs}
\end{equation}
Hence, the QNFs correspond to specific solutions of Eq.~\eqref{eq:homogeneous_RWZ_omega_Sch} which are purely ingoing plane waves at the horizon and purely outgoing plane waves at $\mathcal{I}^+$.
Following Leaver~\cite{leaver1985analytic,Leaver:1986gd}, the homogeneous solution at the QNMs $u^{h}_{\ell m}(\omega,r_*)$ are well represented by a near-horizon expansion rather than an expansion near $r_*\rightarrow\infty$ as in Eq.~\eqref{eq:u_out_full}, in particular,
\begin{equation}
u^{h}(\omega,r_*)= \, e^{i\omega (r_*-4i\omega)} \left(\frac{r-2}{r}\right)^{-4i\omega}\sum_{k=0}^{\infty}a_k(\omega)\left(1-\frac{2}{r}\right)^k \, ,
\label{eq:u_in_near_QNFs}
\end{equation}
with $a_0= 1$ and the remaining coefficients satisfying a recursion relation~\cite{Leaver:1985ax}. 
When computed at the QNFs, the above series is uniformly convergent, i.e. it converges as a function (is at least $C^0$) in $r$~\cite{Leaver:1986gd,leaver1985analytic}.
Substituting Eq.~\eqref{eq:u_out_asympt_expr},~\eqref{eq:u_in_near_QNFs} into Eq.~\eqref{eq:u_in_vs_out_inf-_at_QNFs}, it is possible to find an expression for $A_{\rm out}$ computed at the QNFs 
\begin{equation}
A_{\rm out}(\omega_n)=e^{-4i\omega_n}\sum_{k=0}^{\infty}a_k(\omega_n)\, .
\label{eq:Aout_QNMs}
\end{equation}
Since we are interested in the QNM response, we focus on the second integral in Eq.~\eqref{eq:GF_time_anti_transform_II}, that we will denote as
\begin{equation}
G^{(2)}_{\ell m}(t-t';r_*,r_*')\equiv \int_{-\infty}^{\infty}d\omega \, \frac{i\, e^{-i\omega(t-r_*-t')}}{(2\pi)\cdot 2\omega}\frac{A_{\rm out}(\omega)}{A_{\rm in}(\omega)}u^{\rm out}(\omega,r_*')\equiv \int_{-\infty}^{\infty}d\omega\, I(\omega;t,t',r_*,r_*') \, .
\label{eq:GF2_time_anti_transform}
\end{equation}
We briefly discuss the first integral in Eq.~\eqref{eq:GF_time_anti_transform_II} in Sec.~\ref{subsec:QNMs_propagation_and_regularity}. 
In Fig.~\ref{fig:contour_plot2}, we show the features of the integrand in Eq.~\eqref{eq:GF2_time_anti_transform}, denoted as $I(\omega;t,t',r_*,r_*')$, in the complex $\omega$ plane. The integration  can be performed along one of the two closed contours shown in the figure: either $\Gamma_1+\Gamma_2'$ or $\Gamma_1+\Gamma_2+...+\Gamma_6$.
The contour is selected so that the residue theorem can be applied: the integrand needs to be regular except for isolated poles.
In particular, it needs to be regular for $\mathrm{Im}(\omega)\rightarrow+\infty$ if we integrate along $\Gamma_1+\Gamma_2'$, and for $\mathrm{Im}(\omega)\rightarrow-\infty$ if we integrate along $\Gamma_1+\Gamma_2+...\Gamma_6$.
Since we are interested in the QNMs, i.e. the contribution coming from the poles in the lower half plane, we investigate the values of $(t,r_*,t',r_*')$ for which the integration can be carried over $\Gamma_1+\Gamma_2+...\Gamma_6$.
In the limit $\mathrm{Im}(\omega)\rightarrow-\infty$ the QNFs have small real part and are close to the branch cut~\cite{Leaver:1985ax}. Hence, in this limit, we can deform the contour to encircle the QNFs and approximate $I(\omega)$ with the integrand computed at the QNFs, as shown on the right panel of Fig.~\ref{fig:contour_plot2}.
Near the QNFs, the integrand in Eq.~\eqref{eq:GF2_time_anti_transform} can be computed using Eqs.~\eqref{eq:u_in_near_QNFs},~\eqref{eq:wronskian_near_QNFs}
\begin{equation}
I(\omega\approx\omega_n;t,t',r_*,r_*')\simeq \left[\frac{i\, A_{\rm out}(\omega)}{4\pi\omega A_{\rm in}(\omega)}\cdot \hat{a}(\omega,r')\right]\cdot e^{-i\omega[t-r_*-\mathcal{C}(t',r_*')]} \, ,
\label{eq:integrand_near_QNMs}
\end{equation}
where we have defined the functions
\begin{align}
\hat{a}(\omega,r')\equiv& \frac{\sum_{k=0}^{\infty}a_k(\omega)\left(1-\frac{2}{r'}\right)^k}{\sum_{k=0}^{\infty}a_k(\omega)} \, ,
\label{eq:sum_a_k_def} \\
\mathcal{C}(t',r_*')\equiv & t'+r_*'-4\log\left(\frac{r'-2}{r'}\right) \, ,
\end{align}
We will assume that the following limits are satisfied
\begin{equation}
\lim_{\mathrm{Im}(\omega)\rightarrow-\infty} \frac{1}{\mathrm{Im}(\omega)}\log\bigg|\frac{A_{\rm out}(\omega)}{A_{\rm in}(\omega)}\bigg|=0 \ \ , \ \ \ \lim_{\mathrm{Im}(\omega)\rightarrow-\infty} \frac{\log |\hat{a}(\omega,r')|} {\mathrm{Im}(\omega)}=0  \, , \,  \forall \, r' \, .
\label{eq:basics_assumptions}
\end{equation}
The first limit can be motivated by the results of Andersson~\cite{Andersson:1995zk,Andersson:1996cm} obtained in the limit of large $\omega$. In particular Andersson~\cite{Andersson:1996cm} explicitly computed the ratio $A_{\rm out}/A_{\rm in}$ in the limit $\omega \gg1$, as
\begin{equation}
\frac{A_{\rm out}(\omega)}{A_{\rm in}(\omega)}=\frac{(4i\omega)^{4i\omega}e^{-4i\omega}\Gamma(1/2-4i\omega)}{\sqrt{\pi}e^{i\pi/2}} \, . 
\end{equation}
We expand the above expression in the limit $\mathrm{Im}(\omega)=\omega^{\rm Im}\rightarrow-\infty$ and approximate $\mathrm{Re}(\omega)=\omega^{\rm Re}\approx 0$,
\begin{equation}
\lim_{\omega^{\rm Im}\rightarrow-\infty}\bigg|\frac{A_{\rm out}(i\omega^{\rm Im})}{A_{\rm in}(i\omega^{\rm Im})}\bigg|=
\frac{1}{\sqrt{2}}
|\sec(4\pi |\omega^{\rm Im}|)|= \frac{\sqrt{2}}{|e^{4 i \pi |\omega^{\rm Im}|}+e^{-4 i \pi |\omega^{\rm Im}|}|}\, . 
\end{equation}
We substitute this expression in the first limit in Eq.~\eqref{eq:basics_assumptions}, and find
\begin{equation}
\lim_{|\omega^{\rm Im}|\to\infty} -\frac{1}{2|\omega^{\rm Im}|}\, \log\left[
\frac{1}{2}e^{-8i\pi |\omega^{\rm Im}|} \, \left(1+e^{8i\pi |\omega^{\rm Im}|} \right)^2\right]=0\, .
\label{eq:Lim1_def}
\end{equation}
The second limit in Eq.~\eqref{eq:basics_assumptions} can be shown to be valid for $r'\gg1$, since $\hat{a}(\omega,r'\gg1)\rightarrow1$ regardless of $\omega$. 
The term $\hat{a}(\omega)\rightarrow1$ does not contain contributions $\sim e^{i\omega r'}$, hence we expect the second limit in Eq.~\eqref{eq:basics_assumptions} to be valid everywhere. 
Then, the integrand's behavior will be determined by the exponential term in Eq.~\eqref{eq:integrand_near_QNMs}
\begin{equation}
\lim_{\mathrm{Im}(\omega)\rightarrow-\infty}I(\omega;t,t',r_*,r_*') =
\begin{cases}
& \, <\infty \ \ , \ \ \ t-r_*\geq \mathcal{C}(t',r_*') \, , \\
& \rightarrow \infty \ \ , \ \ \ t-r_*< \mathcal{C}(t',r_*') \, .
\end{cases}
\end{equation}
Hence, the integral in Eq.~\eqref{eq:GF2_time_anti_transform} must be computed through the complex contour $\Gamma_1+...\Gamma_6$ in Fig.~\ref{fig:contour_plot2} (left) for $t-r_*\geq \mathcal{C}(t',r_*')$ only. For $t-r_*< \mathcal{C}(t',r_*')$, the integration must be computed along $\Gamma_1+\Gamma_2'$.
We now have all the necessary ingredients to compute the integral in Eq.~\eqref{eq:GF2_time_anti_transform}.
For $t-r_*< \mathcal{C}(t',r_*')$ we integrate in the upper half plane along $\Gamma_1+\Gamma_2'$, where the single poles corresponding to the QNFs are absent. 
Since we are interested in the GF which propagates the QNMs response only, we neglect this contribution from the present discussion.
For $t-r_*\geq \mathcal{C}(t',r_*')$ we integrate along $\Gamma_1+...+\Gamma_6$ on the left panel of Fig.~\ref{fig:contour_plot2}, yielding 
\begin{equation}
\begin{gathered}
G^{(2),\,\geq}_{\ell m}(t-t';r_*,r_*') = \int_{\Gamma_1} d\omega \, I(\omega;t,t',r_*,r_*') = \\
- \left[\int_{\Gamma_2} d\omega \left(...\right)+\int_{\Gamma_6} d\omega \left(...\right) \right]  - \left[\int_{\Gamma_3} d\omega \left(...\right)+\int_{\Gamma_4} d\omega \left(...\right)+\int_{\Gamma_5} d\omega \left(...\right) \right] \\
+ (-2\pi i) \cdot \sum_{\omega=\{\omega_{np}\}} \mathrm{Res}\left[\, I(\omega;t,t',r_*,r_*')\right] \, .
\end{gathered}
\label{eq:GF_time_three_contribs}
\end{equation}
The integral along $\Gamma_2+\Gamma_6$ give rise to the prompt response~\cite{Leaver:1986gd,Andersson:1996cm}, while the integral along $\Gamma_3+\Gamma_4+\Gamma_5$ originates late-time tails~\cite{Leaver:1986gd,Andersson:1996cm,Asada:1997zu}.
Here, we focus only on the residuals, which propagate the QNMs response.
We substitute the expansion of $A_{\rm in}$ near the QNFs, Eq.~\eqref{eq:wronskian_near_QNFs}, in Eq.~\eqref{eq:integrand_near_QNMs} yielding the GF portion propagating the QNMs for $t-r_*\geq \mathcal{C}(t',r_*')$
\begin{equation}
G^{\rm QNMs,\,\geq}_{\ell m}=(-2\pi i)\sum_{\omega=\{\omega_{np}\}}\mathrm{Res}\left[\, I(\omega;t,t',r_*,r_*')\right] =\sum_{n,p} B_{np}\, \hat{a}(\omega_{np},r') \, e^{-i\omega_{np}\left[t-r_*-\mathcal{C}(t',r_*')\right]} \, ,
\end{equation}
where we have defined the \textit{geometric excitation factors}
\begin{equation}
B_{np}\equiv\frac{e^{-4i\omega_{np}}\sum_{k}a_k(\omega_{np})}{2\omega_{np}\alpha_{np}} \, .
\end{equation}
Since for $t-r_*< \mathcal{C}(t',r_*')$ the QNFs do not give any contribution to the retarded GF, we can write the retarded GF propagating the QNMs response as 
\begin{equation}
G^{\rm QNMs}_{\ell m}(t-t';r_*,r_*')=\theta\left[t-r_*-\mathcal{C}(t',r_*')\right]\cdot \sum_{n,p} B_{np}\, e^{-i\omega_{np}[t-r_*-\mathcal{C}(t',r_*')]}\hat{a}(\omega_{np},r') \, . 
\end{equation}

\section{Standard regularization of activation coefficients}\label{app:regularization}

\begin{figure}[t]
\includegraphics[width=0.5\columnwidth]{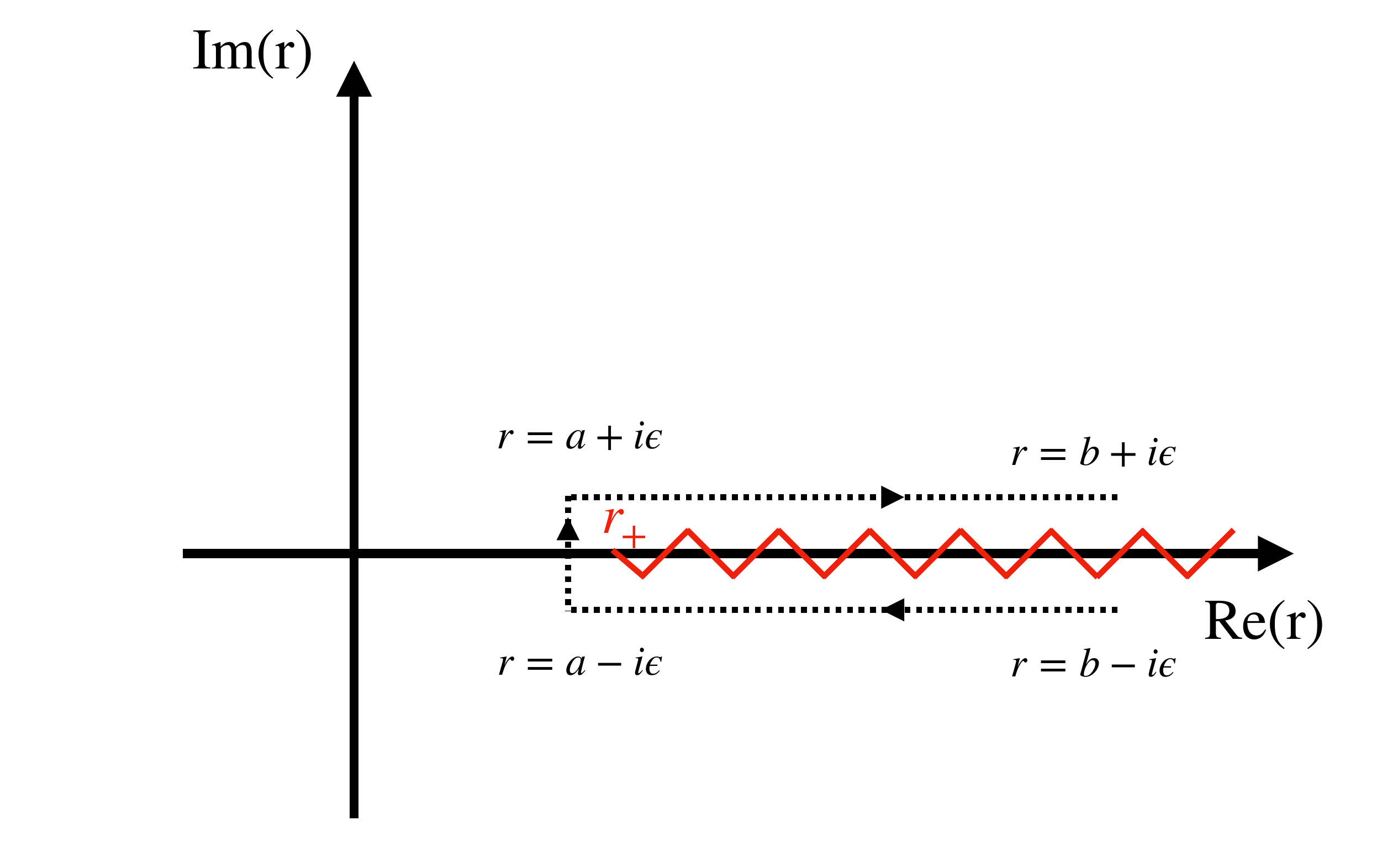} 
\caption{
Analytic continuation of the radial coordinate used to compute the regularized activation coefficients in~\cite{Sun:1988tz}. The zig-zag line represents the branch cut along the real axis, while the dashed line is the deformed integration contour. The integral on such contour matches the one obtained using the counter-term regularization method. 
\label{fig:Price_contour}}
\end{figure}

As discussed in Sec.~\ref{subsec:NearH+_divergence}, the activation coefficients may be divergent should one consider initial data that extends all the way to the horizon, or a source approaching it dynamically. To handle these divergences, different regularization procedures have been introduced in the literature. In Ref.~\cite{Sun:1988tz}, Sun and Price investigated the activation coefficients for initial data extending to the horizon. They introduced a regularization based on an analytic continuation on a deformed contour in the complex $r-$ plane. An analogous approach was used by Leaver~\cite{Leaver:1986gd}. More recently, it was shown that the activation coefficients can be regularized by subtracting suitable counter-terms~\cite{Sberna:2021eui, Cannizzaro:2023jle}.
In this section, we show that the two procedures are equivalent. 

Let us consider the following integral, which describes the near-horizon behavior of the activation coefficient~\eqref{eq:clmn_expression} of the first overtone,
\begin{equation}\label{eq:excit_coeff_non_regul}
    I_H=\int_2^b (r_h-r)^{-4 i \omega_{221}} dr \, ,
\end{equation}
where $b>2$. Such integral is clearly divergent, as $-4|\omega_{221}^{\rm Im}|+1<0$. The regularization procedure by Price, Sun and Leaver amounts to computing the integral along a complex path as the one shown in Fig.~\ref{fig:Price_contour} in the limit $\epsilon \rightarrow 0$. The branch cut of the integrand is placed along the real axis in the interval $r \in (r_h, \infty)$ and the activation coefficient is computed as
\begin{equation}\label{eq:Price_regul}
    I_H= \lim_{\epsilon \rightarrow 0}\frac{\int_a^b dr \Big[ \Big(r_h-r+ i \epsilon \Big)^{-4 i \omega_{221}}- \Big(r_h-r- i \epsilon \Big)^{-4 i \omega_{221}} \Big]}{(1-e^{-8 \omega_{221}})} \, ,
\end{equation}
where $a<2$. The denominator is due the phase change between the two sides of the branch cut. 
The two terms in~\eqref{eq:Price_regul} represent the contributions along the horizontal paths above and below the cut. Note that the contribution of the vertical path is not included, as it becomes infinitesimal in the limit $\epsilon \rightarrow 0$. 

Let us now show how this expression can be rearranged to find the counter-term of regularization method of Refs.~\cite{Sberna:2021eui, Cannizzaro:2023jle}. We introduce a new variable $r_h<\tilde{r}<b$ such that eq.~\eqref{eq:Price_regul} can be rewritten as:
\begin{equation}\label{eq:Price_regul2}
  I_H= \lim_{\epsilon \to 0}\frac{\int_{\tilde{r}}^b dr \Big[ H_+(r, \epsilon)^{-4 i \omega_{221}}- H_-(r, \epsilon)^{-4 i \omega_{221}} \Big]}{(1-e^{-8 \omega_{221}})}-\frac{\Big[ H_+(r, \epsilon)^{1-4 i \omega_{221}}- H_-(r, \epsilon)^{1-4 i \omega_{221}} \Big]_a^{\tilde{r}}}{(1-e^{-8 \omega_{221}})(1-4 i \omega_{221})} \, ,
\end{equation}
where we defined for compactness $H_\pm(r, \epsilon)=(r_h-r \pm i \epsilon)$.
Let us now take the limit $\tilde{r}\rightarrow 2$. Noting that above and below the cut we have $H_-(r, \epsilon)= H_+(r, \epsilon)e^{-8 \omega_{221}}$, the first term in~\eqref{eq:Price_regul2} can be rewritten as:
\begin{align*}
\lim_{\substack{\epsilon\to 0 \\ \tilde{r}\to r_h}}\frac{\int_{\tilde{r}}^b dr \Big[ H_+(r, \epsilon)^{-4 i \omega_{221}}- H_-(r, \epsilon)^{-4 i \omega_{221}} \Big]}{(1-e^{-8 \omega_{221}})}= \lim_{\substack{\epsilon\to 0^+ \\ \tilde{r}\to r_h}}\int_{\tilde{r}}^b dr \Big[ H_+(r, \epsilon)^{-4 i \omega_{221}}\Big]=\lim_{\tilde{r}\to r_h}\int_{\tilde{r}}^b (-r+r_h)^{-4 i \omega_{221}}
\end{align*}
As for the second term, we have two different contributions when the function inside the square bracket is evaluated at $\tilde{r}$ and $a$, respectively. In the first case we have:
\begin{align*}
\lim_{\substack{\epsilon\to 0 \\ \tilde{r}\to r_h}}\frac{\Big[ H_+(\tilde{r}, \epsilon)^{1-4 i \omega_{221}}- H_-(\tilde{r}, \epsilon)^{1-4 i \omega_{221}} \Big]}{(1-e^{-8 \omega_{221}})(1-4 i \omega_{221})}=\lim_{\substack{\epsilon\to 0^+ \\ \tilde{r}\to r_h}} \frac{H_+(\tilde{r}, \epsilon)^{1-4 i \omega_{221}}}{1-4 i \omega_{221}}= \lim_{\tilde{r}\to r_h}\frac{(r_h-\tilde{r})^{1-4 i \omega_{221}}}{1-4 i \omega_{221}}
\end{align*}
When evaluated in $r=a$ instead, this term gives a vanishing contribution. As $a<r_h$, when the limit $\epsilon\to 0$ is performed the branch cut is not crossed and the function is continuous, and therefore $\lim_{\epsilon\to 0} \Big[H_+(r, \epsilon)-H_-(r, \epsilon)\Big]=0$ for $r<r_h$ trivially. Summing up the two terms, we can rewrite the activation coefficient as: 
\begin{equation}
    I_H=\lim_{\tilde{r}\to r_h}\Big[\int_{\tilde{r}}^b dr(r_h-r)^{-4 i \omega_{221}}-\frac{(r_h-\tilde{r})^{1-4 i \omega_{221}}}{1-4 i \omega_{221}}\Big] \, ,
\end{equation}
which is exactly the prescription of the counter-term regularization method. To solidify our conclusions, we verified that such equality holds by computing numerically the integral~\eqref{eq:excit_coeff_non_regul} both with the complex contour and the counter-term regularization method, obtaining the same values up to machine precision.

\section{Additional orbital configurations} %
\label{app:additional_orbits}               %

\begin{figure*}[t]
\captionsetup{justification=centering}
\caption*{$e_0=0.5$}
\includegraphics[width=1.0\textwidth]{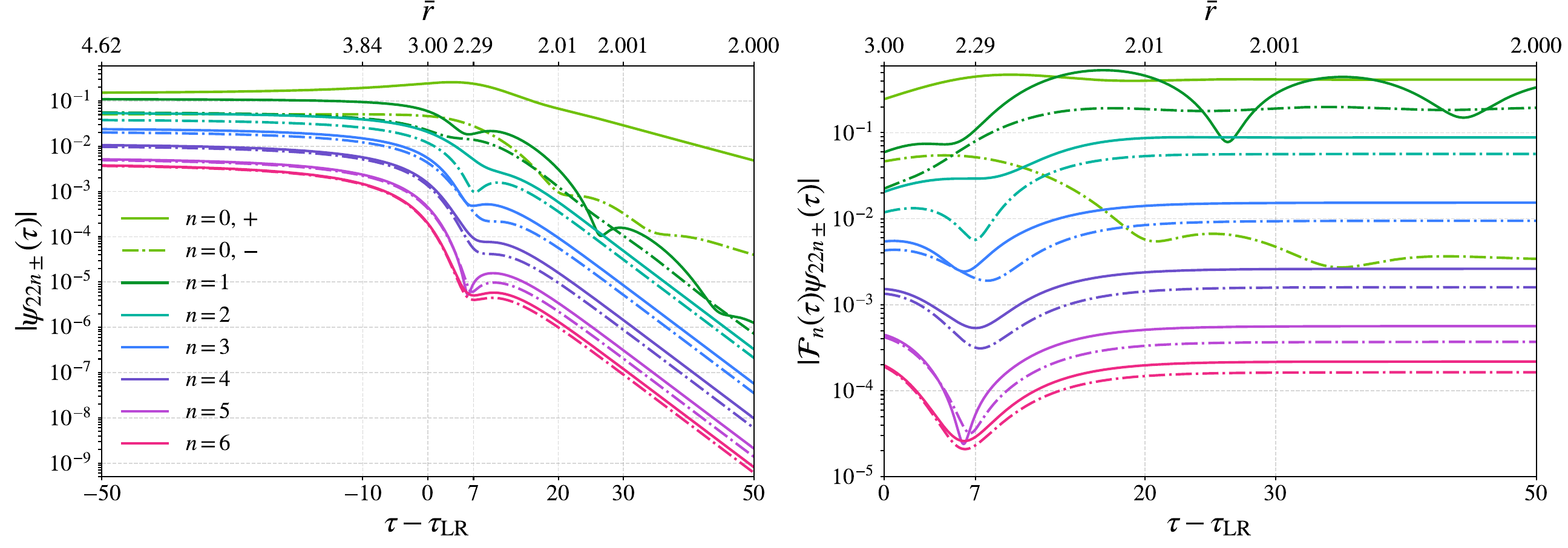}
\includegraphics[width=1.0\textwidth]{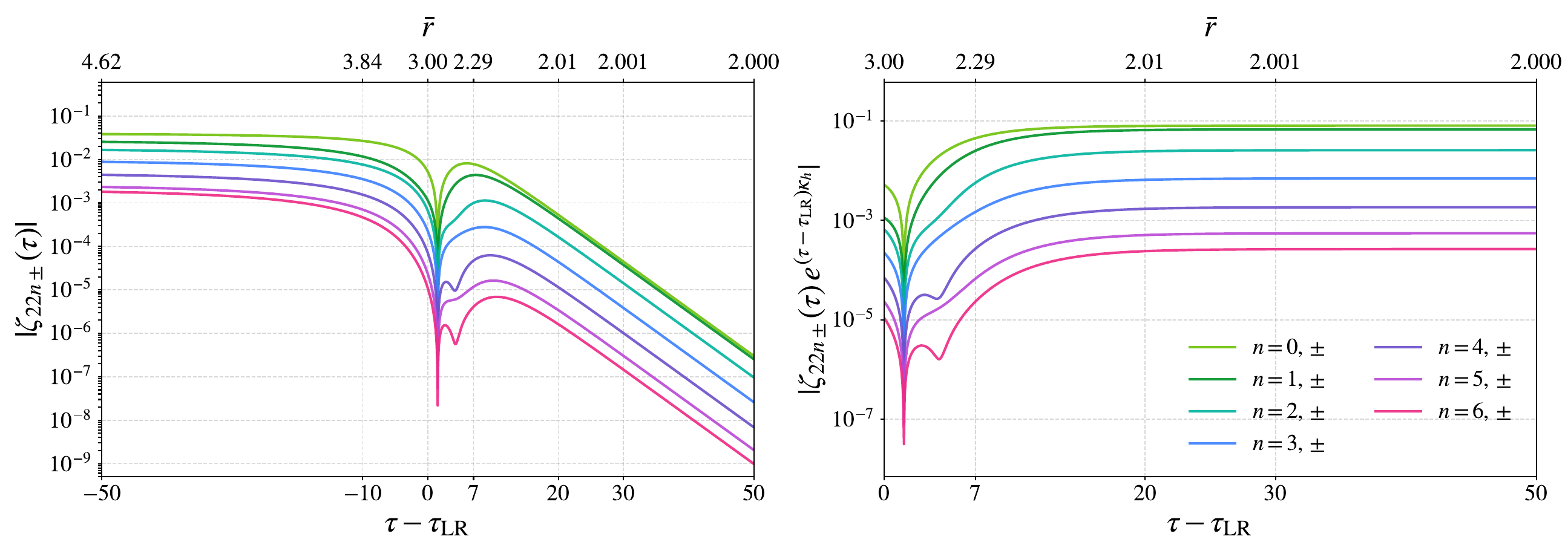}
\caption{Same as Fig.~\ref{fig:EV_clmn_Llmn_ID00mdea}, but for $e_0 = 0.5$.
\label{fig:EV_clmn_Llmn_ID05mdea}}
\end{figure*}
\begin{figure*}[b]
\captionsetup{justification=centering}
\caption*{$e_0=0.9$}
\includegraphics[width=1.0\textwidth]{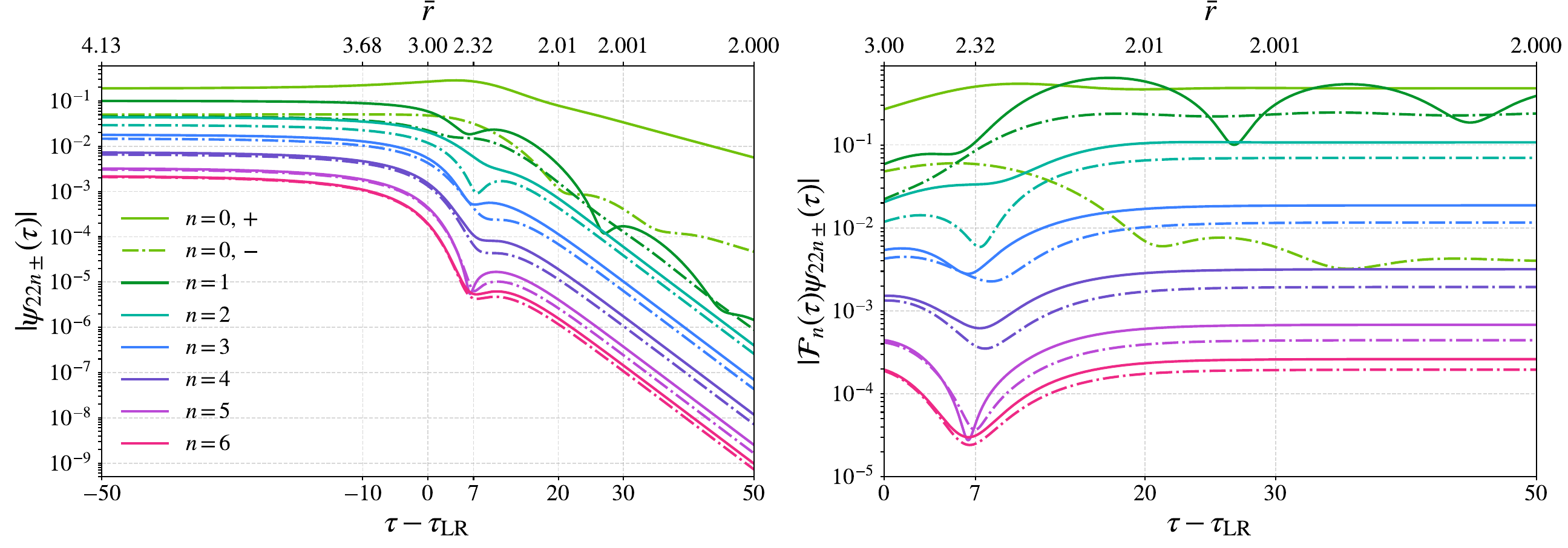}
\includegraphics[width=1.0\textwidth]{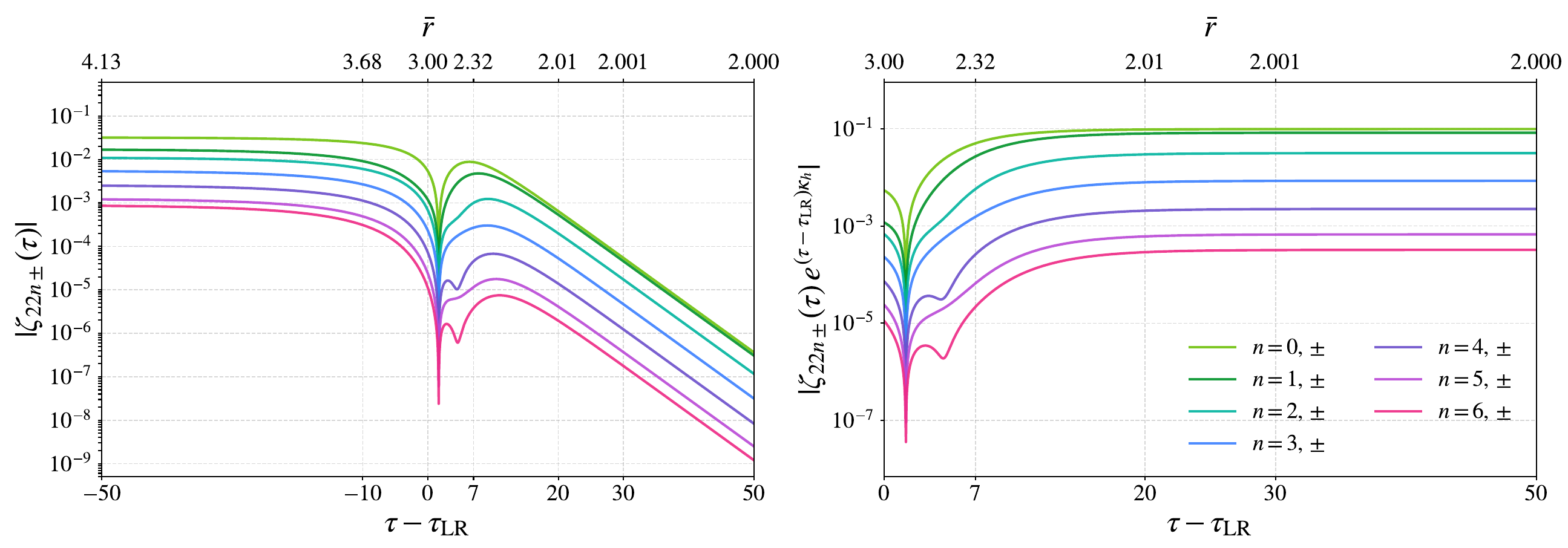}
\caption{Same as Fig.~\ref{fig:EV_clmn_Llmn_ID00mdea}, but for $e_0 = 0.9$.
\label{fig:EV_clmn_Llmn_ID09mdea}}
\end{figure*}

\begin{figure*}[t]
\includegraphics[width=0.49\textwidth]{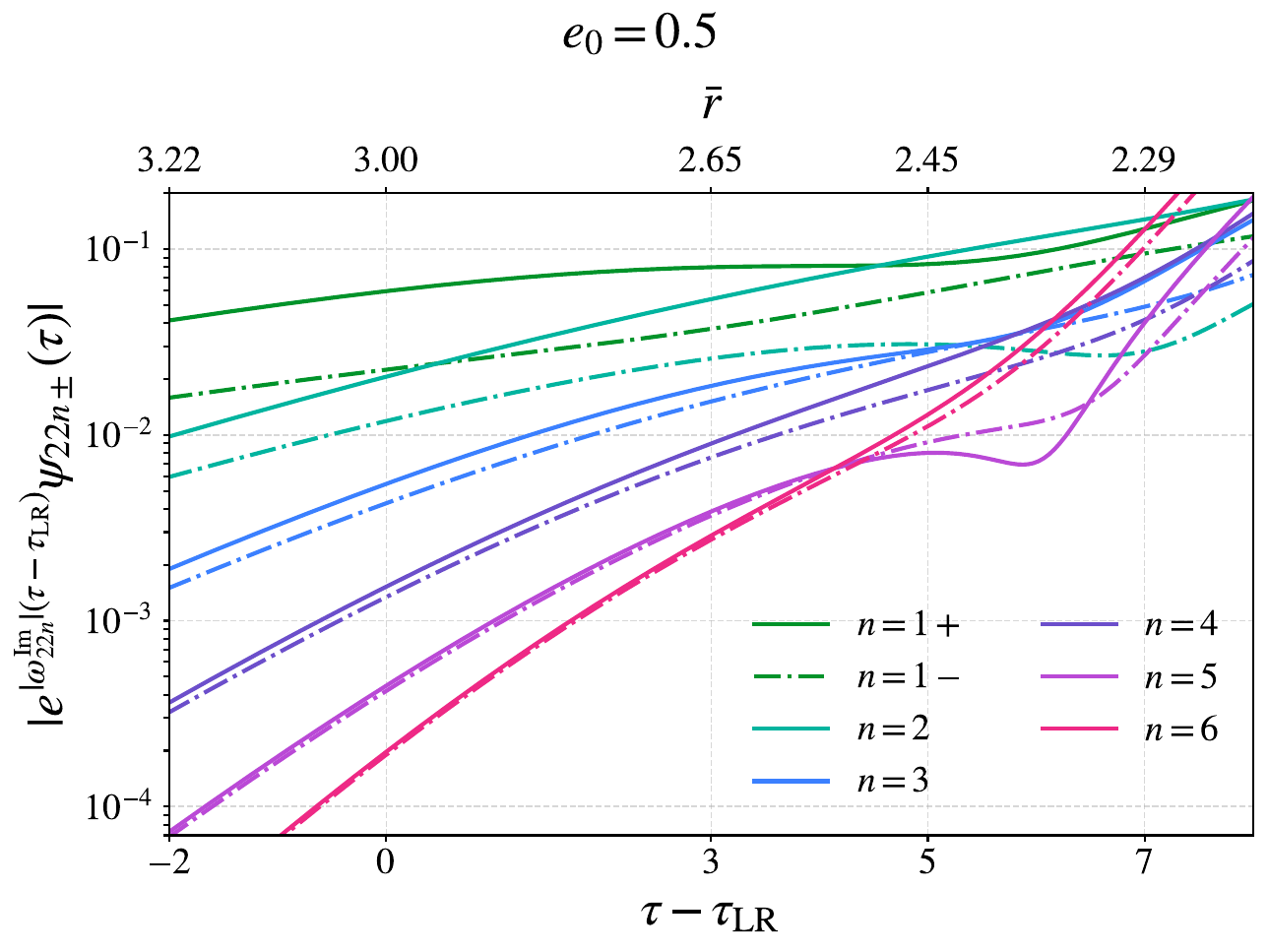}
\includegraphics[width=0.49\textwidth]{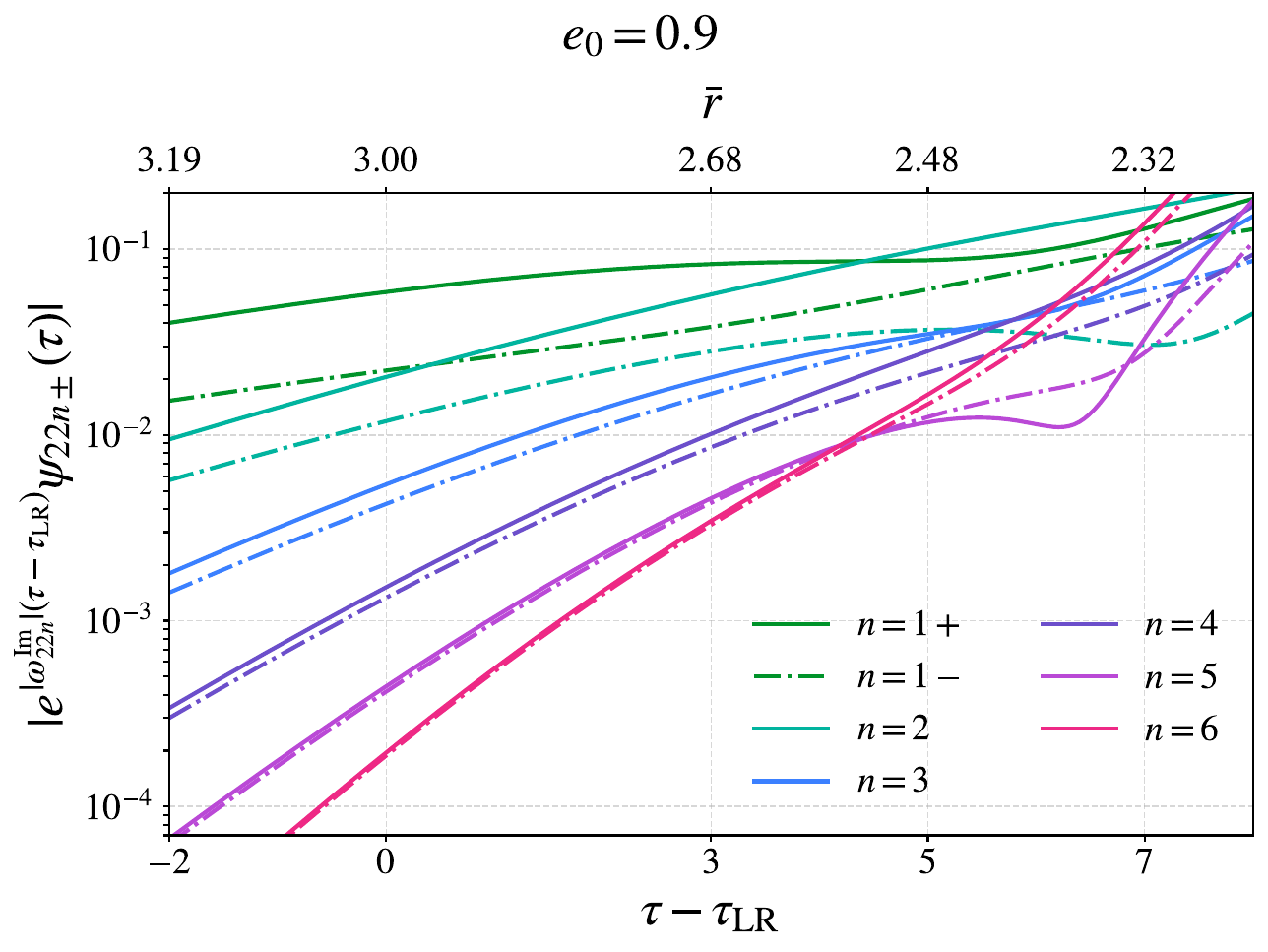}
\caption{
Same as in Fig.~\ref{fig:OVresc_clmn_ID00-01mdea}, but for $e_0=0.5$ (\textit{left}) and $e_0=0.9$ (\textit{right}) of Tab.~\ref{tab:sims_ecc}.
\label{fig:OVresc_clmn_ID05-09mdea}
}
\end{figure*}

The qualitative behavior between redshift factor and QNMs decay seems to not be affected with the initial eccentricity. Figs.~\ref{fig:EV_clmn_Llmn_ID05mdea},~\ref{fig:EV_clmn_Llmn_ID09mdea} and~\ref{fig:OVresc_clmn_ID05-09mdea} analyze eccentric plunges with $e_{0}=0.5$ and $e_{0}=0.9$ of Tab.~\ref{tab:sims_ecc}. In both cases, the redshift factor starts dominating over the QNMs decay for $n>0$ at a retarded time $\tau-\tau_{\rm LR}\approx 7$, in agreement with the quasi-circular case in Fig.~\ref{fig:EV_clmn_Llmn_ID00mdea}. 
The $(221+)$ mode stabilized to a constant amplitude contribution for $\tau-\tau_{\rm LR}\approx 2$, which dominates over the redshift term for $2\lesssim \tau-\tau_{\rm LR}\lesssim 5$. For higher overtones and other mirror modes, the QN behavior cannot be decoupled from the redshift one before the latter dominates the contribution, consistent with the quasi-circular analysis in Fig.~\ref{fig:EV_clmn_Llmn_ID00mdea}.

\section{Simple arguments for redshift terms} %
\label{app:redshift}                          %
%
Redshift terms have been previously identified in the literature as arising from radiation emitted during the plunge phase, when a secondary object approaches the event horizon~\cite{Mino:2008at,Zimmerman:2011dx}. From the viewpoint of a distant observer, the infalling particle never crosses the horizon. Instead, time dilation causes the particle to appear infinitely slowed down, effectively becoming “frozen” near the horizon. As a result, the particle continuously emits redshifted radiation throughout the plunge, which leaks to infinity with an exponentially decaying profile over time. In this appendix, we review a straightforward argument originally proposed by Price, which accounts for the observed exponential decay of signals emitted from the near-horizon region. 
In Ref.~\cite{Price:1971fb}, Price studied the gravitational collapse of a scalar-charged star and showed that, as the star falls inside its gravitational radius, the scalar field $\Phi$ must decay exponentially due to time dilation effects. The key idea behind the argument is that the derivatives of the scalar field must remain finite at the Schwarzschild radius in Kruskal coordinates -- a coordinate system that remains regular across the future horizon. The Kruskal coordinates are
\begin{equation}
    U= -4  e^{-u/4} \ , \ V= -4  e^{-v/4} \, ,
\end{equation}
where $u=t- r_*$, $v=t+r_*$ are null coordinates.
If we impose regularity of $\partial \Phi/ \partial U$, this implies that the field decays exponentially in null coordinates,
\begin{equation}
    \frac{\partial \Phi}{\partial u} = \frac{\partial \Phi}{\partial U}\frac{\partial U}{\partial u}= \frac{\partial \Phi}{\partial U} e^{-u/4} \, , 
\end{equation}
so that $\Phi \approx e^{-u/4}= e^{-(t-r_*)/4}$ as $t \rightarrow \infty$, in agreement with our findings. A similar argument was subsequently used in \cite{Zimmerman:2011dx}, using the Teukolsky formalism. While solutions of the Teukolsky equation that are outgoing at the horizon are divergent and usually considered unphysical, Ref.~\cite{Zimmerman:2011dx} argued that they becomes physical (i.e., well-behaved) when evaluated at the complex frequency of the sub-leading redshift term.

\section{Kerr causality condition} %
\label{app:Kerr_causality}         %

In Kerr, the ingoing solution Eq.~\eqref{eq:u_in_two_terms} at the QNFs can be expanded as follows~\cite{Leaver:1985ax}
\begin{equation}
\begin{split}
\tilde{u}^{\rm in}(\omega,r_*)=\left[(r-r_-)^{3} \left(\frac{r-r_+}{r-r_-}\right)^{2+i\frac{2r_+a}{r_+-r_-}m} \hat{d}(\omega,r)\right] \, \exp\left\lbrace i\omega \left[r_*-\frac{4r_+}{r_+-r_-}\log\left(\frac{r-r_+}{2}\right)+4\log\left(\frac{r-r_-}{2}\right) \right]\right\rbrace  \, ,
\end{split}
\end{equation}
%
%
%
where $a$ is the BH adimensional spin and we have defined
\begin{equation}
\hat{d}(\omega,r)\equiv \sum_{k=0}^{\infty}d_k(\omega)\left(\frac{r-r_+}{r-r_-}\right)^k \, .
\end{equation}
Now $r_*$ is the Kerr tortoise coordinate,
$r_*\equiv r+\frac{2r_+}{r_+-r_-}\log\left(\frac{r-r_+}{2}\right)-\frac{2r_-}{r_+-r_-}\log\left(\frac{r-r_-}{2}\right)$ .
Following the same reasoning of Sec.~\ref{subsec:QNMs_propagation_and_regularity} and assuming an observer at $\mathcal{I}^+$, we can rewrite the causality condition as $t-r_* \geq  \mathcal{C}^{\rm Kerr}(t',r_*')$, with 
\begin{equation}
\mathcal{C}^{\rm Kerr}\equiv t'+r_*'-\frac{4r_+}{r_+-r_-}\log\left(\frac{r-r_+}{2}\right)
+4\log\left(\frac{r-r_-}{2}\right) \, .
\label{eq:causality_kerr}
\end{equation}
It is straightforward to prove that this returns the Schwarzschild causality condition, Eq.~\eqref{eq:QNMs_causality}, in the limit $a\rightarrow 0$.
The horizon redshift $\kappa_{h}$ of a Kerr BH is equal to
\begin{equation}
\kappa_{h}^{\rm Kerr}\equiv\frac{r_+-r_-}{4r_+}\, .
\label{eq:redshift_factor_Kerr}
\end{equation}
Then, the near-horizon behavior of the causality condition in Eq.~\eqref{eq:causality_kerr} is
\begin{equation}
\mathcal{C}^{\rm Kerr}(t(\bar{r}),\bar{r}_*)\simeq - \frac{1}{\kappa_{h}^{\rm Kerr}}\log\left(\frac{\bar{r}-r_+}{2}\right) \ \ , \ \ \ \bar{r}\rightarrow r_+ \, .
\end{equation}
We therefore expect this condition, together with the source, to give rise to redshift terms, as found in~\cite{Mino:2008at,Zimmerman:2011dx}.
As in the latter references, in the Kerr case we expect the redshift terms to acquire a real part proportional to $\Omega_H$.

\newpage


%

\end{document}